\documentclass[prb,
showpacs,
twocolumn,
floats,
10pt,
aps,
citeautoscript,
longbibliography,
superscriptaddress]{revtex4-2}

\usepackage{placeins}
\usepackage{lipsum}
\usepackage{xcolor}
\usepackage[normalem]{ulem}
\usepackage{comment}
\usepackage{tabularx}
\usepackage{graphicx}
\usepackage{dcolumn}
\usepackage{bm}

\usepackage{bbm}
\usepackage{blindtext}
\usepackage{graphics}
\usepackage{verbatim}   
\usepackage{amsfonts}
\usepackage{amsmath}
\usepackage{amssymb}
\usepackage{adjustbox}
\usepackage{microtype} 
\allowdisplaybreaks 
\usepackage{xspace} 
\usepackage{xparse} 
\usepackage{multirow} 
\usepackage{tabstackengine}
\setstackEOL{\cr}

\newcommand{\Eq}[1]{Eq.~\eqref{#1}}

\def\me{\mathrm{e}}   

\newcommand*{\ndots}{\kern-0.075em.\kern-0.05em.\kern-0.05em.}  
\newcommand*{\nidots}{.\kern-0.05em.\kern-0.05em.} 
\newcommand*{\ncdots}{\kern-0.15em\cdot\kern-0.2em\cdot\kern-0.2em\cdot\kern-0.15em}   

\NewDocumentCommand{\doubleI}{O{}}{\mathbbm{1}_{#1}}
\NewDocumentCommand{\doubleIb}{O{}}{{\bar{\mathbbm{1}}_{#1}}}
\NewDocumentCommand{\doubleIk}{O{}}{\mathbbm{1}^\ks_{\! #1}}
\NewDocumentCommand{\doubleId}{O{}}{\mathbbm{1}^\ds_{\! #1}}
\NewDocumentCommand{\doubleIp}{O{}}{\mathbbm{1}^\ps_{\! #1}}
\NewDocumentCommand{\doubleV}{O{}}{\mathbbm{V}_{\! #1}}
\NewDocumentCommand{\doubleVk}{O{}}{\mathbbm{V}^\ks_{\! #1}}
\NewDocumentCommand{\doubleVd}{O{}}{\mathbbm{V}^\ds_{\! #1}}
\NewDocumentCommand{\doubleVp}{O{}}{\mathbbm{V}^\ps_{\! #1}}
\NewDocumentCommand{\doublev}{o}{{\mathbbm{v}_{#1}}}
\NewDocumentCommand{\doubleVb}{o}{{\bar{\mathbbm{V}}_{\! #1}}}
\NewDocumentCommand{\doubleVt}{o}{{\widetilde{\mathbbm{V}}_{\! #1}}}
\NewDocumentCommand{\doubleVh}{o}{\widehat{{\mathbbm{V}}_{\! #1}}}
\NewDocumentCommand{\doubleW}{o}{\mathbbm{W}_{\! #1}}
\NewDocumentCommand{\doubleWk}{o}{\mathbbm{W}^\ks_{\! #1}}
\NewDocumentCommand{\doubleWd}{o}{\mathbbm{W}^\ds_{\! #1}}
\NewDocumentCommand{\doubleWb}{o}{{\bar{\mathbbm{W}}_{\! #1}}}
\NewDocumentCommand{\doubleWt}{o}{{\widetilde{\mathbbm{V}}_{\! #1}}}
\NewDocumentCommand{\doubleWh}{o}{{\widehat{\mathbbm{V}}_{\! #1}}}

\newcommand{\LMUMunich}{Arnold Sommerfeld Center for Theoretical Physics, Center for NanoScience, and Munich Center for Quantum Science and Technology, Ludwig-Maximilians-Universit\"at M\"unchen, 80333 Munich, Germany}

\newcommand{\ITPCologne}{Institute for Theoretical Physics, University of Cologne, 50937 Cologne, Germany}

\makeatletter
\def\l@subsubsection#1#2{}
\makeatother

\usepackage{booktabs,longtable,tabularx,array}

\newcolumntype{Y}{>{\raggedright\arraybackslash}X}

\newcommand{\sym}[3]{\ensuremath{#1} & #2 & \;\;\; #3\\}

\usepackage{multirow}
\usepackage{physics}
\usepackage{hyperref}
\hypersetup{colorlinks=true,breaklinks,linkcolor=blue,urlcolor=blue,citecolor=blue}

\begin{document}

\preprint{}

\title{Composite Fields and Tree Expansions:\\
 A Unified Framework for Renormalized Vertex Decompositions}

\author{Oleksandr Sulyma}
\email{osulyma@uni-koeln.de}
\affiliation{\LMUMunich}
\affiliation{\ITPCologne}

\author{Benedikt Schneider}
\email{Schneider.Benedikt@lmu.de}
\affiliation{\LMUMunich}

\date{August 6, 2026}

\begin{abstract}
One-particle irreducible vertices encode renormalized interactions in quantum field theories, but their practical treatment remains challenging due to their high dimensionality and nontrivial dependence on external variables such as momenta or frequencies. We develop a functional framework that algebraically reorganizes the diagrammatic content of the vertex functions into more efficient building blocks and reveals several established vertex decompositions, such as parquet, single-boson exchange, asymptotic classes and symmetric estimators, as different realizations of a common functional structure. Based on Legendre transforms of the effective action of composite fields, the formalism shows that different vertex representations arise from different choices of composite degrees of freedom. We demonstrate that in our framework higher-order vertices can be systematically obtained via tree expansions, thereby extending the aforementioned decompositions beyond the four-point level. Because our findings are independent of any specific physical setting, the framework applies broadly to quantum field theories in condensed matter physics, particle physics, and beyond.
\medskip
\noindent
\end{abstract}

\maketitle

\begingroup
  \tableofcontents
\endgroup

\section{Introduction}\label{sec:introduction}

\begin{figure*}
    \centering
    \includegraphics[width=1\linewidth]{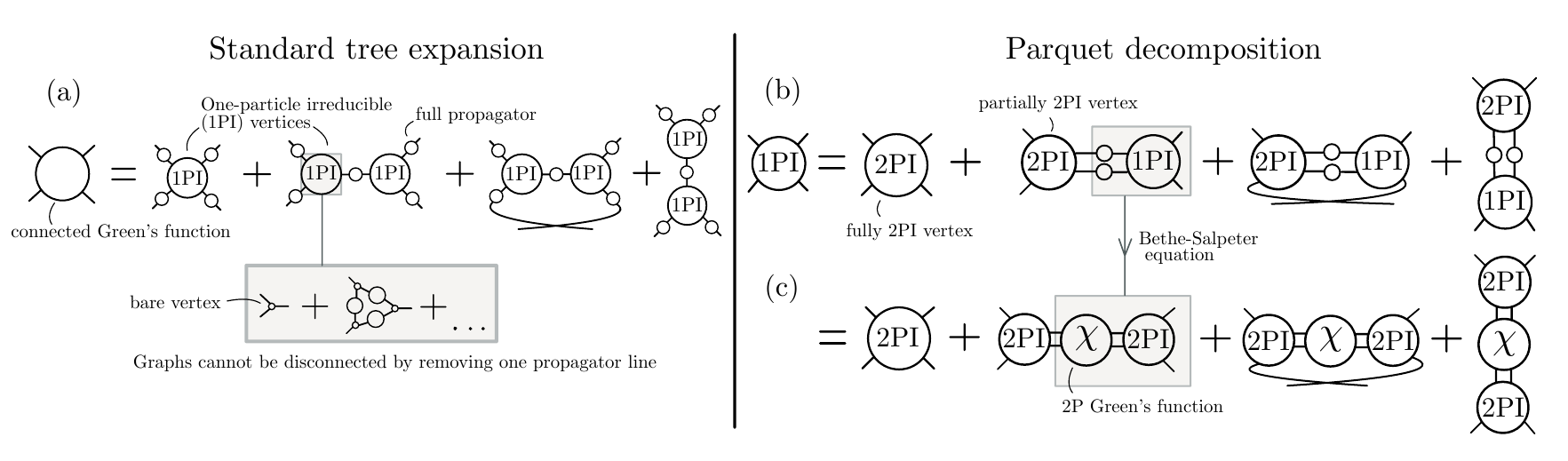}
    \caption{Illustration of the tree structure underlying connected Green's functions (a) and parquet-type decompositions (b,c). In Sec.~\ref{subsec:preliminary}, we show that the standard parquet decomposition (b) can be written as a tree expansion (c) using the Bethe-Salpeter equation. Fermionic signs are not shown.}
    \label{fig:treeParquet}
\end{figure*}

Strongly correlated systems, ranging from electrons in transition-metal oxides to quarks and gluons in Quantum Chromodynamics (QCD), exhibit rich emergent phenomena such as the Mott metal-insulator transition \cite{imada_metal-insulator_1998}, unconventional superconductivity \cite{lee_doping_2006}, and hadron dynamics \cite{gross_50_2023}. These effects arise from the intricate interplay of microscopic degrees of freedom and generally cannot be captured by single-particle descriptions. Instead, the essential physics is encoded in multi-body interaction effects.

In quantum field theory, these interactions are described by renormalized one-particle irreducible (1PI) vertices \cite{Negele1988Quantum,peskin2018introduction}, associated with a fundamental field $\varphi$ (e.g., electrons, quarks, or gluons). Connected $n$-point correlation functions $G^n$ can be systematically expressed in terms of these 1PI vertices and renormalized propagators, forming a hierarchical tree-like structure in which propagation and interaction processes are combined into effective building blocks. (See Fig.~\ref{fig:treeParquet} for the tree expansion of the four-point Green's function $G^4$.)

As shown in the inset of Fig. \ref{fig:treeParquet}, 1PI vertices are defined as the sum of all diagrams that remain connected upon cutting a single propagator line. In weakly interacting regimes, these vertices can be computed perturbatively, for instance, via expansions in the bare interaction using bare or renormalized propagators \cite{baym_self-consistent_1962}. However, such perturbative approaches break down in the strong-coupling regime. In this case, one instead employs self-consistent non-perturbative methods. These include hierarchies of Schwinger-Dyson equations \cite{schwinger_sde_1_1951,dyson_1949,Abrikosov1965, roberts_hadron_1994}, Bethe-Salpeter equations in selected scattering channels \cite{Salpeter1951Bound,Nambu1960Superconductivity}, and functional techniques such as the $n$-particle irreducible effective action \cite{Cornwall1974Action2PI, Berges2004nPI} or the functional renormalization group \cite{MetznerFRG2012, Pawlowski2007FRG}.

While these non-perturbative approaches avoid the explicit combinatorial diagrammatic growth of perturbation theory, they are instead formulated as infinite hierarchies of coupled equations that require truncation. Even with low-order truncations,  the explicit computation and storage of 1PI vertices remain notoriously difficult \cite{li_victory_2019,Krien2019SBE3, kugler_multipoint_2021, lee_multipoint_2021}. Their full momentum and frequency dependence leads to a rapidly increasing numerical cost with vertex order, while strong-coupling regimes are often accompanied by severe numerical instabilities \cite{tam_solving_2013,Schafer2013Divergent,thunstrom_analytical_2018}. 

These challenges have motivated the development of reorganizations of vertex functions into more efficient and physically transparent building blocks. In correlated-electron systems, this has led to approaches such as asymptotic vertex decompositions \cite{wentzell2020high}, single-boson exchange representations \cite{Krien2019SBE2}, and symmetric improved estimators \cite{lihm2024symmetric}, which reduce the complexity of high-dimensional vertices by isolating dominant scattering channels and analytically controlled asymptotic structures. Closely related ideas also appear in non-perturbative QCD, where the operator product expansion separates universal short-distance behavior from infrared dynamics \cite{wilson1969OPE}, and where effective low-energy theories replace complicated quark and gluon multi-point interactions by hadron exchange processes \cite{weinberg1979lagrangians, gies_fRG_mesons,Mitter2015QCD,Braun2016Mesons, Fukushima2022Baryons}. While many of these approaches have proven highly successful, they are often introduced from different perspectives, which obscures their mutual relations and complicates systematic extensions.

The central goal of this work is to provide a functional framework that algebraically reorganizes the diagrammatic content of 1PI vertices into more efficient, robust building blocks, thereby unifying and extending most of the approaches above. The key idea is to reformulate the vertex structure in an extended space of fields that explicitly includes collective degrees of freedom. To this end, we present two main results. The first result establishes a structural reformulation of 1PI vertices in terms of correlations and renormalized interactions of composite fields, while the second provides a compact representation of the 1PI effective action in terms of low-order Green’s functions:

First, we show that the 1PI vertices themselves admit an exact tree expansion in terms of auxiliary fields $\psi^\bullet$, where the lines correspond to propagators of these auxiliary fields, while the vertices describe their interactions with the fundamental field $\varphi$. The fields $\psi^\bullet$ represent collective, \emph{composite} degrees of freedom and are defined in terms of expectation values of products of the fundamental fields $\varphi$. Their renormalized interactions, specifically those entering the tree expansion of 1PI vertices, are generated by functional derivatives of the composite effective action $\Gamma[\bar\varphi, \psi^{\bullet}]$. This functional is defined via a Legendre transform such that its stationarity condition, $\tfrac{\delta\Gamma[\bar\varphi,\psi^\bullet]}{\delta \psi^\bullet}=0$, determines the physical configuration $\psi^\bullet = \psi^\bullet_{\bar\varphi}$ for a fixed background $\bar\varphi$.

The key insight that makes such a tree expansion in terms of the composite fields possible is an inverse Legendre transformation of the composite effective action $\Gamma[\bar\varphi, \psi^\bullet]$,
\begin{align}\label{eq:1PI_via_Gamma}
    \Gamma^\mathrm{1PI}[\bar\varphi]=\Gamma[\bar\varphi, \psi^\bullet_{\bar\varphi}],
\end{align}
where $\Gamma^{\mathrm{1PI}}[\bar\varphi]$ is the standard 1PI effective action whose functional derivatives are exactly 1PI vertex functions \cite{Negele1988Quantum,Berges2004nPI,kopietz2010introduction}. Differentiating Eq.~\eqref{eq:1PI_via_Gamma} $n$ times with respect to $\bar{\varphi}$ then yields the tree expansions of 1PI vertices in terms of composite field propagators and mixed vertices defined by functional derivatives of $\Gamma[\bar\varphi, \psi^\bullet]$.

Second, we show that, for theories whose bare action contains interactions of order no higher than four, the 1PI effective action has the remarkably concise form:
\begin{align}\label{eq:1PIfromGammaIntroduction}
     {\Gamma^{\text{1PI}}[\bar{\varphi}]} &= S[\bar{\varphi}]+{\Omega[{G^2_{\bar\varphi},G^3_{\bar{\varphi}}}]},\\
  \label{eq:OmegafromLambdaIntroduction}
    {\Omega[G^2,G^3]} &=\Lambda[{G^2,G^3}]-G^2\frac{\delta \Lambda}{\delta G^2}   -G^3\frac{\delta \Lambda}{\delta G^3} ,
\end{align}
where $S$ is the classical action and $\Omega$ is the Legendre transformation of the 3PI generalization of the Luttinger-Ward functional  $\Lambda[{G^2,G^3}]$ detailed in the main text below  (see Eqs.~\eqref{eq:classicalAction}--\eqref{eq:OmegafromLambda}).
Equation~\eqref{eq:1PIfromGammaIntroduction} shows that all quantum corrections to the bare interactions can be formally expressed via $\Lambda[G^2, G^3]$.

\begin{figure*}[t]
    \centering
     
    \includegraphics[width=\linewidth]{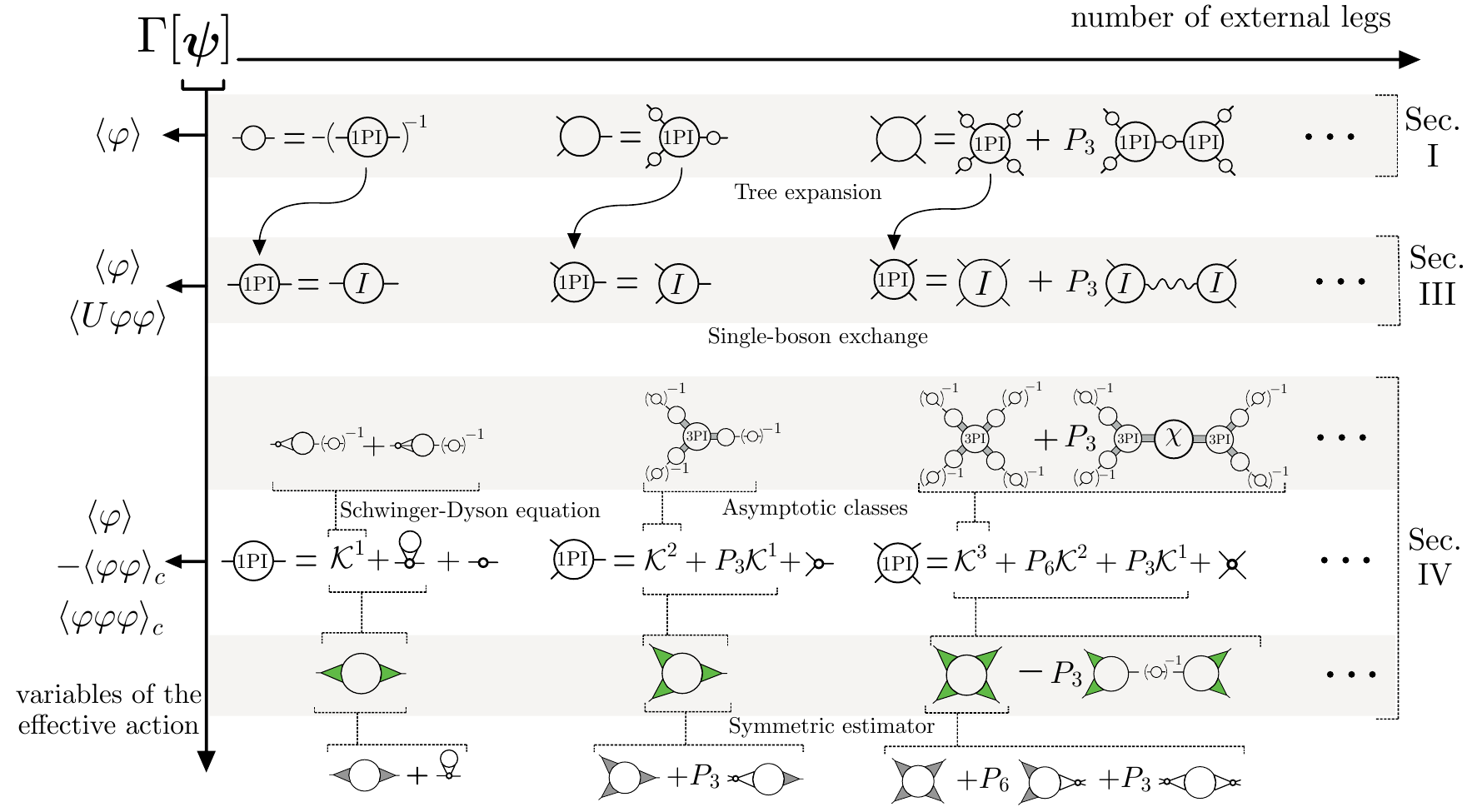}
    
    \caption{Overview of the main results of the manuscript. Rows show different tree expansions of $n$-point objects, with $n$ increasing across the columns. While the structure of the decomposition remains the same, the details depend on the choice of independent variables $\boldsymbol{\psi}$ of the effective action $\Gamma[\boldsymbol{\psi}]$, which are given on the left.
    Row 1: Tree expansion of connected Green's functions in terms of 1PI vertices (Sec.~\ref{subsec:1PImethods}).
    Row 2: Tree expansion of 1PI vertices in terms of interaction-irreducible vertices, yielding the single-boson exchange decomposition (Sec.~\ref{subsec:ExchangeOfCompositeParticles}).
    Row 3: Tree expansion of asymptotic classes in terms of 3PI vertices (Sec.~\ref{subsec:asymptoticClasses}). 
    Row 4: Decomposition of 1PI vertices in terms of asymptotic classes (Sec.~\ref{subsec:asymptoticClasses}). 
    Row 5: Tree expansion of the non-bare part of 1PI vertices in terms of Green's functions of composite fields, yielding symmetric improved estimators (Sec.~\ref{subsec:estimators}). Diagrammatic notation is explained in Figs.~\ref{fig:1PIEffectiveAction}, \ref{fig:compositeFieldEffectiveAction}, \ref{fig:susceptibilities}, \ref{fig:SBEformalism} and \ref{fig:compositeFieldGreensFunctions}.}
    \label{fig:results}
\end{figure*}

Together, Eqs. \eqref{eq:1PI_via_Gamma} and \eqref{eq:1PIfromGammaIntroduction}  allow for a unified interpretation of several vertex constructions that are widely used in correlated-electron systems, while naturally extending to more general interacting field theories such as QCD. In particular, parquet \cite{Dominicis1964EntropyII} and single-boson exchange (SBE) \cite{Krien2019SBE2}  decompositions, high-frequency asymptotic parameterizations \cite{wentzell2020high}, and symmetric improved estimators \cite{kaufmann2019symmetric, lihm2024symmetric}, originally developed from distinct theoretical and practical considerations, emerge as special cases of the same underlying tree expansion for different choices of composite fields $\psi^\bullet$ as illustrated in Fig.~\ref{fig:results}. While these representations address different aspects of vertex calculations, such as computational scaling, high-frequency behavior, or numerical stability, the resulting equations share an identical tree-like structure.

The $n$PI effective action formalism \cite{Dominicis1964EntropyI, Berges2004nPI}, for which the composite fields $\psi^\bullet$ are identified as the connected Green's functions $G^2,\dots,G^n$, represents a special case of our formalism. As shown by Eckhardt et al. \cite{Eckhardt2023functional}, the 2PI effective action can be used to reproduce the standard parquet decomposition, which was historically derived only through combinatorial arguments \cite{Dominicis1964EntropyII, Vasiliev1998Functional}. The tree-expansion form of these parquet equations, which also allows for straightforward generalizations to higher vertex orders, has been previously overlooked. In Fig.~\ref{fig:treeParquet}, we show diagrammatically how the standard parquet decomposition can be rewritten into a tree-expansion form by means of the Bethe-Salpeter equation. 

The SBE decomposition naturally expresses the four-point vertex in terms of lower-order blocks and is derived by grouping vertex contributions based on interaction irreducibility \cite{Krien2019SBE1,Krien2019SBE2,Krien2019SBE3,Krien2020SBE4,Krien2020SBE5,Krien2021SBE6}.  It is equivalent to a particular type of Hubbard-Stratonovich transformation (see Sec.~\ref{subsec:ExchangeOfCompositeParticles}) and is recovered within our framework by choosing local bilinears as composite fields. This construction holds for general theories and naturally extends to higher-order vertices, such as six-point functions. Moreover, the formalism can be generalized to include both bilinear and trilinear composite fields, thereby accounting for the simultaneous exchange of collective bosonic modes and fundamental particles, as encountered, for example, in the $GW$ approximation of the self-energy and QCD-inspired truncations of the quark-gluon vertex \cite{Mitter2015QCD, Eichmann2016Baryons}. While the boson-exchange decomposition leads to a computationally lighter alternative to 2PI and parquet schemes, such trilinear extension plays a role analogous to 3PI approaches \cite{RibicPathIntegral, Williams2016QCD}.

The high-frequency asymptotic parametrization of vertices \cite{wentzell2020high} exploits the observation that their high-frequency structure is governed by diagrammatic contributions with a simpler dependence on external bare vertices. By decomposing the full vertex into distinct asymptotic components (or \emph{asymptotic classes}), this approach captures dominant large-frequency behavior in terms of lower-dimensional objects, significantly reducing the numerical complexity and therefore enabling an efficient and controlled treatment of vertex functions in diagrammatic many-body methods \cite{li_victory_2019, Tagliavini2019MfRG, Schafer2021Hubbard, Ritter2022fRG, Ge2024RFQFT, ritz_keldyshqft_2024, lihm2025finite}.

Meanwhile, symmetric improved estimators \cite{kaufmann2019symmetric, lihm2024symmetric} address the difficulty that conventional vertex estimators require delicate subtractions of disconnected parts and amputations of external legs, which can lead to large numerical errors and instabilities \cite{hafermann2012improved, Hafermann2014Estimator, Gunacker2016Estimator}. These estimators express vertices in terms of higher-order Green's functions \cite{hafermann2012improved, kaufmann2019symmetric, lihm2024symmetric} and can be viewed as a many-body generalization of a classic result from quantum electrodynamics (QED)~\footnote{This result can be found in many QFT textbooks; for example, see Sec. 14.8.3 in Ref.~\cite{Schwartz2014Quantum}.}: when computing the connected part of the scattering matrix involving photons, one can use a current operator in place of the product of the inverse bare propagator (with an external index) and a photon field operator in the connected correlation function.

Crucially, both the high-frequency asymptotic parametrization and the symmetric improved estimators are naturally recovered within our formalism by choosing the composite fields to be $\psi^\bullet = (G^2,G^3)$ and exhibit an underlying tree structure as presented in Fig.~\ref{fig:results}.

The remainder of this paper is organized as follows. The introduction continues in the next subsection with a review of the standard tree expansion. In Sec.~\ref{sec:compositeFieldMethods}, we use Eq.~\eqref{eq:1PIfromGammaIntroduction} to derive self-energy estimators, Bethe-Salpeter equations, as well as parquet and asymptotic class decompositions of the four-point vertex. The subsequent sections develop these results in different ways and can be read independently. With more explicit notation, Sec.~\ref{sec:1PIEffectiveActionViaCompositeFields} presents self-consistent schemes (inspired by the SBE formalism) that allow one to renormalize 1PI vertices while keeping the computational scaling under control. Sec.~\ref{sec:SymmetricEstimators} follows a different path to study 1PI vertices, using Eq.~\eqref{eq:1PIfromGammaIntroduction}: we show how to express asymptotic classes and 1PI vertices in terms of Green's functions of composite fields. 
For convenience, a table of symbols can be found in Appendix~\ref{App:TableOfSymbols}.

Throughout this article we use a superfield notation, so all results can be applied to any fermionic, bosonic or mixed theory. However, to make the derivations simpler to follow, we choose to do the calculations using a fundamental field with fixed statistics, except for in Secs.~\ref{subsec:ExchangeOfCompositeParticles}, \ref{subsec:calculationEffectiveAction} and App.~\ref{app:beyond_SBE}.

\subsection{1PI effective action formalism}
\label{subsec:1PImethods}
Analogous to how bare interactions are defined through derivatives of the classical action, renormalized interactions are determined by the quantum effective action $\Gamma^\mathrm{1PI}$. Perturbatively, it is given by a sum of all connected one-particle irreducible (1PI) diagrams with external lines contracted by $\bar{\varphi}^a$, where one-particle irreducible diagram means that one cut of any internal line cannot make the diagram disconnected. In this section, we adopt the non-perturbative definition of the effective action via a Legendre transform and review a standard result of the 1PI formalism: the tree expansion, which expresses arbitrary connected Green’s functions in terms of derivatives of the 1PI effective action. The material presented here can be found in many textbooks, such as \cite{Negele1988Quantum, kopietz2010introduction, peskin2018introduction, DeWitt1966Fields}.

Consider a theory described by the classical action $S[\varphi]$, where the field $\varphi$ has fixed statistics. We explicitly retain a factor $\zeta$ with $\zeta = 1$ for bosonic systems and $\zeta = -1$ for fermionic systems. The generating functional  $W$ of the connected Green's functions is defined as
\begin{equation}\label{eq:GeneratingFunctionalW}
    \me^{-W[J]}=\int \mathrm{D}\varphi \, \me^{-S[\varphi]-J_a\varphi^a}.
\end{equation}
Here, the Einstein summation convention is used for each contraction of one lower and one upper index. We use DeWitt's notation \cite{DeWitt1966Fields}: each index $a$ is a multi-index that can consist of multiple discrete and continuous quantum numbers. For example, $J_a\varphi^a = \sum_i \int \mathrm{d}^4x J_i(x)\varphi^i(x)$. 
From $W$ (Eq.~\eqref{eq:GeneratingFunctionalW}) we can calculate connected correlation functions of the fundamental field $\varphi^a$ via functional derivatives:
\begin{align}\label{eq:correlationFunction}
G^{a_1...a_n}\equiv\frac{\delta}{\delta J_{a_1}}...\frac{\delta}{\delta J_{a_n}}W[J]=(-1)^{n-1}\langle\varphi^{a_1}...\varphi^{a_n}\rangle_c,
\end{align}
where the average of an arbitrary functional $A[\varphi]$ built out of the $\varphi^a$ is defined as
\begin{align}\label{eq:Def_Expval}
\langle{A[\varphi]}\rangle\equiv\me^W\int \mathrm{D}\varphi\, A[\varphi] \me^{-S[\varphi]-J_a\varphi^a},
\end{align}
and a subscript $c$ in $\langle \dots \rangle_c$ indicates that only the connected part is taken.
Calculating the right side of Eq.~\eqref{eq:correlationFunction} for $G^{ab}$, we obtain
\begin{align}\label{eq:2pCorrelationFunction}
G^{ab}=\bar\varphi^a\bar\varphi^b    -\langle{\varphi^a\varphi^b}\rangle,
\end{align}
where $\bar{\varphi}^a\equiv \frac{\delta W}{\delta J_a}=\langle\varphi^a\rangle$.

\begin{figure}
    \centering
    
    \includegraphics[width=1\linewidth]{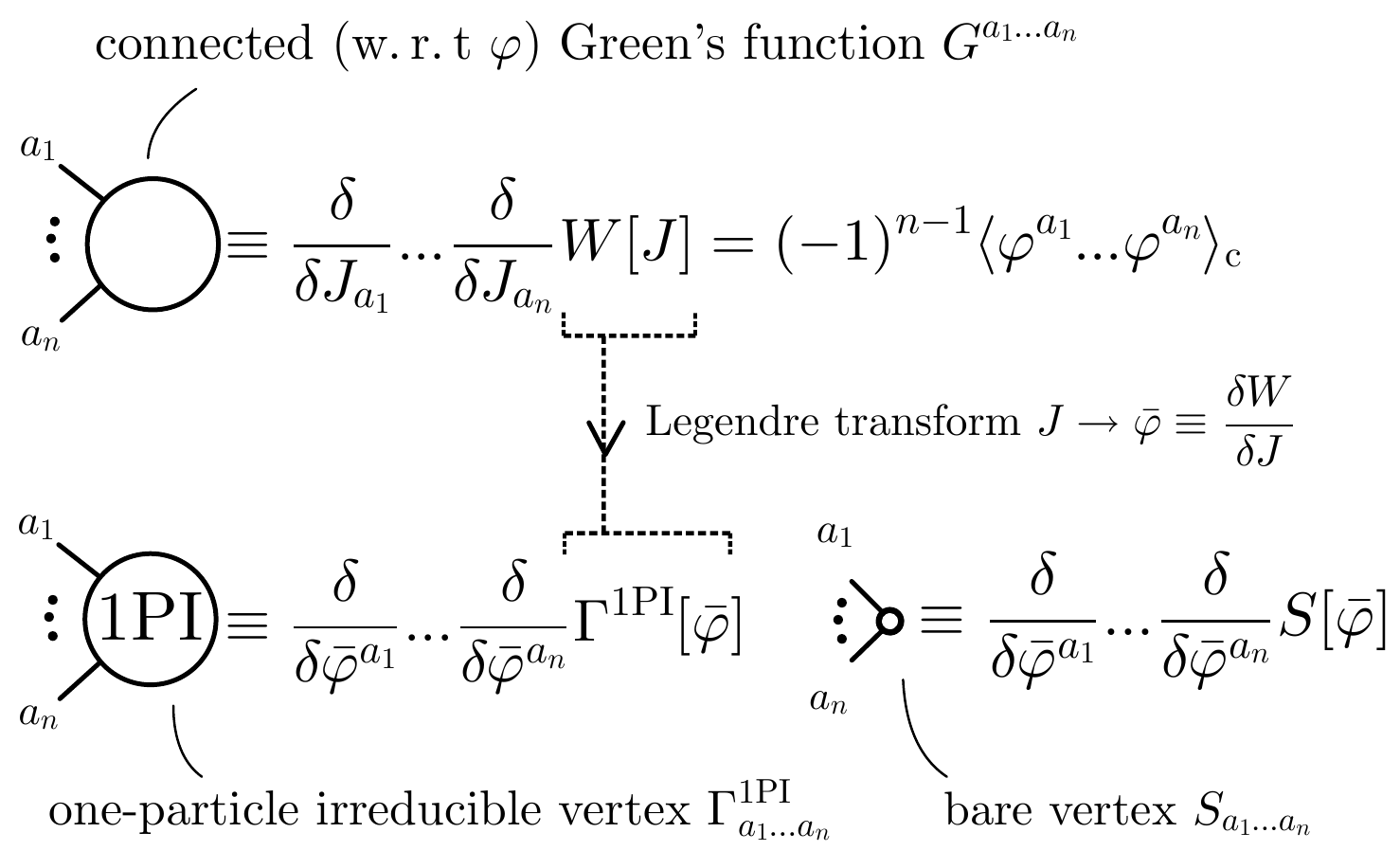}
    \caption{Diagrammatic definitions of the connected Green's function $G^{a_1...a_n}$ (see Eq.~\eqref{eq:correlationFunction}), the 1PI vertex $\Gamma_{a_1...a_n}^\mathrm{1PI}$ (see Eq.~\eqref{eq:1PIvertexDefinition}) and the bare vertex $S_{a_1\ldots a_n}$ (see Eq.~\eqref{eq:classicalAction}).}
    \label{fig:1PIEffectiveAction}
\end{figure}
The 1PI effective action $\Gamma^\mathrm{1PI}[\bar{\varphi}]$ is defined as the Legendre transformation of $W$ with respect to the source $J_a$ \footnote{We could also use $W'=-W$ and define Green's functions as derivatives of $W'$. The definition $\Gamma^{\mathrm{1PI}}=J_a\bar\varphi^a-W'$ then leads to additional minus signs in the tree expansion for our main results (Fig.~\ref{fig:results}).},
\begin{align} \label{eq:1PI_Definition}
 \Gamma^\mathrm{1PI} [\bar{\varphi}] =  W - J_a \bar{\varphi}^a,
\end{align}
where $J_a$ as a functional of $\bar{\varphi}^a$ is the solution of $\frac{\delta W}{\delta J_a}=\bar{\varphi}^a$.
The fundamental result of the 1PI effective action formalism is that any connected correlation function $G^{a_1...a_n}$ of order $n>2$ can be written as a sum of tree diagrams, where the internal lines  are one-particle  propagators $G^{a_1a_2}$ and the vertices are functional derivatives of $\Gamma^\mathrm{1PI}$ (called 1PI vertices). This result is the so-called \textit{tree expansion}. Although the derivation of the tree expansion is textbook knowledge (see \cite{kopietz2010introduction}, for example), we include it here to emphasize the recursive structure that generates the expansion and that we use in the next sections.

We start with the equation of motion that follows from  the definition of $\Gamma^\mathrm{1PI}$ (Eq.~\eqref{eq:1PI_Definition}),
\begin{align} \label{eq:1PI_Motion}
\frac{{\delta}}{\delta \bar{\varphi}^a}\Gamma^\mathrm{1PI}=-\zeta J_a,
\end{align} 
where $\zeta$ comes from commuting $J_a\bar\varphi^a=\zeta\bar\varphi^aJ_a$. The general 1PI vertex is defined as
\begin{align}\label{eq:1PIvertexDefinition}
{\Gamma^\mathrm{1PI}_{a...b}}\equiv\frac{\delta}{\delta \bar{\varphi}^a}...\frac{\delta}{\delta \bar{\varphi}^b}\Gamma^\mathrm{1PI}.    
\end{align}
Note that the order of indices is the same on both sides. Diagrammatic definitions of the bare and 1PI vertices as well as the connected Green's function are shown in Fig.~\ref{fig:1PIEffectiveAction}.
The transformation matrix $\frac{\delta J_a}{\delta\bar\varphi^b}$ and its inverse  $\frac{\delta\bar\varphi^a}{\delta J_b}$ (in which $\bar\varphi^a$ is a functional of $J_b$) give the identity
\begin{align}
    \frac{\delta \bar{\varphi}^b}{\delta J_a}\frac{\delta J_c}{\delta \bar{\varphi}^b}=\delta^a_c,
\end{align}
 where $\delta_{c}^a$ is a product of Kronecker-deltas for discrete indices and Dirac-deltas for continuous indices. Using $G^{ab}=\frac{\delta \bar{\varphi}^b}{\delta J_a}$ and the equation of motion \eqref{eq:1PI_Motion}, the identity becomes
\begin{align}\label{eq:1PI_Invertibility}
    G^{ab}{\Gamma^\mathrm{1PI}_{bc}}=-\zeta\delta^a_c.
\end{align}
 As another consequence of $G^{ab}=\frac{\delta \bar{\varphi}^b}{\delta J_a}$, the chain rule $\frac{\delta}{\delta J_a}=\frac{\delta\bar\varphi^b}{\delta J_a}\frac{\delta}{\delta\bar{\varphi}^b}$ can be written as
\begin{align}\label{eq:1PI_Jderiv}
    \frac{\delta}{\delta J_a}=G^{ab}\frac{\delta}{\delta\bar{\varphi}^b}. 
\end{align}

\subsubsection*{Tree expansion}
Now, let us find expressions for higher-order connected correlation functions $G^{a_1...a_n}$ in terms of 1PI vertices. To emphasize the recursive structure of the derivation, it is sufficient to consider the bosonic case. Therefore, we set $\zeta=1$ for the rest of this subsection. For a general derivation of the tree expansion, including all possible signs and indices, see Chapter 6.2.2 of Ref.~\onlinecite{kopietz2010introduction}.

It will be useful to further condense our notation and write, instead of the sequence of indices $a_1...a_n$, only their number $n$ (and omit them completely for $\varphi^a$ and $G^{ab}$), so, in particular, $G^{ab}{\Gamma^\mathrm{1PI}_{bc}} = G\Gamma^\mathrm{1PI}_2$. To reduce ambiguity, we place contracting indices as close together as possible. For example, $ (G)^2\Gamma^{\mathrm{1PI}}_3G$ stands for $G^{aa'}G^{bb'}\Gamma^{\mathrm{1PI}}_{a'b'c'}G^{c'c}$ \footnote{Here we implicitly used that the expression is connected, so $G^{ab}G^{b'a'}\Gamma^{\mathrm{1PI}}_{a'b'c'}G^{c'c}$ is not possible.}. Equations without indices effectively describe a zero-dimensional system. However, recovering the proper index structure in the end is straightforward.

This notation allows us to write Eq.~\eqref{eq:correlationFunction} for $n>2$ as  
\begin{align}\label{eq:correlationFunction_condensed}
    G^n=\Big (\frac{\delta}{\delta J_1} \Big )^{n-2}G=\Big((-\Gamma^\mathrm{1PI}_2)^{-1}\frac{\delta}{\delta\bar\varphi}\Big)^{n-2}(-\Gamma^\mathrm{1PI}_2)^{-1},
\end{align} 
where for the second equality we used Eq.~\eqref{eq:1PI_Jderiv} and 
\begin{align}\label{eq:1PI_tree_expansion_g}
    G=(-\Gamma^\mathrm{1PI}_2)^{-1}
\end{align} (as follows from Eq.~\eqref{eq:1PI_Invertibility} above for $\zeta=1$). For $n=3$ one finds
\begin{align}\label{eq:treeExpansion_G^3}
G^3 = (G)^3\Gamma^{\mathrm{1PI}}_3.
\end{align} 

Apply Eq.~\eqref{eq:1PI_Jderiv} one more time to get the four-point correlation function
\begin{align}\label{eq:treeExpansion_G^4}
G^4=(G)^4\Gamma^\mathrm{1PI}_4+P_3(G)^2\Gamma^\mathrm{1PI}_3G\Gamma^\mathrm{1PI}_3(G)^2,
\end{align}
where $P_3$ indicates three terms coming from the derivative of $(G)^3$. More specifically, $P_n$ denotes a sum over all $n$ distinct permutations of the external indices in the associated term (with a factor $\zeta$ inserted if the permutation involves an interchange of an odd number of indices). As the simplest example, consider $P_2\varphi^a\varphi^b=\varphi^a\varphi^b+\zeta\varphi^b\varphi^a$. Although there are $4!$ possible permutations of the four external indices in Eq.~\eqref{eq:treeExpansion_G^4}, symmetries of the tensors involved reduce this number to only three distinct terms which are often referred to as \textit{channels}, sometimes labeled by the Mandelstam variables $s$, $t$ and $u$ \cite{peskin2018introduction}.

\begin{figure}
    \centering
    
    \includegraphics[width=1\linewidth]{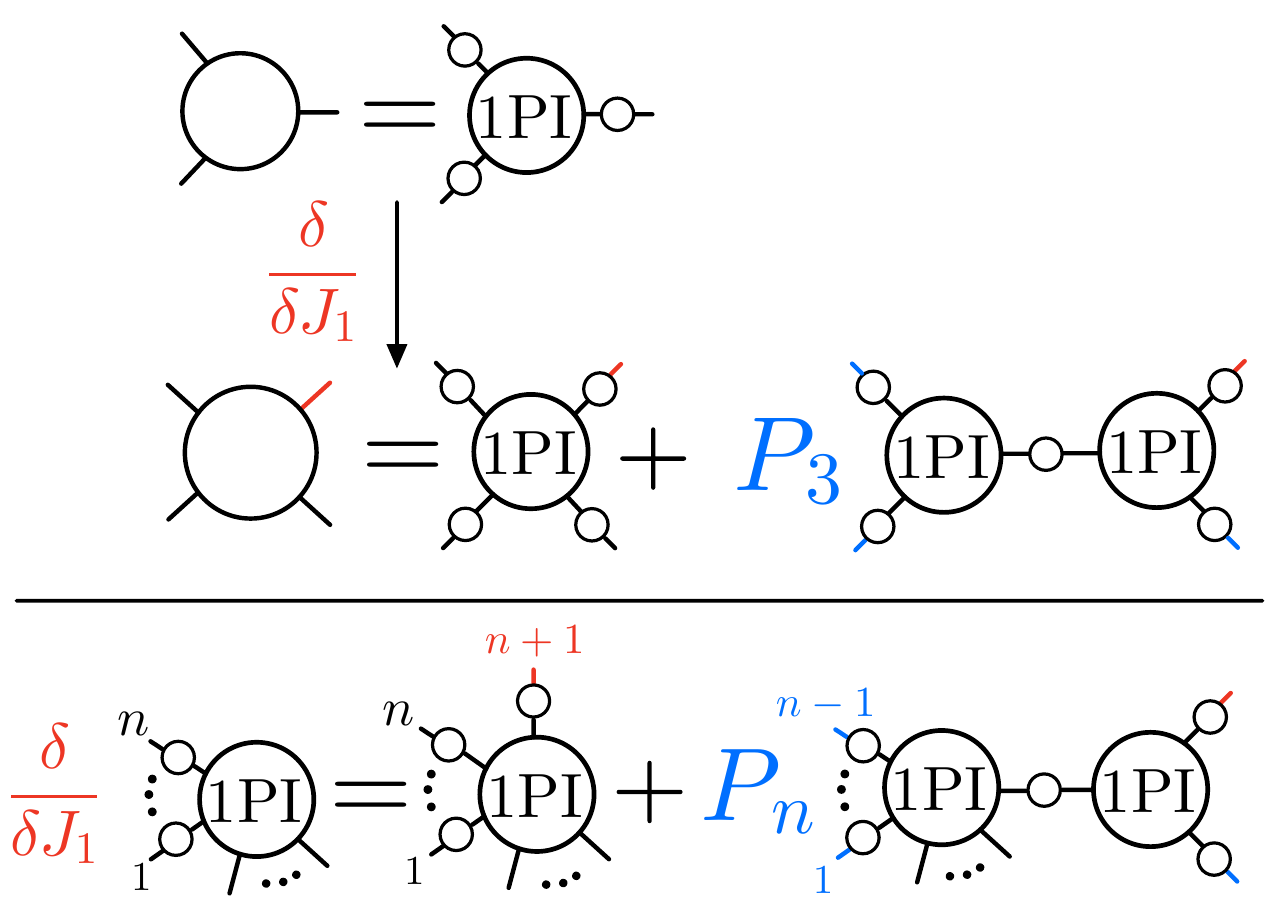}
    \caption{Tree expansions of the connected Green’s functions $G^3$ and $G^4$ (see Eqs.~\eqref{eq:treeExpansion_G^3} and \eqref{eq:treeExpansion_G^4}), using the recursive rule in Eq.~\eqref{eq:treeExpansion_rule}.}
    \label{fig:1PItreeExpansion}
\end{figure}

As shown in Fig.~\ref{fig:1PItreeExpansion}, Eqs.~\eqref{eq:treeExpansion_G^3} and \eqref{eq:treeExpansion_G^4} can be represented as tree diagrams where an open circle with two legs denotes $G$ and a circle with $n$ legs, labeled ``1PI'', represents the vertex $\Gamma^\mathrm{1PI}_n$.
To perform further differentiations with respect to $J_1$, it is convenient to employ the following recursive rules: the derivative of an internal propagator (i.e., one without external indices) yields $\frac{\delta}{\delta J_1}G=(G)^3\Gamma^\mathrm{1PI}_3$ (recall Eq.~\eqref{eq:treeExpansion_G^3})
and the derivative of a product $(G)^n \Gamma^\mathrm{1PI}_{n+m}$ with $n$ external indices in $(G)^n$ and $m$ internal indices (i.e., ones that contract with internal propagators) gives
\begin{align} \frac{\delta}{\delta J_1} [ (G)^n \Gamma^\mathrm{1PI}_{n+m}]= 
& (G)^{n+1}\Gamma^\mathrm{1PI}_{1+n+m} \nonumber\\
&+P_n (G)^{n-1} \Gamma^\mathrm{1PI}_{n+m} G\Gamma^\mathrm{1PI}_3(G)^2,\label{eq:treeExpansion_rule}
\end{align}
where $P_n$ comes from the derivative of $(G)^n$.
A diagrammatic representation of Eq.~\eqref{eq:treeExpansion_rule} is shown at the bottom of Fig.~\ref{fig:1PItreeExpansion}. 
The figure highlights that differentiation of a diagram with respect to $J_1$ corresponds to the insertion of an external line with propagator in all possible ways both at already existing vertices and into each propagator, thereby yielding a new three-point vertex attached to that propagator. Iterating these rules, starting with Eq.~\eqref{eq:treeExpansion_G^3}, shows that every correlation function is expressible as a sum of all tree graphs with a fixed number of external lines, where the indices at the free ends are permuted (via $P_n$) such that the sum is (anti)-symmetric with respect to the exchange of any two external indices.

This concludes the derivation of the 1PI tree expansion.
In Section \ref{subsec:1PIVerticesViaGeneralCompositeFields}, we show how the same 1PI vertices can be further expressed using a similar tree expansion but in terms of derivatives of the composite effective action. This is illustrated in Fig.~\ref{fig:ItreeExpansion}. Before that, however, Section~\ref{sec:compositeFieldMethods} discusses the connection between 1PI and composite effective actions via the inverse Legendre transform.

\section{Inverse Legendre transform of the composite effective action to the 1PI effective action}\label{sec:compositeFieldMethods}
As we have seen in the previous section, the Legendre transformation allows us to define the 1PI effective action as a functional of the fundamental field average $\bar\varphi$. By introducing additional sources coupled to nonlinear combinations of fundamental fields (e.g., $\varphi^a\varphi^b$), one can define a more general functional $\Gamma[\bar\varphi,\psi^\bullet]$ depending on composite fields $\psi^\bullet$ (the details of this are presented in Sec.~\ref{subsec:generalEffectiveAction}). The $n$PI effective action \cite{Berges2004nPI} is a special case of this construction, where the composite fields $\psi^\bullet=(G^2,\dots,G^n)$ are connected Green’s functions up to $n$th order \footnote{For a historical sketch of the development of functional Legendre transforms including composite fields, see the end of Sec. 6.2.1 in \cite{Vasiliev1998Functional}}.

In this section, we study the inverse Legendre transformation of $\Gamma[\bar\varphi,\psi^\bullet]$ to $\Gamma^\mathrm{1PI}[\bar\varphi]$, using the 3PI effective action $\Gamma[\bar\varphi,G^2, G^3]$ as an illustrative example. In \ref{subsec:preliminary} we demonstrate the simplicity and utility of our formalism by explicitly computing the first four functional derivatives of Eq.~\eqref{eq:1PIfromGammaIntroduction} while postponing formal definitions to Sec.~\ref{subsec:generalEffectiveAction}, where the generalizations of the 1PI relations presented in \ref{subsec:1PImethods} are developed systematically for general composite effective actions.

For nonrelativistic systems with cubic and quartic interactions the structure of $\Gamma[\bar\varphi,G^2, G^3]$ was first investigated by De Dominicis and Martin in \cite{Dominicis1964EntropyI} and \cite{Dominicis1964EntropyII}. Their results can be extended \cite{Cornwall1974Action2PI,KazVas1972Legendre} to general classical actions of the form \footnote{The order of indices in Eq.~\eqref{eq:classicalAction} can be confirmed by taking functional derivatives of both sides w.r.t.\ $\varphi$ and then setting $\varphi=0$.}
\begin{align}
\label{eq:classicalAction}
S[\varphi]= \sum_{n=2}^{N+1}\frac{1}{n!}\varphi^{a_n}...\varphi^{a_1}S_{a_1...a_n}[0].
\end{align}
 For a theory with cubic and quartic interactions ($N=3$) the 3PI effective action ${\Gamma}[\bar\varphi,G^2,G^3]$ has a simple dependence on $\bar{\varphi}$ \cite{Dominicis1964EntropyII,KazVas1972Legendre}
\begin{align}
{\Gamma}[\bar\varphi,G^2,G^3]&=S[\bar\varphi]+\sum^3_{n=2} \mathcal{S}_n[\bar{\varphi}]G^n+\Lambda[G^2,G^3],\label{effActionStructure} \\
\mathcal{S}_{a_1...a_n}[\bar\varphi] &\equiv \zeta^n S_{a_n...a_1}[\bar\varphi](-1)^{n-1}/n!\label{eq:definition_F},
\end{align}
where $\mathcal{S}_n G^n$ stands for $\mathcal{S}_{a_1...a_n}G^{a_1...a_n}$. The same $\mathcal{S}_n$ holds for the choice $\psi^\bullet=G^{ab}$ when only cubic interactions are present (for $S_3[0]\ne 0$ the system with fixed statistics can only be bosonic, $\zeta=1$).

Equation~\eqref{effActionStructure} shows that the full dependence of the functional ${\Gamma}[\bar\varphi,G^2,G^3]$ on $\bar\varphi$ is captured by its first two terms which we know exactly. At the same time, Eq.~\eqref{effActionStructure} can be viewed as the definition of $\Lambda[G^2,G^3]$. It was shown in \cite{Dominicis1964EntropyII} that $\Lambda$ consists of $\frac{-1}{2}\ln\det G$ plus the sum of all possible 3PI diagrams \footnote{Such graphs can be disconnected by cutting three lines (each represented by a full propagator) into exactly two parts, one of which must be a three-point 1PI vertex}. In that sense, $\Lambda$, modulo the $\ln\det G$ part, can be seen as a generalization of the Luttinger-Ward functional and can be exactly identified with the Luttinger-Ward functional under the choice $\psi^{ab}=G^{ab}$. For completeness, we derive Eq.~\eqref{effActionStructure} from the path integral definition of the theory in Appendix~\ref{app:effActionStructure}.

$\Gamma[\bar\varphi,G^2,G^3]$ can be related to the 1PI effective action $\Gamma^{\mathrm{1PI}}[\bar{\varphi}]$ by an inverse Legendre transform \eqref{eq:1PI_via_Gamma}, where all composite field sources are set to zero.
After a short calculation (see Eqs.~\eqref{eq:1PIfromGammaPhi}--\eqref{eq:firstGamma_bullet} for details), we find the remarkably simple result shown in Eqs.~\eqref{eq:1PIfromGammaIntroduction} and \eqref{eq:OmegafromLambdaIntroduction}, reprinted here for convenience:
\begin{align}
\label{eq:1PIfromGamma}
     {\Gamma^{\text{1PI}}[\bar{\varphi}]} &= S[\bar{\varphi}]+{\Omega[{G^2_{\bar\varphi},G^3_{\bar{\varphi}}}]},\\
\label{eq:OmegafromLambda}     
{\Omega[G^2,G^3]}&=\Lambda[{G^2,G^3}]-\sum_{n=2}^3G^n\frac{\delta \Lambda}{\delta G^n},
 \end{align}
where $\Omega[G^2,G^3]$ is the Legendre transform of $\Lambda$ with respect to $G^2$ and $G^3$. The subscript $\bar\varphi$ in $G^n_{\bar\varphi}$ means that the latter is a functional of $\bar\varphi$ obtained from the equations of motion $\frac{\delta \Gamma}{\delta G^2}=0$ and $\frac{\delta \Gamma}{\delta G^3}=0$. This will be explained in detail in Section~\ref{subsec:generalEffectiveAction}. We see from \eqref{eq:1PIfromGamma} that ${\Gamma^{\text{1PI}}}$ expressed in terms of $\bar{\varphi}$ and Green's functions has an even simpler dependence on $\bar{\varphi}$ than the 3PI effective action $\Gamma[\bar{\varphi},G^2,G^3]$ in \Eq{effActionStructure}. Let us emphasize that the definition of $\Omega[G^2,G^3]$ in terms of a Legendre transform directly implies that its functional derivatives generate tree expansions -- something we use extensively throughout the rest of the manuscript.

\subsection{Preliminary results: self-energy estimator, Bethe-Salpeter equations, parquet and asymptotic class decompositions from 3PI effective action}\label{subsec:preliminary}

The functional $\Gamma[\bar\varphi, G^2, G^3]$ possesses two properties that lead to useful non-perturbative results for 1PI vertices, as derived in this section.

First, three-particle irreducibility of the functional $\Gamma[\bar\varphi,G^2,G^3]$ allows us to classify contributions to the four-point 1PI vertex and derive Bethe-Salpeter-type equations for its reducible part. For an even theory (i.e., one with vanishing odd-order Green's and vertex functions) we recover the well-known parquet formalism \cite{Dominicis1964EntropyII}. These results are extended to local composite fields and higher-order vertices in Sec.~\ref{sec:1PIEffectiveActionViaCompositeFields}.

Second, the structure \eqref{effActionStructure} allows us to classify contributions to 1PI vertices by the connectivity of their external legs to the bare interactions. These groupings are referred to as asymptotic classes \cite{wentzell2020high}, as they govern the high-frequency behavior of the vertex functions. In Sec.~\ref{sec:SymmetricEstimators} this result is used to derive representations for asymptotic classes and 1PI vertices via Green's functions of composite fields (the so-called symmetric estimators \cite{kaufmann2019symmetric, lihm2024symmetric}). Here, we reproduce such a formula for the self-energy~\cite{Kugler2022Improved}.

For simplicity, we restrict ourselves to bosonic systems (with up to quartic interactions).

\subsubsection*{Step 1: First functional derivative of the effective action}
Let us start by taking the first functional derivative of Eq.~\eqref{eq:1PIfromGamma} with respect to $\bar{\varphi}$
\begin{multline}\label{eq:1p1PIfromOmega}
    {\Gamma_1^{\text{1PI}}[\bar{\varphi}]}={S_1[\bar{\varphi}]}+\sum_{n=2}^3\frac{\delta G^n_{\bar{\varphi}}}{\delta \bar\varphi}\frac{\delta \Omega}{\delta G^n}\\=S_1[\bar\varphi]-\sum_{n,m=2}^3G^{-1}_{\bar\varphi}G^{1+n}_{\bar\varphi}G^m_{\bar\varphi}\frac{\delta^2\Lambda}{\delta G^m\delta G^n},
\end{multline}
where we used $\tfrac{\delta G^n_{\bar{\varphi}}}{\delta\bar\varphi}=G_{\bar\varphi}^{-1}G^{1+n}_{\bar\varphi}$ with the help of Eq.~\eqref{eq:1PI_Jderiv}. The dependence of $G^4_{\bar\varphi}$ on $\bar\varphi$ is defined via the tree-expansion formula \eqref{eq:treeExpansion_G^4}. All derivatives of $\Lambda$ are evaluated at $G^n=G^n_{\bar\varphi}$.

To proceed, we need functional relations that connect derivatives of $\Gamma$ (or $\Lambda$) to Green's functions, similar to Eq.~\eqref{eq:1PI_Invertibility}. One such relation is easily obtained by differentiation of the equation of motion, $0=\frac{\delta \Gamma}{\delta G^m}=\mathcal{S}_m+\frac{\delta\Lambda}{\delta G^m}$, at $G^m=G^m_{\bar\varphi}$ w.r.t. $\bar\varphi$, which gives
\begin{align}\label{eq:derivativeOfG^bullet}
G^{-1}\sum_{n=2}^3G^{1+n}\frac{\delta^2\Lambda}{\delta G^n\delta G^m}=-\frac{\delta\mathcal{S}_{m}}{\delta\bar\varphi}.
\end{align}
\noindent
Equation~\eqref{eq:1p1PIfromOmega} becomes
\begin{align}\label{eq:1pVertex}
    {\Gamma_1^{\text{1PI}}}={S_1}+\sum_{m=2}^3G^m_{\bar\varphi}\frac{\delta\mathcal{S}_m}{\delta\bar\varphi}.
\end{align}
\noindent
The last term in Eq.~\eqref{eq:1pVertex} contains an implicit $\bar{\varphi}$-dependence in $G^m_{\bar\varphi}$ as well as an explicit one in $\frac{\delta\mathcal{S}_{m}[\bar\varphi]}{\delta\bar\varphi}$ (see Eq.~\eqref{eq:definition_F}). 

By taking derivatives $\frac{\delta}{\delta\bar\varphi}=G^{-1}\frac{\delta}{\delta J_1}$ of Eq.~\eqref{eq:1pVertex}, we reproduce Schwinger-Dyson (SD) equations. Indeed, the first differentiation gives
\begin{align}\label{eq:SD_selfEnergy}
-G_{\bar\varphi}^{-1}=S_2-\frac{1}{2}G_{\bar\varphi}S_4+G_{\bar\varphi}^{-1}\sum_{m=2}^3G^{1+m}_{\bar\varphi}\frac{\delta\mathcal{S}_m}{\delta\bar\varphi},
\end{align} 
where we used Eq.~\eqref{eq:1PI_Motion} and $G^{ba} \frac{\delta \mathcal{S}_{ba}}{ \delta \bar{\varphi}^c \delta\bar\varphi^d}=-\frac{1}{2}G^{ba}S_{abcd}$ in condensed form (note that $\frac{\delta^2 \mathcal{S}_m}{\delta\bar\varphi^2}$ vanishes for $m>2$ ).

\subsubsection*{Step 2: Chain rule and general formula for $n$-point 1PI vertex}
Another way to generate 1PI vertices (first proposed in \cite{Vasilev1973VertexFromNPI}) is to use the chain rule $\frac{\delta}{\delta\bar{\varphi}}=\frac{\delta}{\delta\psi}+\sum_{l=2}^3\frac{\delta G^l_{\bar{\varphi}}}{\delta \bar\varphi}\frac{\delta}{\delta G^{l}}$. We use $\psi$ in place of $\bar\varphi$  to distinguish $\bar\varphi$, the variable of $\Gamma^{\mathrm{1PI}}[\bar\varphi]$, from $\psi$, the first variable of the effective action $\Gamma[\psi,G,G^3]$ (which is the same functional as \eqref{effActionStructure}). 
Equation~\eqref{eq:derivativeOfG^bullet} gives for $\frac{\delta G_{\bar\varphi}^l}{\delta\bar\varphi}=G_{\bar\varphi}^{-1}G^{1+l}_{\bar\varphi}$,
\begin{align}\label{eq:derivativeOfG^bullet'}
\frac{\delta G^l_{\bar{\varphi}}}{\delta \bar\varphi}=\sum_{k=2}^3{\frac{\delta\mathcal{S}_k}{\delta\bar\varphi}}\chi^{k|l},
\end{align}
where we defined $\chi^{k|l}$ as the inverse of $(-\frac{\delta^2\Lambda}{\delta G^l\delta G^m})$:
\begin{align}\label{eq:Def_K^{k|l}}
    \sum_{l=2}^3\chi^{k|l}\frac{\delta^2\Lambda}{\delta G^l\delta G^m}=-\delta^k_m.
\end{align}
With the help of Eq.~\eqref{eq:derivativeOfG^bullet'}, the chain rule becomes
\begin{align}\label{J_1Derivative}
  \frac{\delta}{\delta \bar{\varphi}} =\frac{\delta}{\delta\psi}+\sum_{k,l=2}^3{\frac{\delta\mathcal{S}_k}{\delta\bar\varphi}}\chi^{k|l}\frac{\delta}{\delta G^{l}}.
\end{align}

To calculate $\Gamma^\mathrm{1PI}_n$, Eq.~\eqref{eq:1pVertex} is differentiated $(n-1)$ times using \eqref{J_1Derivative} for all terms except $S_1$,
\begin{align}\label{eq:npVertex_fromGamma}
    \Gamma^\mathrm{1PI}_n=S_n+\Big (\frac{\delta}{\delta \psi} +\sum_{k,l=2}^3{\frac{\delta\mathcal{S}_k}{\delta\bar\varphi}} \chi^{k|l}\frac{\delta}{\delta G^{l}} \Big)^{n-1}\sum_{m=2}^3G^m\frac{\delta\mathcal{S}_m}{\delta\bar\varphi}.
\end{align} 
After taking all functional derivatives, we substitute $G^n=G^n_{\bar\varphi}$.
Evaluating only the $\psi$-derivatives in Eq.~\eqref{eq:npVertex_fromGamma} generates groups of terms with different connectivities of the external legs to the bare vertices (contained in $\frac{\delta^m\mathcal{S}_{n}[\bar\varphi]}{\delta\bar\varphi^m}$); these groups can be identified with asymptotic classes. To see this, we calculate Eq.~\eqref{eq:npVertex_fromGamma} for $n=2,3,4$ in the next steps.

\subsubsection*{Step 3: Second derivative of the effective action --- self-energy estimator}
For $n=2$ we get
\begin{align}\label{2pVertex'}
    {\Gamma_2^{\text{1PI}}}={S_2}+G_{\bar\varphi}\frac{\delta^2\mathcal{S}_2}{\delta\bar\varphi^2}+\sum_{k,l=2}^3{\frac{\delta\mathcal{S}_k}{\delta\bar\varphi}}\chi^{k|l}{\frac{\delta\mathcal{S}_l}{\delta\bar\varphi}}.
\end{align}
For an even theory, $S_2=S_2[0]$ and the last term
simplifies to ${\frac{\delta\mathcal{S}_3}{\delta\bar\varphi}}\chi^{3|3}{\frac{\delta\mathcal{S}_3}{\delta\bar\varphi}}$. When expressed in terms of composite fields, Eq.~\eqref{2pVertex'}
then reduces to the symmetric estimator for the self-energy established by Kugler \cite{Kugler2022Improved}. To show this, we need an analog of Eq.~\eqref{eq:1PI_Invertibility} for the composite effective action (proved in Section~\ref{subsec:generalEffectiveAction}),
\begin{align}\label{eq:K^3|3}
    \chi^{3|3}=-\langle\varphi^3\varphi^3\rangle-\langle\varphi^3\varphi\rangle G^{-1}\langle\varphi\varphi^3\rangle.
\end{align}
\noindent
The contraction $\frac{\delta\mathcal{S}_3}{\delta\bar\varphi}\langle\varphi^3...\rangle$ in Eq.~\eqref{2pVertex'} defines the following composite fields in correlation functions: 
$\frac{\delta\mathcal{S}_{bcd}}{\delta\bar\varphi^a}\varphi^b\varphi^c\varphi^d=\frac{\delta}{\delta\varphi^a}S_\mathrm{int}$, where $S_\mathrm{int}$ is the interacting (quartic) part of the action in an even theory. Applying this identity to Eq.~\eqref{2pVertex'} yields the self-energy estimator
\begin{multline}\label{eq:selfEnergyEstimatorBosonic}
    -G^{-1}-S_2[0]=\langle\frac{\delta^2 S_\mathrm{int}}{\delta \varphi^2}\rangle \\-\langle \frac{\delta S_\mathrm{int}}{\delta\varphi}\frac{\delta S_\mathrm{int}}{\delta\varphi}\rangle-\langle \frac{\delta S_\mathrm{int}}{\delta\varphi}\varphi\rangle G^{-1}\langle\varphi \frac{\delta S_\mathrm{int}}{\delta\varphi}\rangle,
\end{multline} 
where we also used Eqs.~\eqref{eq:1PI_Invertibility} and $G\frac{\delta^2\mathcal{S}_2}{\delta\bar\varphi^2}=\frac{1}{2}\langle\varphi^2\rangle S_4=\langle\frac{\delta^2 S_\mathrm{int}}{\delta \varphi^2}\rangle$. Equation~\eqref{eq:selfEnergyEstimatorBosonic} corresponds to Eq.~(12) of \cite{Kugler2022Improved}.

\subsubsection*{Step 4: Derivatives of the equation of motion, $\frac{\delta\Gamma}{\delta G^m}=0$, at $G^m=G^m_{\bar\varphi}$ --- Bethe-Salpeter equations}
\begin{figure*}[t]
    \centering
    \includegraphics[width=1.0\linewidth]{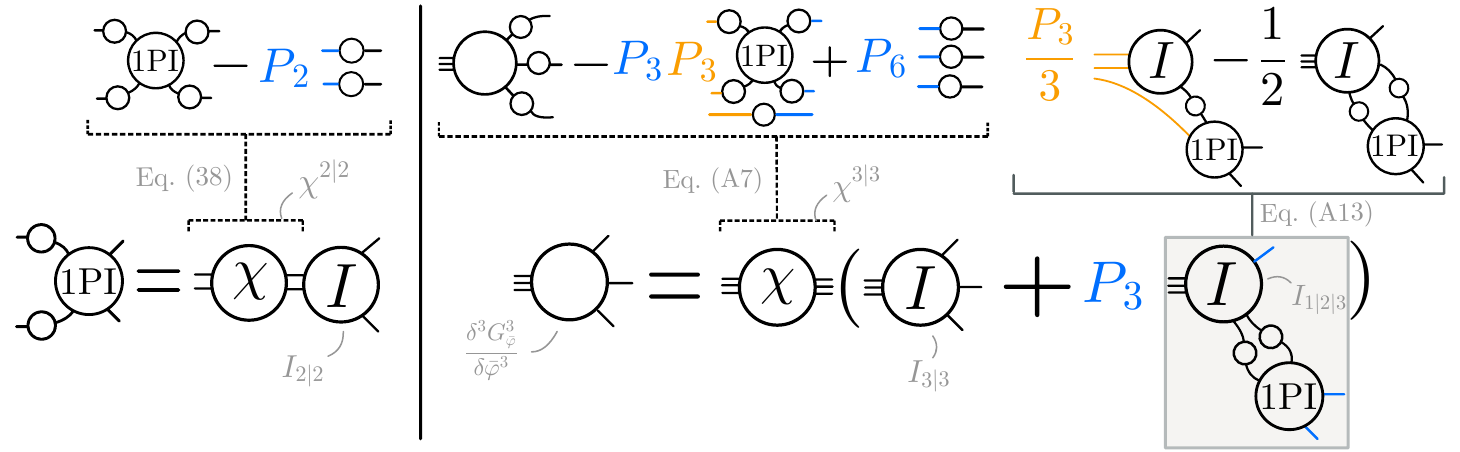}
    \caption{Left: Bethe-Salpeter (BS) equation for the vertex $I_{ab|cd}$ as given by Eq.~\eqref{eq:BS_0'}. Right: BS equation for the vertex $I_{abc|def}$ as given by Eq.~\eqref{eq:BS_n=3}.}
    \label{fig:BSEs_nPI}
\end{figure*}

For $n>2$, the combination $\sum_{k=2}^3\chi^{k|l}\frac{\delta\mathcal{S}_k}{\delta\bar\varphi}$ appears frequently in Eq.~\eqref{eq:npVertex_fromGamma} due to the relation \eqref{eq:derivativeOfG^bullet'}, which was obtained by differentiating the equation of motion, $\frac{\delta\Gamma}{\delta G^m}=0$ at $G^m=G^m_{\bar\varphi}$. It is convenient to evaluate its $\bar\varphi$-derivative using the chain rule~\eqref{J_1Derivative} and Eqs.~\eqref{eq:derivativeOfG^bullet'}--\eqref{eq:Def_K^{k|l}},
\begin{align}
\frac{\delta}{\delta\bar\varphi}&
  \sum_{k=2}^3 \chi^{k|l}   \frac{\delta\mathcal{S}_k}{\delta\bar\varphi}
  = \biggl(
       \frac{\delta}{\delta\psi}
       + \sum_{n=2}^3 \frac{\delta G_{\bar\varphi}^n}{\delta\bar\varphi}
         \frac{\delta}{\delta G^{n}}
     \biggr)
     \sum_{k=2}^3 \chi^{k|l} \frac{\delta\mathcal{S}_k}{\delta\bar\varphi}
     \nonumber\\[0.3em]
  &= \sum_{k=2}^3 \chi^{k|l} \frac{\delta^2\mathcal{S}_k}{\delta\bar\varphi^2}
     + \sum_{k,n=2}^3 \frac{\delta G_{\bar\varphi}^n}{\delta\bar\varphi}
       \biggl( \frac{\delta}{\delta G^{n}} \chi^{k|l} \biggr)
       \frac{\delta\mathcal{S}_k}{\delta\bar\varphi}
     \nonumber\\[0.3em]
  &= \sum_{k=2}^3 \chi^{k|l} I_{k|2},
\label{eq:BS_0_rhs}
\\[0.3em]
I_{k|2} &\equiv
    \frac{\delta^2\mathcal{S}_k}{\delta\bar\varphi^2}
    + \sum_{n,m=2}^3
      \frac{\delta G_{\bar\varphi}^n}{\delta\bar\varphi}
      \frac{\delta^3 \Lambda}
           {\delta G^{n} \delta G^{k} \delta G^{m}}
      \frac{\delta G_{\bar\varphi}^m}{\delta\bar\varphi}.
\label{eq:definitionI_22}
\end{align}
Now, consider Eq.~\eqref{eq:derivativeOfG^bullet'}. Differentiating both sides with respect to $\bar\varphi$, and using Eq.~\eqref{eq:BS_0_rhs}, we obtain
\begin{align}\label{eq:BS_0}
    \frac{\delta^2 G_{\bar\varphi}^l}{\delta\bar\varphi^2}=\sum_{k=2}^3\chi^{k|l}I_{k|2},
\end{align}
 where the right side is a functional of $\bar\varphi$ after substituting $\psi=\bar\varphi$ and $G^n=G^n_{\bar\varphi}$ (recall that $\chi^{m|n}$ and $I_{k|2}$ are defined in terms of derivatives of $\Gamma[\psi,G,G^3]$).
For an even theory and $l=2$, Eq.~\eqref{eq:BS_0} is precisely the Bethe-Salpeter equation for $I_{2|2}$ after we relate $\chi^{2|2}$ to the two-particle Green's function similar to Eq.~\eqref{eq:K^3|3} (for details see Section \ref{subsec:generalEffectiveAction}). This leads to
\begin{align}\label{eq:2partGreenFunction}
\chi^{ab|cd}=G^{aa'}G^{bb'}\Gamma^\mathrm{1PI}_{a'b'c'd'}G^{c'c}G^{d'd}-G^{ac}G^{bd}-G^{ad}G^{bc}.
\end{align}
Using Eqs.~\eqref{eq:1PI_Jderiv} and \eqref{eq:treeExpansion_G^4}, the left side of \eqref{eq:BS_0} for $l=2$ becomes $G_{\bar\varphi}^4(G_{\bar\varphi})^{-2}=G_{\bar\varphi}G_{\bar\varphi}\Gamma^\mathrm{1PI}_4$. We get
\begin{align}\label{eq:BS_0'}
G_{\bar\varphi}G_{\bar\varphi}\Gamma^\mathrm{1PI}_4=\chi^{2|2}I_{2|2},
\end{align}
which is illustrated on the left in Fig.~\ref{fig:BSEs_nPI}.

Notably, this framework naturally generates Bethe-Salpeter equations formulated directly in terms of connected kernels (such as $I_{2|2}$), closely resembling Weinberg's hierarchy for multi-particle scattering~\cite{Weinberg1964Scattering}. Indeed, by further differentiating both sides of Eq.~\eqref{eq:BS_0} for $l=3$ with respect to $\bar\varphi$, and using \eqref{eq:BS_0'}, we obtain an equation for the connected three-particle irreducible vertex, $I_{3|3}\equiv  (\frac{\delta G^3_{\bar\varphi}}{\delta\bar\varphi})^3\frac{\delta^4\Lambda}{\delta G^3\delta G^3\delta G^3\delta G^3}$, 
as shown on the right side of Fig.~\ref{fig:BSEs_nPI} (see also Appendix~\ref{app:cluster_formula}). The resulting equation describes three-particle scattering processes and is equivalent to Eq.~(82) of Ribic~\cite{RibicPathIntegral}.

\subsubsection*{Step 5: Calculation of the 3p and 4p 1PI vertices --- Parquet decomposition}
To obtain the decomposition for the three-point vertex, we need to  differentiate \eqref{2pVertex'} using Eqs.~\eqref{eq:derivativeOfG^bullet'}--\eqref{eq:Def_K^{k|l}}:
\begin{multline}\label{eq:3pvertex}
    {\Gamma_3^{\text{1PI}}}={S_3}+P_3 \Big (\frac{\delta G_{\bar\varphi}}{\delta\bar\varphi}{\frac{\delta^2\mathcal{S}_2}{\delta\bar\varphi^2}} \Big )\\+\sum_{n,m,l=2}^3\frac{\delta G_{\bar\varphi}^n}{\delta\bar\varphi}\frac{\delta G_{\bar\varphi}^m}{\delta\bar\varphi}\frac{\delta^3\Lambda }{\delta G^{n}\delta G^{m}\delta G^{l}}\frac{\delta G_{\bar\varphi}^l}{\delta\bar\varphi}.
\end{multline}
As a check that \eqref{eq:3pvertex} is consistent with the equation of motion $\tfrac{\delta \Gamma}{\delta G^3}=0$, we can substitute the simplest one-loop approximation $\Lambda\approx \tfrac{-1}{2}\ln\det G$. The resulting equation coincides with $\tfrac{\delta\Gamma}{\delta G^3}=0$, in which $\Lambda$ is approximated up to three loops~\cite{Williams2016QCD}.

Next, we evaluate the four-point 1PI vertex by taking the derivative of Eq.~\eqref{eq:3pvertex}. All terms involving derivatives of $\frac{\delta G^n_{\bar\varphi}}{\delta\bar\varphi}=\sum_{n'=2}^3\chi^{n|n'}\frac{\delta \mathcal{S}_{n'}}{\delta\bar\varphi}$ are evaluated using Eq.~\eqref{eq:BS_0}; their sum yields
\begin{multline}
P_3\sum_{k,l=2}^3\Big (\frac{\delta^2\mathcal{S}_l}{\delta\bar\varphi^2}+\sum_{n,m=2}^3\frac{\delta G_{\bar\varphi}^n}{\delta\bar\varphi}\frac{\delta G_{\bar\varphi}^m}{\delta\bar\varphi}\frac{\delta^3\Lambda }{\delta G^{n}\delta G^{m}\delta G^{l}} \Big )\chi^{k|l}I_{k|2}\\=P_3\sum_{k,l=2}^3I_{l|2}\chi^{k|l}I_{k| 2}.
\end{multline}
We get
\begin{align}\label{eq:parquet}
\Gamma^\mathrm{1PI}_4&=I_4+P_3\sum_{k,l=2}^3I_{k|2}\chi^{k|l}I_{l|2},\\ \label{eq:I4_def}
I_4 & \equiv S_4+\sum_{n,m,l,k=2}^3\frac{\delta G_{\bar\varphi}^n}{\delta\bar\varphi}\frac{\delta G_{\bar\varphi}^m}{\delta\bar\varphi}\frac{\delta^4\Lambda }{\delta G^{n}\delta G^{m}\delta G^{l}\delta G^k}\frac{\delta G_{\bar\varphi}^l}{\delta\bar\varphi}\frac{\delta G_{\bar\varphi}^k}{\delta\bar\varphi} .
\end{align}
For even theories, one recalls the Bethe-Salpeter equation \eqref{eq:BS_0'} to obtain $I_{2|2}\chi^{2|2}I_{2|2}=I_{2|2}G_{\bar\varphi}G_{\bar\varphi}\Gamma^\mathrm{1PI}_4$. Equation~\eqref{eq:parquet} then reduces to the parquet decomposition \cite{Dominicis1964EntropyII}. This step is visualized on the right side of Fig.~\ref{fig:treeParquet}.

Two more important points about Eq.~\eqref{eq:parquet} should be noted.
First, Eq.~\eqref{eq:parquet} has the tree expansion form given by Eq.~\eqref{eq:treeExpansion_G^4}. This result will be extended to vertices of arbitrary order and local composite fields in Sec.~\ref{sec:1PIEffectiveActionViaCompositeFields}.

Second, \Eq{eq:I4_def} implies the relation for an even theory:
\begin{align}\label{eq:I_4_via_I_3|3}    I_{abcd}=S_{abcd}+I_{abc|efh}G^{ee'}G^{ff'}G^{hh'}\Gamma^\mathrm{1PI}_{e'f'h'd},
\end{align}
where $I_{3|3}\equiv  (\frac{\delta G^3_{\bar\varphi}}{\delta\bar\varphi})^3\frac{\delta^4\Lambda}{\delta G^3\delta G^3\delta G^3\delta G^3}$ satisfies the Bethe-Salpeter equation on the right side of Fig.~\ref{fig:BSEs_nPI}. It is possible to similarly relate composite field irreducible vertices, using local two- and three-point Green's functions as explained in Appendix~\ref{app:beyond_SBE}.

\subsubsection*{Step 6: Asymptotic class decomposition of the 4p 1PI vertex}
Finally, let us expand Eq.~\eqref{eq:parquet} using the definition in Eq.~\eqref{eq:definitionI_22}, as follows
\begin{align}\label{eq:Decomposition4p1PIviaClasses}
\Gamma^{\mathrm{1PI}}_4=S_4+P_3\mathcal{K}^1_4+P_6\mathcal{K}^2_4+\mathcal{K}^3_4,
\end{align}
where we defined asymptotic classes 
\begin{align}
\mathcal{K}^1_4 &\equiv \frac{\delta^2\mathcal{S}_2}{\delta\bar\varphi^2}\chi^{2|2}\frac{\delta^2\mathcal{S}_2}{\delta\bar\varphi^2}\label{eq:K^1_4}\\
\mathcal{K}^2_4 &\equiv \frac{\delta^2\mathcal{S}_2}{\delta\bar\varphi^2}\sum_{n,m,l=2}^3 \chi ^{2|n}\frac{\delta^3\Lambda }{\delta G^{n}\delta G^{m}\delta G^{l}}\frac{\delta G_{\bar\varphi}^m}{\delta\bar\varphi}\frac{\delta G_{\bar\varphi}^l}{\delta\bar\varphi}\label{eq:K^2_4}\\
\mathcal{K}^3_4 &\equiv \sum_{n,m,l,k=2}^3\frac{\delta G_{\bar\varphi}^n}{\delta\bar\varphi}\frac{\delta G_{\bar\varphi}^m}{\delta\bar\varphi}\frac{\delta G_{\bar\varphi}^l}{\delta\bar\varphi}\frac{\delta G_{\bar\varphi}^k}{\delta\bar\varphi}\Big(\frac{\delta^4\Lambda }{\delta G^{n}\delta G^{m}\delta G^{l}\delta G^k} +\nonumber
\\&P_3\sum_{n',m'=2}^3\frac{\delta^3\Lambda }{\delta G^{n}\delta G^{m}\delta G^{n'}}\chi^{n'|m'}\frac{\delta^3\Lambda }{\delta G^{m'}\delta G^{l}\delta G^{k}}\Big).\label{eq:IKI_3PI}
\end{align}
In Eq.~\eqref{eq:IKI_3PI}, $P_3$ permutes the derivatives $\frac{\delta}{\delta G^{\dots}}$ that contain external indices \footnote{It agrees with the definition of $P_n$ if we consider external indices in $\frac{\delta}{\delta G^{\dots}}$ as a single composite external index. Alternatively, one can expand the brackets in Eq.~\eqref{eq:IKI_3PI} and then apply $P_3$ to the external indices in products of $\frac{\delta G^{\dots}_{\bar\varphi}}{\delta\bar\varphi}$.}. The upper index $m$ in $\mathcal{K}^m_4$ denotes the number of independent time (or frequency) arguments, assuming time-translational invariance and time-local bare interactions. For example, in the class $\mathcal{K}^2_4$, two indices are contained in $\frac{\delta^2\mathcal{S}_2}{\delta\bar\varphi^2}=-\frac{1}{2}S_4$, which forces their time components to be equal; the total number of independent time arguments is then $4-1-1=2$, where the second reduction by one comes from time-translational invariance. 
Equations~\eqref{eq:K^1_4} and \eqref{eq:K^2_4} correspond to Eqs.~(B1) and (B2) of \cite{wentzell2020high}, respectively \footnote{Note that the definition in Eq.~\eqref{eq:definitionI_22} and the BS equation \eqref{eq:BS_0'} imply $\mathcal{K}^2_4=\frac{\delta^2\mathcal{S}_2}{\delta\bar\varphi^2}GG\Gamma^\mathrm{1PI}_4-\frac{\delta^2\mathcal{S}_2}{\delta\bar\varphi^2}\chi^{2|2}\frac{\delta^2\mathcal{S}_2}{\delta\bar\varphi^2}$ for an even theory. The first term in Eq.~\eqref{eq:IKI_3PI} is not included in $\mathcal{K}_4^3$ and is denoted as $\mathcal{R}$ in Ref.~\cite{wentzell2020high} (see also their Fig.~5).}.

Notably, the asymptotic classes $\mathcal{K}^2_4$ and $\mathcal{K}^3_4$ contain the tree expansion (see Eqs.~\eqref{eq:treeExpansion_G^3} and \eqref{eq:treeExpansion_G^4}) with derivatives of $\Lambda$ as vertices and $\chi^{n|m}$ as internal lines. This structure allows us to express asymptotic classes in terms of Green's functions of composite fields (such as $\frac{\delta S_{\mathrm{int}}}{\delta\varphi}$ in Eq.~\eqref{eq:selfEnergyEstimatorBosonic}), similarly to how the tree expansion in derivatives of $\Gamma^\mathrm{1PI}$ yields the Green's functions of the fundamental field $\varphi$. In Sec.~\ref{sec:SymmetricEstimators}, we demonstrate this result for general asymptotic classes and apply it to 1PI vertices.

\subsection{General formulation with condensed notation}\label{subsec:generalEffectiveAction}

Before further developing the ideas presented above, we clarify and extend two key concepts. First, we introduce the composite effective action and derive the functional relations it satisfies. Then, we perform its inverse Legendre transform to recover the 1PI effective action. To this end, let us
 define the generating functional $W[J_{\mathbf{a}}]$ more generally from 
\begin{align}\label{eq:W_definition}
\me^{-W[J_\mathbf{a}]}=\int \! \mathrm{D}\varphi\,  \me^{-S[\varphi, J_{(\bullet)}]-J_{\mathbf{b}}\phi^{\mathbf{b}}[\varphi]},
\end{align}
where the classical action $S[\varphi, J_{(\bullet)}]$ in the presence of external sources $J_{(\bullet)}$ does not contain terms linear in $J_{(\bullet)}$, i.e., $\frac{\delta S}{\delta J_{(\bullet)}}\underset{J_{(\bullet)}\rightarrow0}{\longrightarrow} 0$.

The bold index $\mathbf{a}$ means a vector containing the fundamental-field index $a$ and multi-indices of the composite fields, i.e.,  $\mathbf{a} =\begin{pmatrix} a, (\bullet) \end{pmatrix}$. We define $\phi^a\equiv \varphi^a$, while keeping the composite index $(\bullet)$ and the corresponding field $\phi^{(\bullet)}$ unspecified. The contraction $J_{\mathbf{a}}{\phi}^{\mathbf{a}}$ can then be written as 
\begin{align}\label{eq:Ja_barvarphia}
J_{\mathbf{a}}{\phi}^{\mathbf{a}}=J_a{\varphi}^a+J_{(\bullet)}{\phi}^{(\bullet)}.
\end{align}
As an example, let us consider two composite fields $\phi^{(a_1\dots a_n)}=\varphi^{a_1}\dots\varphi^{a_n}$ for $n=2,3$. In this case $(\bullet)$ is a vector of $(ab)$ and $(abc)$  (in condensed notation $(2)$ and $(3)$, respectively), and the contraction becomes $J_{(\bullet)}{\phi}^{(\bullet)}=J_{(ab)}{\phi}^{(ab)}+J_{(abc)}{\phi}^{(abc)} = J_{(2)}{\phi}^{(2)}+J_{(3)}{\phi}^{(3)}$.
We use a bold symbol $\boldsymbol{\phi}$ as a condensed notation for $\phi^{\mathbf{a}}$ (to avoid confusion with $\phi=\varphi$ for $\phi^a=\varphi^a$) and similarly for the sources.
With the definition \eqref{eq:W_definition}, one can compute the connected correlation function of the fundamental field $\varphi$ as well as the composite ones \footnote{We assume that terms arising from $J_{(\bullet)}$-derivatives of $S[\varphi, J_{(\bullet)}]$ generate Green's functions of lower order than the leading term in Eq.~\eqref{eq:1PcorrelationFunctions}, so that $\langle\phi^{\mathbf{a_1}}...\phi^{\mathbf{a_n}}\rangle_c$ can be straightforwardly expressed in terms of the derivatives of $W[J_\mathbf{a}]$.},
\begin{multline}\label{eq:1PcorrelationFunctions}
G^{\mathbf{a_1}\dots\mathbf{a_n}}\equiv\frac{\delta}{\delta J_{\mathbf{a_1}}}\dots\frac{\delta}{\delta J_{\mathbf{a_n}}}W[J_\mathbf{a}]\\=(-1)^{n-1}\langle\phi^\mathbf{a_1}\dots\phi^\mathbf{a_n}\rangle_c +\ldots,
\end{multline}
where the ellipsis denotes terms originating from $J_{(\bullet)}$-derivatives of $S[\varphi, J_{(\bullet)}]$. We similarly define $\bar{\phi}^{\mathbf{a}}\equiv \frac{\delta W}{\delta J_{\mathbf{a}}}\underset{J_{(\bullet)}\rightarrow0}{\longrightarrow}\langle\phi^{\mathbf{a}}\rangle$.

The effective action $\Gamma[\boldsymbol{\bar{\phi}}]$ is defined via generalized Legendre transformation
\begin{align} \label{eq:Gamma_Definition}
 \Gamma[\boldsymbol{\bar{\phi}}] =  W - J_{\mathbf{a}} \bar{\phi}^{\mathbf{a}},
\end{align}
where $J_{\mathbf{a}}$ as a functional of $\bar{\phi}^{\mathbf{a}}$ is the solution of $\frac{\delta W}{\delta J_{\mathbf{a}}}=\bar{\phi}^{\mathbf{a}}$.
The $n$PI effective action is constructed from $\Gamma[\boldsymbol{\bar\phi}]$ by choosing $S[\varphi, J_{(\bullet)}]=S[\varphi]$ and $\phi^{(a_1\dots a_m)}=\varphi^{a_1}\dots\varphi^{a_m}$ with $m=2,\dots,n$. The standard Hubbard-Stratonovich transformation corresponds to the choice $S[\varphi, J_{(\bullet)}]=S[\varphi]+\tfrac{1}{2}J_{(\bullet)}\chi_0^{\bullet|\bullet}J_{(\bullet)}$, where $\phi^{(\bullet)}$ is a local bilinear and $\chi_0^{\bullet|\bullet}$ is associated with a bare composite field propagator (see \ref{subsec:ExchangeOfCompositeParticles} for details in the context of the SBE). To account for simultaneous exchange of bosonic modes and fundamental particles, more elaborate choices are required (see Appendix \ref{app:beyond_SBE}).

Vertices are defined as derivatives of the effective action $\Gamma[\boldsymbol{\bar{\phi}}]$:
\begin{align}\label{eq:CompositeVertexDefinition}
\Gamma_{\mathbf{a_1}\dots\mathbf{a_n}}\equiv \frac{\delta}{\delta \bar{\phi}^\mathbf{a_1}}\dots\frac{\delta}{\delta \bar{\phi}^\mathbf{a_n}}\Gamma.
\end{align}
$\Gamma[\boldsymbol{\bar{\phi}}]$ obeys relations similar to those of the 1PI effective action $\Gamma^\mathrm{1PI} [\bar{\varphi}]$. In particular, the equation of motion becomes
\begin{align} \label{eq:Gamma_Motion}
\frac{{\delta}}{\delta \bar{\phi}^{\mathbf{a}}}\Gamma=-\gamma^{\mathbf{b}}_\mathbf{a} J_{\mathbf{b}} \;,
\end{align}
where we interchanged $J_{\mathbf{a}} \bar{\phi}^{\mathbf{a}}=\bar{\phi}^{\mathbf{a}}\gamma^{\mathbf{b}}_\mathbf{a}J_{\mathbf{b}}$ using the matrix \footnote{The notation for the composite index $\mathbf{a}$ and the matrix $\gamma^\mathbf{a}_\mathbf{b}$ is taken from  Ref.~\cite{Pawlowski2007FRG}}
\begin{align}\label{eq:definition_metric}
\gamma^{\mathbf{b}}_\mathbf{a}=\zeta^{N_\mathbf{a}}\delta^{\mathbf{b}}_\mathbf{a},
\end{align}
where $N_\mathbf{a}$ is the number of field components denoted by $\mathbf{a}$ (recall that $\zeta$ keeps track of fermionic signs).
 Another relation, similar to Eq.~\eqref{eq:1PI_Invertibility}, is
\begin{align}\label{eq:Gamma_Invertibility}
    G^{\mathbf{ab}}\Gamma_{\mathbf{bc}}=-\gamma^{\mathbf{a}}_{\mathbf{c}}.
\end{align}
From the definition \eqref{eq:definition_metric} we have $\gamma^{\mathbf{a}}_\mathbf{b}\gamma^{\mathbf{b}}_\mathbf{c}=\delta^{\mathbf{a}}_\mathbf{c}$. We can then multiply Eq.~\eqref{eq:Gamma_Invertibility} from the left by $\gamma^{\mathbf{d}}_\mathbf{a}$ to get $(\gamma^{\mathbf{d}}_{\mathbf{a}}G^{\mathbf{ab}})\Gamma_{\mathbf{bc}}=-\delta^{\mathbf{d}}_{\mathbf{c}}$, or more explicitly for $\Gamma_{\mathbf{bc}}$
\begin{align}
\label{eq:invertibilityexplicit}
\begin{pmatrix}
    {\Gamma_2}&&{\Gamma_{1(\bullet)}}\\
    {\Gamma_{(\bullet)1}}&&{\Gamma_{(\bullet)(\bullet)}}
\end{pmatrix}=-\begin{pmatrix}
     \zeta G &&  \zeta G^{1(\bullet)}\\
   \gamma^{(\bullet)}_{(\bullet)}G^{(\bullet) 1} && \gamma^{(\bullet)}_{(\bullet)}G^{(\bullet)(\bullet)}
\end{pmatrix}^{-1}.
\end{align}
To avoid ambiguities, contracted bullet indices (one lower and one upper) are placed as close together as possible.
Using a block-matrix identity for the right side of Eq.~\eqref{eq:invertibilityexplicit}
\begin{align}\label{blockMatrixIdentity}
\begin{pmatrix}
    A&&B\\C&&D
\end{pmatrix}^{-1}=\begin{pmatrix}
    ...&&...\\
    ...&&(D-CA^{-1}B)^{-1}
\end{pmatrix},
\end{align}
we get
\begin{align}\label{eq:K2}
    (-\Gamma_{(\bullet)(\bullet)})^{-1}=\gamma^{(\bullet)}_{(\bullet)}(G^{(\bullet)(\bullet)}-G^{(\bullet) 1}G^{-1}G^{1(\bullet)}),
\end{align}
which is the analog of Eq.~\eqref{eq:1PI_tree_expansion_g}. 

One may want to use other variables $\psi^{\mathbf{a}}[\boldsymbol{\bar{\phi}}]$ instead of $\boldsymbol{\bar{\phi}}\equiv\begin{pmatrix}
    \bar\varphi , \bar\phi^{(\bullet)}
\end{pmatrix}$, while keeping the fundamental field $\bar{\varphi}^a$ as one of the components. This implies that  $\psi^a=\bar\phi^a=\bar{\varphi}^a$, and for the remaining components, we choose $\psi^\bullet$, where the bullet index $\bullet$ takes the same values as in $\phi^{(\bullet)}$ \footnote{Placing parentheses around $\bullet$ allows us to distinguish the field $\bar\phi^{(\bullet)}$ from $\psi^\bullet$ in the vertex $\Gamma_{(\bullet)\dots}$, separate a Green's function $G^{(\bullet)}$ from its connected part $G^\bullet$ for $\bullet=(2,\dots,n)$, and unambiguously specify the contractions (such as in $\psi^\bullet\Gamma_\bullet=\psi^\bullet\frac{\delta \bar{\phi}^{(\bullet)}}{\delta \psi^\bullet}\Gamma_{(\bullet)}$).}. For example, when the composite fields are connected correlation functions, we set $\psi^\bullet=G^\bullet$ with $\bullet=\begin{pmatrix}2,3\end{pmatrix}$, which corresponds to $(\bullet)=\begin{pmatrix}(2),(3)\end{pmatrix}$ in $\phi^{(\bullet)}$. The chain rules are $\frac{\delta}{\delta \bar{\phi}^{\mathbf{a}}}=\frac{\delta\psi^{\mathbf{b}}}{\delta \bar\phi^{\mathbf{a}}}\frac{\delta}{\delta\psi^{\mathbf{b}}}$ and $\frac{\delta}{\delta \bar{\phi}^{(\bullet)}}=\frac{\delta \psi^\bullet}{\delta \bar\phi^{(\bullet)}}\frac{\delta}{\delta\psi^\bullet}$, using $\frac{\delta \psi}{\delta \bar{\phi}^{(\bullet)}}=0$.
\begin{figure}
    \centering
    
    \includegraphics[width=\linewidth]{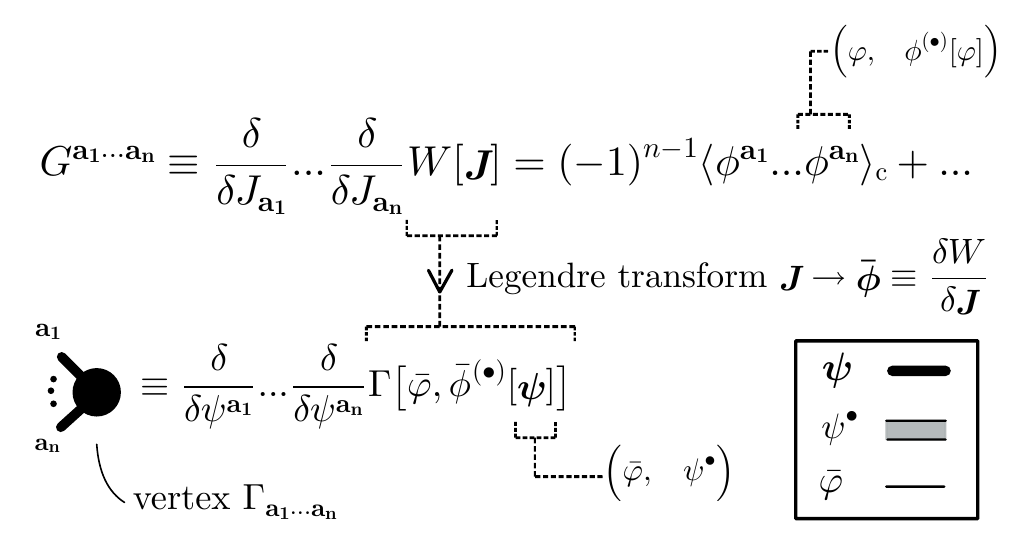}
    
    \caption{Diagrammatic definitions of the connected composite field Green's function $G^{\mathbf{a_1}\dots \mathbf{a_n}}$  (see Eq.~\eqref{eq:1PcorrelationFunctions}) and the corresponding vertex $\Gamma_{\mathbf{a_1}\dots \mathbf{a_n}}$ (see Eq.~\eqref{eq:CompositeVertexDefinition}). The line styles of the vertex indices depend on their type (bold line: general field; gray line: composite field; single line: fundamental field), as seen in the inset as well as in Fig.~\ref{fig:ItreeExpansion}.}
    \label{fig:compositeFieldEffectiveAction}
\end{figure}
Expressing $\bar{\phi}^{(\bullet)}$ in $\Gamma\big[\bar\varphi,{\bar{\phi}^{(\bullet)}}\big]$ in terms of $\boldsymbol{\psi}\equiv\begin{pmatrix}
    \bar\varphi,\psi^\bullet
\end{pmatrix}$, yields a new functional 
\begin{equation}\label{eq:Gamma_change_of_basis}
    \Gamma[\boldsymbol{\psi}] \equiv \Gamma\big[\bar\varphi,{\bar{\phi}^{(\bullet)}}[\boldsymbol{\psi}]\big],
\end{equation} which we denote by the same symbol. The definition and graphical representation of the composite effective action and its vertices are summarized in Fig.~\ref{fig:compositeFieldEffectiveAction}.

The functional relation \eqref{eq:K2} can be expressed via the derivatives of $\Gamma[\bar\varphi,\psi^\bullet]$ using $\frac{\delta}{\delta \bar{\phi}^{(\bullet)}}=\frac{\delta \psi^\bullet}{\delta \bar\phi^{(\bullet)}}\frac{\delta}{\delta\psi^\bullet}$. We have
\begin{multline}\label{eq:K2_general}
    \chi^{\bullet|\bullet}\equiv(-\gamma^\bullet_\bullet\Gamma_{\bullet\bullet})^{-1}=\\\frac{\delta \psi^\bullet}{\delta \bar\phi^{(\bullet)}}(G^{(\bullet)(\bullet)}-G^{(\bullet) 1}G^{-1}G^{1(\bullet)})\frac{\delta \psi^\bullet}{\delta \bar\phi^{(\bullet)}},
\end{multline}
where the last equality holds for either $J_{(\bullet)}=0$ or a linear relation between $\psi^\bullet$ and $\bar\phi^{(\bullet)}$. 

For an even theory and the choice $\psi^\bullet=G^\bullet$ with $\bullet=(2,3)$, the diagonal components of the matrix $ \chi^{\bullet|\bullet}$ in Eq.~\eqref{eq:K2_general} reduce to Eqs.~\eqref{eq:K^3|3} and \eqref{eq:2partGreenFunction}, provided we use $\Gamma_{\bullet\bullet}=\Lambda_{\bullet\bullet}$ for a bosonic system with $S[\varphi, J_{(\bullet)}]=S[\varphi]$ (see \eqref{effActionStructure}) and $\frac{\delta G^n}{\delta G^{(m)}}=(-1)^{n-1}\delta^n_m$. For general theories with fixed statistics the right side of Eq.~\eqref{eq:2partGreenFunction}  acquires additional contributions as shown in the second row of Fig.~\ref{fig:susceptibilities} (and derived in Appendix~\ref{app:cluster_formula}).
If $\psi^\bullet$ corresponds to a local bilinear $\langle U\varphi\varphi\rangle$ with a bare vertex $U$, then $\chi^{\bullet|\bullet}$ becomes a bosonic propagator. It can be obtained from the two-particle Green's function by contracting two pairs of external legs with bare vertices as depicted in the first row of Fig.~\ref{fig:susceptibilities}.

\begin{figure}
    \centering
    
    \includegraphics[width=1\linewidth]{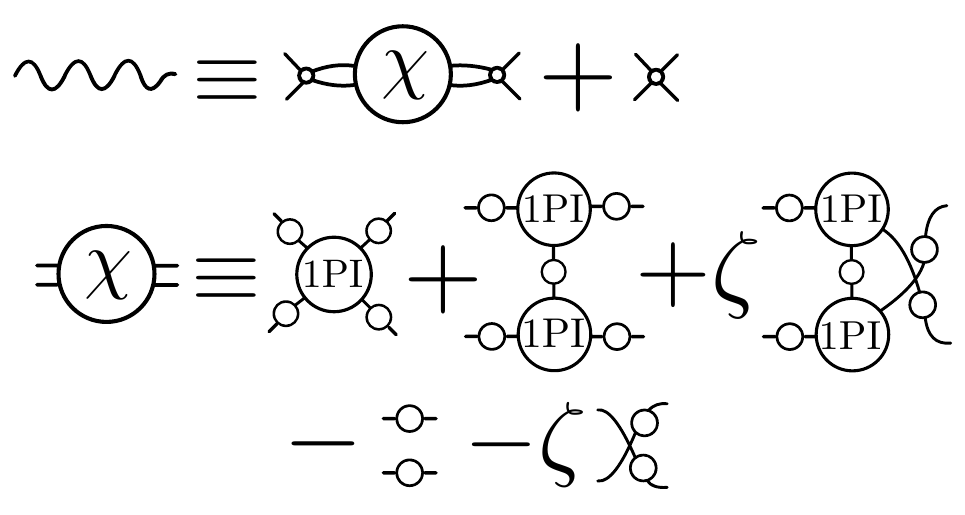}
    
    \caption{Composite field propagator $\chi^{\bullet|\bullet}\equiv (-\gamma^\bullet_\bullet\Gamma_{\bullet\bullet})^{-1}$, as given by Eq.~\eqref{eq:K2_general}, for different composite fields: local bilinears (first row) and the full propagator (second row).}
    \label{fig:susceptibilities}
\end{figure}
\subsubsection*{Inverse Legendre transformation to 1PI effective action}
Since we are interested in 1PI vertices, defined as functional derivatives of $\Gamma^{\mathrm{1PI}}[\bar{\varphi}]$, we need to establish a connection between $\Gamma^{\mathrm{1PI}}[\bar{\varphi}]$ and the composite effective action $\Gamma[\boldsymbol{\bar{\phi}}]$.  We can obtain the 1PI effective action $\Gamma^{\mathrm{1PI}}[\bar{\varphi}]$ from $\Gamma[\boldsymbol{\bar{\phi}}]$ by performing an inverse Legendre transformation from $\bar{\phi}^{(\bullet)}$ to $J_{(\bullet)}$ and then setting $J_{(\bullet)}=0$. This procedure is possible only if we retain $\bar\varphi$ as one of the variables in $\boldsymbol{\bar\phi}$.

To demonstrate this more explicitly, we temporarily keep $J_{(\bullet)}$ nonzero. Substituting $W$ from the definition of $\Gamma[\boldsymbol{\bar{\phi}}]$ (Eq.~\eqref{eq:Gamma_Definition}) into the definition of 1PI effective action (Eq.~\eqref{eq:1PI_Definition}), we perform an inverse Legendre transform of the composite effective action $\Gamma[\bar\varphi,\bar{\phi}^{(\bullet)}]$ to the 1PI action
\begin{align}\label{eq:1PIfromGammaPhi}
    {\Gamma^{\text{1PI}}[\bar{\varphi}]}={\Gamma[\bar\varphi,\bar{\phi}_{\bar{\varphi}}^{(\bullet)}]}+J_{(\bullet)}\bar{\phi}_{\bar{\varphi}}^{(\bullet)}.
\end{align}
Here, $\bar{\phi}_{\bar{\varphi}}^{(\bullet)}$ is the solution of $\frac{\delta \Gamma[\bar{\varphi},\bar{\phi}_{\bar{\varphi}}^{(\bullet)}]}{\delta \bar{\phi}_{\bar{\varphi}}^{(\bullet)} } = \Gamma_{(\bullet)}=-\gamma^{(\bullet)}_{(\bullet)}J_{(\bullet)}$ for $\bar{\phi}^{(\bullet)}$; it is a functional of $\bar{\varphi}$ (denoted by the subscript) and $J_{(\bullet)}$. Note that the dependence of ${\Gamma^{\text{1PI}}[\bar{\varphi}]}$ on $J_{(\bullet)}$ is not explicitly shown.

Let us return to the example $\phi^{(n)}[\varphi]=\varphi^n$ and $\psi^n=G^n$ with $n=2,3$. There exists a linear relation between $G^\bullet$ and $\bar{\phi}^{(\bullet)}$ (see Eqs.~\eqref{2pCorrelationFunction} and \eqref{eq:3pCorrelationFunction})
\begin{align}\label{eq:linearRelation_phi_psi}
    \bar{\phi}^{(\bullet)}=\phi^{(\bullet)}[\bar{\varphi}]+G^\bullet\frac{\delta \bar{\phi}^{(\bullet)}}{\delta G^\bullet}.
\end{align}

Equation~\eqref{eq:linearRelation_phi_psi}, together with $J_{(\bullet)}=-\gamma^{(\bullet)}_{(\bullet)}{\Gamma_{(\bullet)}}$, then yields $J_{(\bullet)}\bar{\phi}^{(\bullet)}[\bar\varphi,G^\bullet_{\bar\varphi}]=J_{(\bullet)}{\phi^{(\bullet)}}[\bar\varphi]-G^\bullet_{\bar\varphi}\frac{\delta \bar{\phi}^{(\bullet)}}{\delta G^\bullet}{\Gamma_{(\bullet)}}$ \footnote{When moving $\gamma^{(\bullet)}_{(\bullet)}\Gamma_{(\bullet)}$ to the right through $G^\bullet_{\bar{\varphi}}\frac{\delta \bar{\phi}^{(\bullet)}}{\delta G^\bullet}$, another $\gamma^{(\bullet)}_{(\bullet)}$ appears from the permutation of contracted $(\bullet)$ indices. The product of two $\gamma^{(\bullet)}_{(\bullet)}$ then yields $\delta^{(\bullet)}_{(\bullet)}$, so that $\gamma^{(\bullet)}_{(\bullet)}\Gamma_{(\bullet)}G^\bullet_{\bar{\varphi}}\frac{\delta \bar{\phi}^{(\bullet)}}{\delta G^\bullet}=G^\bullet_{\bar{\varphi}}\frac{\delta \bar{\phi}^{(\bullet)}}{\delta G^\bullet}\Gamma_{(\bullet)}$} and allows us to change the basis in $\frac{\delta \bar{\phi}^{(\bullet)}}{\delta G^\bullet}{\Gamma_{(\bullet)}}={\Gamma_{\bullet}}$, such that Eq.~\eqref{eq:1PIfromGammaPhi} becomes
\begin{align}\label{eq:firstGamma_bullet}
    {\Gamma^{\text{1PI}}[\bar{\varphi}]}=\Gamma-G^\bullet_{\bar\varphi}{\Gamma_{\bullet}}+J_{(\bullet)}{\phi^{(\bullet)}}[\bar\varphi],
\end{align}
where $\Gamma=\Gamma[\bar\varphi,G^\bullet_{\bar\varphi}]$. Using Eq.~\eqref{effActionStructure} for the first two terms on the right side and setting $J_{(\bullet)}=0$ (so the last term vanishes), we arrive at the result given in Eqs.~\eqref{eq:1PIfromGammaIntroduction}--\eqref{eq:OmegafromLambdaIntroduction}.

Generally, the 1PI effective action can be calculated from $\Gamma[\bar\varphi,\psi^\bullet_{\bar\varphi}]$ by substituting $\bar\phi_{\bar\varphi}^{(\bullet)}=\bar\phi^{(\bullet)}[\bar\varphi,{\psi^\bullet_{\bar\varphi}}]$ in Eq.~\eqref{eq:1PIfromGammaPhi}, where $\psi^\bullet_{\bar\varphi}$ is the solution of $\Gamma_\bullet=0$ for $J_{(\bullet)}=0$ (as follows from $\Gamma_\bullet=\frac{\delta \bar{\phi}^{(\bullet)}}{\delta \psi^\bullet}\Gamma_{(\bullet)}$ and $\Gamma_{(\bullet)}=0$). In this case, \eqref{eq:1PIfromGammaPhi} reduces to $\Gamma^\mathrm{1PI}[\bar\varphi]=\Gamma[\bar\varphi, \psi^\bullet_{\bar\varphi}]$, which is Eq.~\eqref{eq:1PI_via_Gamma}.
Equation~\eqref{eq:1PI_via_Gamma} will be used in Sec.~\ref{sec:1PIEffectiveActionViaCompositeFields} to decompose 1PI vertices into contributions whose irreducibility is determined by the choice of composite fields $\psi^\bullet$, and in Sec.~ \ref{sec:SymmetricEstimators}, focusing on $\psi^\bullet=G^\bullet$ with $\bullet=(2,3)$, to derive a representation of 1PI vertices via the Green’s functions of composite fields. Remarkably, all the results can be expressed using simple tree diagrams, as shown in Fig.~\ref{fig:results}. 

\section{Decomposition of the 1PI vertices via composite fields}\label{sec:1PIEffectiveActionViaCompositeFields}

 It is computationally advantageous to use local composites to decompose renormalized interactions: their field indices involve fewer continuous degrees of freedom, thereby reducing the dimensionality of composite field Green's and vertex functions. In this section, we use the effective action to study this idea in detail and show how various self-consistent schemes can be derived that allow one to renormalize 1PI vertices with a reduced numerical scaling of their building blocks.

First, we consider local fermion bilinears, which may represent density, magnetization, pairing fields, or chiral four-vectors, among others. For a purely fermionic system with only quartic interactions, the single-boson exchange (SBE) formalism is recovered (and summarized in Fig.~\ref{fig:SBEformalism}), which treats different channels on an equal footing. The functional approach allows us to extend it to more general systems such as QCD, for which diagrammatic arguments are complicated by mixed interactions and the inapplicability of the Hubbard-Stratonovich trick. In Section \ref{subsec:1PIVerticesViaGeneralCompositeFields}, general composite fields are considered, so one can include more exchange processes, such as fermion-boson exchange as explained in Appendix \ref{app:beyond_SBE}. 

In the following subsection, we immediately present the results, using local analogs of the functional relations from Section~\ref{sec:compositeFieldMethods}. Readers who are comfortable with the condensed notation and prefer to see the derivation first may wish to start with \ref{subsec:1PIVerticesViaGeneralCompositeFields}--\ref{subsec:calculationEffectiveAction}, and then return to \ref{subsec:ExchangeOfCompositeParticles} afterward.

\subsection{1PI vertices via exchange of composite bosons}\label{subsec:ExchangeOfCompositeParticles}

Let us consider a theory containing complex fermion fields (e.g., electrons or quarks), $f_i(\tau)$ and $f_i^\dagger(\tau)$, where $i$ represents the remaining spatial, spin, and other indices. We construct a local (in time) composite field
\begin{align}\label{eq:definition_local_phi}
    \phi^{(\alpha)}(\tau)\equiv f_i^\dagger(\tau)f_j(\tau) U^\alpha_{ij},
\end{align}
where the index $\alpha$ enumerates different composite field components. For this quantity to be real, the coefficients must satisfy $(U^\alpha_{ij})^*=U^\alpha_{ji}$. In the condensed notation, we write $\phi^{(\bullet)}$, where the composite index is $\bullet=\alpha\tau$. 

The fundamental field $\varphi^a$ may contain additional components (e.g., photons or gluons and ghosts), whose statistics will not be important here. We will only assume that correlators containing an odd number of fields $f_i^\dagger(\tau)$ and $f_j(\tau)$ vanish.

Motivated by the results of Section \ref{sec:compositeFieldMethods}, we are interested in decomposing 1PI vertices in terms of lower-order blocks by differentiating the relation $\Gamma^{\mathrm{1PI}}[\bar\varphi]=\Gamma[\bar\varphi,\psi_{\bar\varphi}^\bullet]$, where $\psi^\bullet$ is some function of $\bar f_i^\dagger(\tau)$, $\bar f_j(\tau)$ and $\bar\phi^{(\alpha)}(\tau)$ to be chosen, whose physical configuration $\psi^\bullet_{\bar\varphi}$ is determined by the equation of motion, $\frac{\delta\Gamma}{\delta\psi^\bullet}=0$. To this end, we first specify our generating functional in \eqref{eq:W_definition}.

We can achieve the effect of the Hubbard-Stratonovich transformation with the following modification of the classical action \footnote{Indeed, representation of the last term in Eq.~\eqref{eq:S_for_SBE} via the Gaussian path integral leads to the Hubbard-Stratonovich transformed theory, in which all sources appear coupled linearly to the fields. },
\begin{align}\label{eq:S_for_SBE}
{S}[\varphi,J_{(\bullet)}]=S[\varphi]+\frac{1}{2}\int \mathrm{d}\tau J_{(\alpha)}(\tau)\chi_0^{\alpha|\beta}J_{(\beta)}(\tau),
\end{align}
where a real field $J_{(\alpha)}(\tau)$ couples linearly to $\phi^{(\alpha)}(\tau)$ in the path integral \eqref{eq:W_definition} and $\chi^{\alpha|\beta}_0$ is a free parameter alongside $U^\alpha_{ij}$. The results of this section are easily generalized to the case of time-nonlocal $\chi_0^{\alpha|\beta}$. By performing the Legendre transform of the generating functional $W[J_a, J_{(\bullet)}]$ in \eqref{eq:W_definition} w.r.t. $J_a$ and $J_{(\bullet)}$, we define $\Gamma[\bar\varphi,\bar\phi^{(\bullet)}]$, and then $\Gamma[\bar\varphi,\psi^\bullet]$ after a possible change of basis (as explained in Section \ref{subsec:generalEffectiveAction}).

The simplest choice for the composite variable in the effective action is $\psi^\alpha(\tau)\equiv \bar\phi^{(\alpha)}(\tau)$, where $\bar\phi^{(\alpha)}(\tau)\equiv \frac{\delta W}{\delta J_{(\alpha)}(\tau)}\overset{J_{(\bullet)}\rightarrow0}{\longrightarrow}\langle\phi^{(\alpha)}(\tau)\rangle$. Written more explicitly,
\begin{align}\label{eq:definition_local_psi}
 \psi^\alpha(\tau)\overset{J_{(\bullet)}\rightarrow0}{\longrightarrow} U^\alpha_{ij}\bar f_i^\dagger(\tau)\bar f_j(\tau)+U^\alpha_{ij}G^{ji}(\tau,\tau),
\end{align}
where we have introduced the conventional fermionic Green's function $G^{ji}(\tau,\tau')\equiv G^{ab}$, with $a,b$ corresponding to $f_j(\tau)$ and $f^\dagger_i(\tau')$, respectively. The results of this section can be straightforwardly modified for less trivial choices of $\psi^\bullet$, such as the amplitude and phase components of $\bar\phi^{(\bullet)}$.

In analogy to the derivations of 1PI vertices \eqref{eq:3pvertex} and \eqref{eq:parquet} from the 3PI effective action, we will show in the next subsection that 1PI vertices of arbitrary order admit a tree expansion in terms of derivatives of the composite effective action $\Gamma[\bar\varphi,\psi^\bullet]$. This implies, in particular, that $\Gamma^\mathrm{1PI}_{nklm}(\tau_n,\tau_k,\tau_l,\tau_m)\equiv \tfrac{\delta\vphantom{\delta\Gamma^\mathrm{1PI}}}{\delta \bar f^\dagger_n(\tau_n)}
\tfrac{\delta\vphantom{\delta\Gamma^\mathrm{1PI}}}{\delta \bar f^\dagger_k(\tau_k)}
\tfrac{\delta\vphantom{\delta\Gamma^\mathrm{1PI}}}{\delta \bar f_l^{\vphantom{\dagger}}(\tau_l)}
\tfrac{\delta\Gamma^\mathrm{1PI}}{\delta \bar f_m^{\vphantom{\dagger}}(\tau_m)}$ can be written as the fourth derivative of $\Gamma[\bar\varphi, \psi^\bullet]$ w.r.t. the fermionic fields $\bar f_i(\tau)$ and $\bar f_j^\dagger(\tau')$, plus the sum of tree diagrams constructed from vertices $I_{ij|\alpha}(\tau,\tau'|\tau'')\equiv \frac{\delta}{\delta\bar f^\dagger_i(\tau)}\frac{\delta^2\Gamma}{\delta\bar f_j(\tau')\delta\psi^\alpha(\tau'')}$ with the bosonic propagator $\chi^{\alpha|\beta}$ as internal lines \footnote{This result can also be obtained by amputating the four-point Green's function when expressed via the standard tree expansion in terms of derivatives of $\Gamma[\bar\varphi,\bar\phi^{(\bullet)}]$.}: 
\begin{align}\label{eq:4p1PIvertex_via_boson_exchange}
    \Gamma_{nklm}^\mathrm{1PI}=\Gamma_{nklm}+(\int \mathrm{d} \tau _\alpha \mathrm{d}\tau _\beta I_{kl|\alpha}\chi^{\alpha|\beta}I_{nm|\beta}-n\leftrightarrow k),
\end{align}
where time arguments are suppressed and $\Gamma_{nklm}$ is defined analogously to $\Gamma^\mathrm{1PI}_{nklm}$.
The bosonic propagator is defined by $\int \mathrm{d}\tau' \chi^{\alpha|\beta}(\tau,\tau')\frac{\delta^2\Gamma}{\delta\psi^\beta(\tau')\delta\psi^\gamma(\tau'')}=-\delta^\alpha_\gamma\delta^{\tau}_{\tau''}$, which, with the help of Eq.~\eqref{eq:K2}, relates it to the two-point correlator of $\phi^{(\bullet)}$.

By varying $\chi^{\alpha|\beta}_0$ and $U^\alpha_{ij}$, we aim to simplify the structure of the residual interactions (such as $\Gamma_{nklm}$), which can be obtained by differentiating the equation of motion considered as a functional of $\boldsymbol{\psi}=(\bar\varphi,\psi^{\bullet})$ (see Section \ref{subsec:calculationEffectiveAction}):
\begin{align}\label{eq:SDE_npGamma}
    \frac{\delta}{\delta\bar f^\dagger_n}...\frac{\delta\Gamma}{\delta\bar f_l}=\frac{\delta}{\delta\bar f^\dagger_n}...\Big (\langle \frac{\delta S_{\mathrm{int}}[\varphi]}{\delta f_l}\rangle-I^0_{il|\beta}\bar f^\dagger_i\langle \phi^{(\beta)}\rangle \Big )_{\boldsymbol{\psi}},
\end{align}
where $I^0_{il|\beta}\equiv -U_{il}^{\alpha}\chi^{-1}_{0,\alpha|\beta}$ and the ellipsis denotes two or more derivatives w.r.t. $\boldsymbol{\psi}$. One way to increase the number of free parameters is to introduce additional composites, such as the pairing fields $f_i(\tau)f_j(\tau)$ and $f_i^\dagger (\tau)f_j^\dagger(\tau)$, which make an additional exchange term explicit in the decomposition of the four-point 1PI vertex \eqref{eq:4p1PIvertex_via_boson_exchange}.
\begin{figure*}
    \centering
    
    \includegraphics[width=\linewidth]{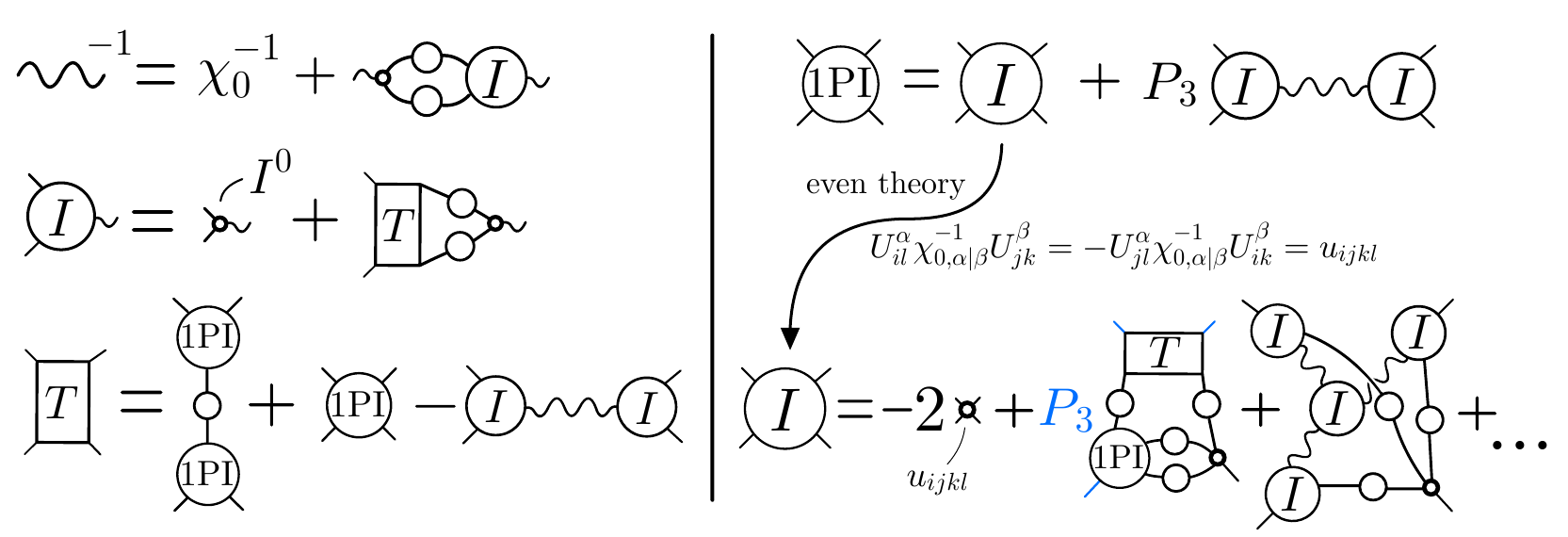}
    
    \caption{Left: Schwinger-Dyson equation \eqref{eq:propagator_of_composite_boson} and Bethe-Salpeter-type equation \eqref{eq:BS_gU_n=2}. Right: SBE decomposition \eqref{eq:decomposition_4p1PI} and Schwinger-Dyson equation \eqref{eq:SD_I_4} for the residual interaction (for an even theory choosing $U^\alpha_{il}\chi^{-1}_{0,\alpha|\beta}U^\beta_{jk}=-U^\alpha_{jl} \chi_{0,\alpha|\beta}^{-1}U^\beta_{ik}=u_{ijkl}$ and analogously for the pairing coefficients). The ellipsis denotes terms involving the vertices $\Gamma_{ab\bullet\bullet}$, $\Gamma_{abcd\bullet}$ and $\Gamma_{abcdef}$.}
    \label{fig:SBEformalism}
\end{figure*}

For example, consider a fermionic theory with only quartic interactions,
\begin{align}\label{eq:quartic_fermion_interaction}
    S_\mathrm{int}=\tfrac{1}{4}u_{ijkl}\int \mathrm{d}\tau f^\dagger_i(\tau)f^\dagger_j(\tau)f_k(\tau)f_l(\tau),
\end{align}
where the bare vertex satisfies the antisymmetries $u_{ijkl}=-u_{jikl}$ and $u_{ijkl}=-u_{ijlk}$. One can choose $U^\alpha_{ij}$ and $\chi^{\alpha|\beta}_0$ together with the corresponding coefficients for the pairing fields, such that the partially disconnected parts of $\langle \frac{\delta S_{\mathrm{int}}[\varphi]}{\delta f_l}\rangle$, given by $ -u_{ijkl}\bar f^\dagger_i\langle f^\dagger_j f_k\rangle $ and $-\tfrac{u_{ijkl}}{2}\bar f_k\langle f^\dagger_if^\dagger_j\rangle$, are canceled in Eq.~\eqref{eq:SDE_npGamma} (with inclusion of pairing components), thereby simplifying the evaluation of the residual interactions. In particular, one can show that $\Gamma_{ijkl}$ simplifies to a sum of interaction-irreducible diagrams, which is illustrated by the Schwinger-Dyson equation sketched on the right of Fig.~\ref{fig:SBEformalism} and obtained from Eq.~\eqref{eq:SDE_npGamma} in Section \ref{subsec:calculationEffectiveAction}. Here, nonlinear response functions (depicted as an irreducible vertex with three wavy lines) are also needed to partially reconstruct $\Gamma_{ijkl}$.

For theories with both fermionic and fundamental bosonic degrees of freedom, such as QCD, the term $\langle \frac{\delta S_{\mathrm{int}}}{\delta f_l}\rangle$ contains mixed fermion-boson correlators. Using the equations of motion of the fundamental bosonic fields, these can be rewritten in terms of higher-order correlators, including fermion trilinears \footnote{To see this in QCD, we substitute the Schwinger-Dyson equation $-J_A(\tau')=\langle\tfrac{\delta S_0[\varphi]}{\delta A(\tau')}\rangle+\langle \tfrac{\delta S_\mathrm{int}[\varphi]}{\delta A(\tau')}\rangle$ for the gluon field $A(\tau)$ (where spatial, Lorentz and color indices are suppressed) into $0=\tfrac{\delta J_{A}(\tau')}{\delta J_l(\tau)}$ to get an exact identity: $\langle A (\tau'')f_l(\tau)\rangle_c=\int \mathrm{d}\tau'G^A_0 (\tau'',\tau')\langle \tfrac{\delta S_\mathrm{int}[\varphi]}{\delta A(\tau')}f_l(\tau)\rangle_c$, where $G^A_0$ is a bare gluon propagator. The derivative $\frac{\delta S_{\mathrm{int}}[\varphi]}{\delta A(\tau')}$ is equal to the sum of fermion currents and gluon self-interaction terms.}. Consequently, one can similarly simplify the structure of the residual interactions to efficiently express 1PI vertices of the fundamental fields via the exchange of composite bosons.

Following the arguments of Section~\ref{subsec:preliminary}, we can obtain equations for all the building blocks appearing in the vertex decompositions such as the Schwinger-Dyson (SD) equation \eqref{eq:propagator_of_composite_boson} for $\chi^{\bullet|\bullet}$ and the Bethe-Salpeter equation \eqref{eq:BS_gU_n=2} for $I_{ab|\bullet}$, which are depicted on the left of Fig.~\ref{fig:SBEformalism} and are derived below. We see that for a purely fermionic system (for which $\Gamma^\mathrm{1PI}_{abe}=0$), the set of equations for $\chi^{\alpha|\beta}$, $I_{ij|\alpha}$ and $\Gamma^\mathrm{1PI}_{ijkl}$ is closed \footnote{For the full propagator we can use the Schwinger-Dyson equation~\eqref{eq:SD_selfEnergy}.} provided the residual interaction, $\Gamma_{ijkl}$, is known. After Fourier transformation, Eqs.~\eqref{eq:propagator_of_composite_boson}, \eqref{eq:BS_gU_n=2} (with the appropriate modifications for the pairing-field components) and the vertex decomposition \eqref{eq:4p1PIvertex_via_boson_exchange} correspond to the single-boson exchange (SBE) equations~(6), (B3) and (8) of \cite{Krien2019SBE2}, respectively~\footnote{Our $I_4$, $\chi^{\bullet|\bullet}$ and $I_{2|\bullet}$ correspond to $\varphi^\mathrm{firr,\alpha}-2U^\alpha$, $\omega^\alpha$ and $\lambda^\alpha$ of \cite{Krien2019SBE2}, respectively}.

This functional framework suggests a systematic way to improve the SBE scheme by reconstructing $\Gamma_{ijkl}$ with more physically transparent building blocks. For example, nonlinear response functions (or three-point bosonic vertices) provide further resummation but require the calculation of six-point functions. We can then apply the tree expansion to the six-point 1PI vertex, which allows us to approximate it in terms of lower-order blocks we already have \footnote{To ensure that diagrams are not overcounted when applying approximations, the diagrammatic content of high-order vertices, such as $\Gamma_{ab\bullet\bullet}$, $\Gamma_{abcd\bullet}$ and $\Gamma_{abcdef}$, can be tracked using Eq.~\eqref{eq:SDE_npGamma}. Explicit evaluation confirms that the lowest-order contributions to $\Gamma_4$ originating from $\Gamma_{ab\bullet\bullet}$ are distinct from the terms shown in Fig.~\ref{fig:SBEformalism}.}. 

It therefore remains to be explained how a closed description can eventually be obtained by relating such building blocks back to the 1PI vertices. We first focus on the equations for $\chi^{\bullet|\bullet}$ and $I_{ab|\bullet}$, and consider more general composite Green's and vertex functions in the next subsection.

The starting point is the equation of motion, $\frac{\delta \Gamma}{\delta \psi^\alpha(\tau)}=0$ for $J_{(\alpha)}(\tau)=0$, which defines the functional $\psi^\alpha=\psi^\alpha_{\bar\varphi}$. In direct analogy to the derivation of the Bethe-Salpeter-type equation~\eqref{eq:BS_0},  we take the derivative of $\frac{\delta \Gamma}{\delta \psi^\alpha(\tau)}=0$ at $\psi^\alpha=\psi^\alpha_{\bar\varphi}$ w.r.t. $\bar f_l(\tau_l)$. Using the chain rule, this gives $\int d\tau_\beta\frac{\delta\psi^\beta_{\bar\varphi}}{\delta\bar f_l}\frac{\delta^2\Gamma}{\delta\psi^\beta\delta\psi^\alpha}+\frac{\delta^2\Gamma}{\delta \bar f_l\delta\psi^\alpha}=0$, which leads to $\frac{\delta\psi^\alpha_{\bar\varphi}}{\delta \bar f_l}=\int d\tau_\beta\frac{\delta^2\Gamma}{\delta\bar f_l\delta\psi^{\beta}}\chi^{\beta|\alpha}$. Taking one more derivative of both sides w.r.t. $\bar f^\dagger_k(\tau_k)$ then yields
\begin{align}\label{eq:BS_n=2_local}
    \tfrac{\delta^2\psi^{\alpha}_{\bar\varphi}(\tau)}{\delta\bar f^\dagger_k(\tau_k)\delta\bar f_l(\tau_l)}=\int \mathrm{d}\tau_\beta I_{kl|\beta}(\tau_k,\tau_l|\tau_\beta)\chi^{\beta|\alpha}(\tau_\beta,\tau).
\end{align}

The left side of \eqref{eq:BS_n=2_local} contains a four-point object. On the one hand, it can be related to the four-point Green's function using $G\tfrac{\delta}{\delta\bar\varphi}=\tfrac{\delta}{\delta J_1}$ and $\tfrac{\delta}{\delta J_a}\dots\tfrac{\delta}{\delta J_b}\bar\phi^{(\bullet)}=G^{a\dots b(\bullet)}$, which after contraction with a bare vertex, can be related to the two-point correlator of $\phi^{(\bullet)}$, i.e., $\chi^{\bullet|\bullet}$. Equation~\eqref{eq:BS_n=2_local} then reduces to the SD equation \footnote{More explicitly, we have:
\begin{align}
G^{bd}G^{ac}\tfrac{\delta^2\bar\phi^{(\bullet)}_{\bar\varphi}}{\delta\bar\varphi^{c}\delta\bar\varphi^{d}}=G^{ab(\bullet)}+G^{abc}\Gamma^\mathrm{1PI}_{cd}G^{d(\bullet)},    
\end{align}
where the indices $a,b$ correspond to fermionic fields $\varphi^a=f^\dagger_i(\tau)$ and $\varphi^b=f_j(\tau)$. With Eq.~\eqref{eq:K2_general}, the right side of this equation reduces to $\chi_0^{\alpha|\beta}\delta^\tau_{\tau'}-\chi^{\alpha|\beta}(\tau,\tau')$ if $\bullet=\beta\tau'$ and indices $i,j$ are contracted by $U^\alpha_{ij}$. For the left-hand side, we use $\bar\phi^{(\bullet)}=\psi^\bullet$ and Eq.~\eqref{eq:BS_n=2_local}, which leads to the SD equation \eqref{eq:propagator_of_composite_boson} for $\chi^{\alpha|\beta}(\tau,\tau')$.} for $\chi^{\bullet|\bullet}$,
\begin{multline}\label{eq:propagator_of_composite_boson}
    \chi^{\alpha|\beta}(\tau,\tau')=\chi_0^{\alpha|\beta}\delta^\tau_{\tau'}+\chi_0^{\alpha|\gamma}\int \mathrm{d}\tau''\Pi_{{\gamma\rho}}(\tau,\tau'')\chi^{\rho|\beta}(\tau'',\tau'),\\
    \Pi_{\gamma\rho}(\tau,\tau'')\equiv I^0_{ij|\gamma}\int \mathrm{d}\tau_n \mathrm{d}\tau_m G^{mi}(\tau_m,\tau)G^{jn}(\tau,\tau_n)\\\times I_{nm|\rho}(\tau_n,\tau_m|\tau''),
\end{multline}
where $\chi_0^{\alpha|\beta}$ and $I^0_{ij|\gamma
    }\equiv -U^{\gamma'}_{ij}\chi^{-1}_{0,\gamma'|\gamma}$ play the roles of the bare propagator and the bare vertex, respectively.
On the other hand, the left side, $\tfrac{\delta^2\psi^{\alpha}_{\bar\varphi}(\tau)}{\delta\bar f^\dagger_k(\tau_k)\delta\bar f_l(\tau_l)}$,  of \eqref{eq:BS_n=2_local} can be expressed in terms of 1PI vertex functions using Eqs.~\eqref{eq:definition_local_psi} and \eqref{eq:1PI_Invertibility} (extended to a mixed system). Substituting Eq.~\eqref{eq:propagator_of_composite_boson} once for $\chi^{\beta|\alpha}(\tau,\tau')$ on the right side of \eqref{eq:BS_n=2_local}, we then obtain a Bethe-Salpeter-type equation for $I_{kl|\beta}(\tau_k,\tau_l|\tau)$,
\begin{multline}\label{eq:BS_gU_n=2}
I_{kl|\beta}(\tau_k,\tau_l|\tau)=I^0_{kl|\beta}\delta^\tau_{\tau_k}\delta^\tau_{\tau_l}+\\\int \mathrm{d}\tau_n \mathrm{d}\tau_m T_{nklm}(\tau_n,\tau_k,\tau_l,\tau_m)G^{mi}(\tau_m,\tau)G^{jn}(\tau,\tau_n)I^0_{ij|\beta},\\
T_{abcd}\equiv \Gamma^\mathrm{1PI}_{cae}G^{ee'}\Gamma^\mathrm{1PI}_{e'bd}+\Gamma^\mathrm{1PI}_{abcd}- I_{bc|\bullet }\chi^{\bullet|\bullet}I_{ad|\bullet},
\end{multline}
 where $T_{nklm}(\tau_n,\tau_k,\tau_l,\tau_m)\equiv T_{abcd}$ with the multi-indices matched according to their order \footnote{Note that the field types are fixed by the definitions of $I_{kl|\beta}(\tau_k,\tau_l|\tau)$ and the fermionic propagator in Eq.~\eqref{eq:BS_gU_n=2}. For example, the first index $n$ appears in $G^{jn}(\tau,\tau_n)$, which implies that the corresponding field is $\varphi^a=f^\dagger_n(\tau_n)$. }. We use a similar explicit notation for $\Gamma^\mathrm{1PI}_{abcd}$.

An alternative way to simplify the structure of the residual interaction is to include correlators of the fermions $f_i(\tau)$, $f_j^\dagger(\tau')$, and the corresponding local bilinears, as explained in Appendix~\ref{app:beyond_SBE}. Such a choice allows one to go beyond the single-boson exchange by incorporating fermion-boson exchange processes. Representations of this type are well established, for example, in the $GW$ approximation for the self-energy and in QCD truncations of the quark-gluon vertices \cite{Mitter2015QCD, Eichmann2016Baryons}.

\subsection{Tree expansions for general composite fields and vertex order}\label{subsec:1PIVerticesViaGeneralCompositeFields}

Here, we prove the tree expansions of the previous subsection and extend them (as well as the results of \ref{subsec:preliminary}) to general composite effective actions and higher vertex orders. Specifically, differentiating Eq.~\eqref{eq:1PI_via_Gamma}, we obtain the following simple rule for decomposing a general $n$-point 1PI vertex:

Sum over all possible tree diagrams built from vertices $I_n$ and $I_{n|\bullet|...|\bullet}$ ($n>1$), connected by $\chi^{\bullet|\bullet}$ as internal lines, where $I_n$ and $I_{n|\bullet|...|\bullet}$ are defined through derivatives of $\Gamma[\bar\varphi,{\psi}^\bullet]$ (see Eqs.~\eqref{eq:definition_In}, \eqref{eq:definition_I_bullet_n} and \eqref{eq:derivativeOfpsi^bullet}). All Bethe-Salpeter-type equations are obtained via similar tree expansions, in which some of the external indices are composite.

In Section \ref{subsec:preliminary},  $\psi^\bullet$ were chosen to be connected correlation functions $G^{ab}$ and $G^{abc}$. In this case, $I_n$ acquires interpretation in terms of 3PI diagrams, and $\chi^{\bullet|\bullet}$ becomes a multi-particle Green’s function. In Section \ref{subsec:ExchangeOfCompositeParticles}, $\psi^\bullet$ are bilinear fields local in time. There, $I_n$ corresponds to the sum of interaction–irreducible diagrams, and $\chi^{\bullet|\bullet}$ becomes the propagator of the composite boson particle. Now, we consider general $\psi^\bullet$.

To keep things simple, we set $J_{(\bullet)}=0$ and assume that $\phi^{(\bullet)}$ and $\psi^\bullet$ are bosonic fields, so that $\gamma^{(\bullet)}_{(\bullet)}=\delta^{(\bullet)}_{(\bullet)}$ and $\gamma^\bullet_\bullet=\delta^\bullet_\bullet$. For fermionic systems ($\zeta=-1$), we further assume that only even correlation and vertex functions are nonzero. 

We begin by differentiating Eq.~\eqref{eq:1PI_via_Gamma} with respect to $\bar\varphi^a$:
\begin{align}\label{eq:firstDerivative1PIfromGammaPsi}
\Gamma^{\text{1PI}}_a=\frac{\delta \psi^{\mathbf{a}}_{\bar\varphi} }{\delta \bar\varphi^a}{\Gamma_{\mathbf{a}}},
\end{align}
 where $\psi^{\mathbf{a}}_{\bar\varphi}\equiv\begin{pmatrix} \bar\varphi , \psi^\bullet_{\bar\varphi}\end{pmatrix}$. To express $\frac{\delta\psi_{\bar\varphi}^{\bullet}}{\delta\bar\varphi^a}$ via derivatives of $\Gamma[\bar\varphi,\psi^\bullet]$, consider the equation of motion $\Gamma_\bullet=0$ at $\psi^\bullet=\psi^\bullet_{\bar\varphi}$. Applying the derivative $\frac{\delta}{\delta\bar\varphi^a}$ to both sides of this equation and multiplying from the right by $\chi^{\bullet|\bullet}=-(\Gamma_{\bullet\bullet})^{-1}$, we get
\begin{align}\label{eq:derivativeOfpsi^bullet}
\frac{\delta\psi_{\bar\varphi}^{\bullet}}{\delta\bar\varphi^a}=\Gamma_{a \bullet} \chi^{\bullet|\bullet}.
\end{align}
 
 When differentiating Eq.~\eqref{eq:firstDerivative1PIfromGammaPsi} with respect to $\bar\varphi^b$, the term $\Gamma_{\mathbf{a}}\frac{\delta^2 \psi^{\mathbf{a}}_{\bar\varphi} }{\delta\bar\varphi^b\delta \bar\varphi^a}\sim \Gamma_{\bullet}$ vanishes due to $\Gamma_\bullet=0$. We thus obtain
\begin{align}\label{eq:secondDerivative1PIfromGammaPsi}
\Gamma^{\text{1PI}}_{ba}=\frac{\delta \psi^{\mathbf{b}}_{\bar\varphi} }{\delta \bar\varphi^b}\frac{\delta \psi^{\mathbf{a}}_{\bar\varphi} }{\delta \bar\varphi^a}{\Gamma_{\mathbf{ba}}},
\end{align}
where in the fermionic case, we treat $\frac{\delta \psi^{\mathbf{a}}_{\bar\varphi} }{\delta \bar\varphi^a}$ and $\frac{\delta \psi^{\mathbf{b}}_{\bar\varphi} }{\delta \bar\varphi^b}$ as bosonic because they are nonzero only when the indices $\mathbf{a}$ and $\mathbf{b}$ are not composite. 
Note that we can also write $\Gamma^{\mathrm{1PI}}_{ab}=\frac{\delta \psi^{\mathbf{a}}_{\bar\varphi} }{\delta \bar\varphi^a}\Gamma_{b\mathbf{a}}$ because of $\frac{\delta \psi^{\mathbf{a}}_{\bar\varphi} }{\delta \bar\varphi^a}\Gamma_{\bullet\mathbf{a}}=0$, which follows from differentiation of the equation of motion $\Gamma_\bullet=0$ at $\psi^\bullet=\psi^\bullet_{\bar\varphi}$ with respect to $\bar\varphi^a$ (analogous to the proof of Eq.~\eqref{eq:derivativeOfpsi^bullet}).  
As another result of $\frac{\delta \psi^{\mathbf{a}}_{\bar\varphi} }{\delta \bar\varphi^a}\Gamma_{\bullet\mathbf{a}}=0$, the derivative of Eq.~\eqref{eq:secondDerivative1PIfromGammaPsi} with respect to $\bar\varphi$ acts only on $\Gamma_{\mathbf{ab}}$ (as in Eq.~\eqref{eq:firstDerivative1PIfromGammaPsi}), yielding
\begin{align}\label{eq:firstDerivatives1PIfromGammaPsi}
    \Gamma^\mathrm{1PI}_{a_1...a_n}=I_{a_1...a_n}, \qquad n=1,2,3,
\end{align}
where $I_{a_1...a_n}$, the \textit{fully irreducible vertex} with respect to the composite fields, is defined as
\begin{align}\label{eq:definition_In}
I_{a_1...a_n}=\frac{\delta \psi^{\mathbf{a_1}}_{\bar\varphi} }{\delta \bar\varphi^{a_1}}...\frac{\delta \psi^{\mathbf{a_n}}_{\bar\varphi} }{\delta \bar\varphi^{a_n}}\Gamma_{\mathbf{a_1}...\mathbf{a_n}}.
\end{align}

Applying $\frac{\delta}{\delta\bar{\varphi^a}}$ to  Eq.~\eqref{eq:firstDerivatives1PIfromGammaPsi} for $n=3$ yields the four-point 1PI vertex
\begin{align}\label{eq:decomposition_4p1PI}
\Gamma^\mathrm{1PI}_{abcd}=I_{abcd}+P_3\frac{\delta^2 \psi^\bullet_{\bar\varphi} }{\delta\bar\varphi^a\delta \bar\varphi^b}I_{cd|\bullet}
\end{align}
where $I_{a_1...a_n|\bullet|...|\bullet}$ is defined as
\begin{align}\label{eq:definition_I_bullet_n}
I_{ a_1...a_n|\bullet|...|\bullet}=\frac{\delta \psi^{\mathbf{a_1}}_{\bar\varphi} }{\delta \bar\varphi^{a_1}}...\frac{\delta \psi^{\mathbf{a_n}}_{\bar\varphi} }{\delta \bar\varphi^{a_n}}\Gamma_{\mathbf{a_1}...\mathbf{a_n}\bullet...\bullet},
\end{align}
with the same number of external bullet indices on both sides. We recall that a bullet index $\bullet$ denotes differentiation with respect to $\psi^\bullet$. We refer to $I_{a_1...a_n|\bullet}$ simply as the \textit{irreducible vertex} with respect to the composite fields.

As the simplest example, let us consider ${S}[\varphi, J_{(\bullet)}]={S}[\varphi]$ and $\psi^\bullet=G^{ab}$ for a moment.  Considering an even theory, (with vanishing odd Green's and vertex functions) we have $\frac{\delta^2\psi^{a b}_{\bar\varphi}}{\delta\bar\varphi^{ c}\delta\bar\varphi^{ d}}=\Gamma^\mathrm{1PI}_{c d a' b'}G^{a' a}G^{b' b}$, using Eq.~\eqref{eq:1PI_Invertibility}  for $\psi^{\bullet}$. Eq.~\eqref{eq:decomposition_4p1PI} then reduces to the parquet decomposition \cite{Dominicis1964EntropyII} with $I_{4}=\frac{\delta^4\Gamma}{\delta\bar\varphi^4}$ being fully 2PI vertex.

Returning to the general case, Eq.~\eqref{eq:decomposition_4p1PI} can be cast into a tree expansion form analogous to Eq.~\eqref{eq:treeExpansion_G^4} independently of the choice of $\psi^\bullet$. To achieve this, let us substitute $\psi^\bullet=\psi^\bullet_{\bar\varphi}$ into $\Gamma_{\bullet}=0$ and differentiate it w.r.t. $\bar\varphi$ to get again $\frac{\delta \psi^{\mathbf{ a}}_{\bar\varphi}}{\delta\bar\varphi^{a}}\Gamma_{\mathbf{a}\bullet}=0$. Taking one more derivative, we find $\tfrac{\delta^2\psi^{\bullet}_{\bar\varphi}}{\delta\bar\varphi^{a}\delta\bar\varphi^{b}}\Gamma_{\bullet \bullet}+I_{ab|\bullet}=0$, or
\begin{align}\label{eq:BS_n=2}
\frac{\delta^2 \psi^\bullet_{\bar\varphi} }{\delta\bar\varphi^a\delta \bar\varphi^b}=I_{ab|\bullet}\chi^{\bullet|\bullet}.
\end{align}

Inserting Eq.~\eqref{eq:BS_n=2} into Eq.~\eqref{eq:decomposition_4p1PI} casts the decomposition into the symmetric, tree expansion form
\begin{align}\label{eq:treeExpansion_4p1PI}
\Gamma^\mathrm{1PI}_4=I_4+P_3I_{2|\bullet}\chi^{\bullet|\bullet}I_{2|{\bullet}}.
\end{align}

Now, just as 
\Eq{eq:treeExpansion_4p1PI} has the same structure as 
\Eq{eq:treeExpansion_G^4}, we can derive a recursion relation in the form of \Eq{eq:treeExpansion_rule} to express a general $n$-point 1PI vertex in terms of tree diagrams.
To derive the recursion relation, we need to know how $\frac{\delta}{\delta \bar\varphi}$ acts on  $I_{n...}$. 

Recalling the definition~\eqref{eq:definition_I_bullet_n}, the derivative acts on $\Gamma_{\mathbf{a_1}..\mathbf{a_n}...}$, producing $I_{n+1...}$, and on $n$ factors of $\frac{\delta \psi^\bullet_{\bar\varphi} }{\delta\bar\varphi}$, giving
\begin{align}\label{eq:derivativeOfI}
    \frac{\delta}{\delta \bar\varphi}I_{n...}=I_{n+1...}+P_n\frac{\delta^2 \psi^\bullet_{\bar\varphi} }{\delta\bar\varphi^2}I_{n-1|\bullet...}.
\end{align}

\begin{figure}
    \centering
    
    \includegraphics[width=\linewidth]{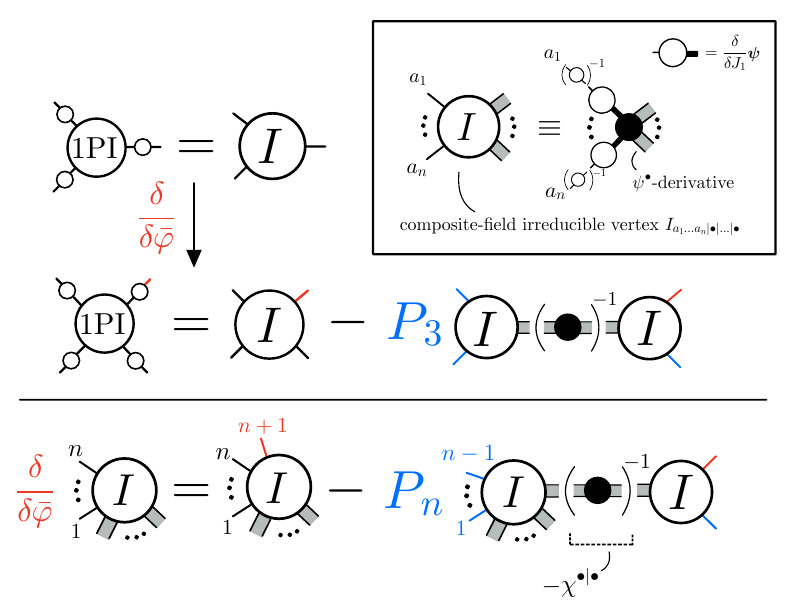}
    
    \caption{Tree expansions of the 1PI vertices $\Gamma_3^\mathrm{1PI}$ and $\Gamma_4^\mathrm{1PI}$ (see Eqs.~\eqref{eq:firstDerivatives1PIfromGammaPsi} and \eqref{eq:treeExpansion_4p1PI}), using the rule in Eq.~\eqref{eq:treeExpansion_I}, with internal lines given by $\chi^{\bullet|\bullet} = -(\Gamma_{\bullet\bullet})^{-1}$. The definition of $I_{a_1\ldots a_n|\bullet|...|\bullet}$ is provided in Eq.~\eqref{eq:definition_I_bullet_n}.
    }
    \label{fig:ItreeExpansion}
\end{figure}
Using Eq.~\eqref{eq:BS_n=2}, we obtain 
\begin{align}\label{eq:treeExpansion_I}
    \frac{\delta}{\delta \bar\varphi}I_{n...}=I_{n+1...}+P_nI_{2|\bullet }\chi^{\bullet|\bullet}I_{n-1|\bullet|...}.
\end{align}

Together with 
\begin{align}
    \frac{\delta}{\delta \bar\varphi}\chi^{\bullet|\bullet}=\chi^{\bullet|\bullet}I_{1|\bullet|\bullet}\chi^{\bullet|\bullet},
\end{align} Eq.~\eqref{eq:treeExpansion_I} generates the tree expansion (see Fig.~\ref{fig:ItreeExpansion}), in direct analogy to Eq.~\eqref{eq:treeExpansion_rule}. Applying the rule \eqref{eq:treeExpansion_I} to $\Gamma^\mathrm{1PI}_n=(\frac{\delta}{\delta \bar\varphi})^{n-3}I_3$ (see Eq.~\eqref{eq:firstDerivatives1PIfromGammaPsi}) proves that any $n$-point 1PI vertex can be expressed as the sum of all tree diagrams constructed from $I_m$ and $I_{m|\bullet|...|\bullet}$, with internal bullet indices connected by $\chi^{\bullet|\bullet}$.

For $n=4$, this reproduces \Eq{eq:treeExpansion_4p1PI}, while for $n=5$, one finds
\begin{align}\label{eq:treeExpansion_5p1PI}
\Gamma^{\mathrm{1PI}}_5=I_5+P_{10}I_{3|\bullet}\chi^{\bullet|\bullet}I_{2|\bullet}+P_{15}I_{2|\bullet}\chi^{\bullet|\bullet}I_{1|\bullet|\bullet}\chi^{\bullet|\bullet}I_{2|\bullet}.
\end{align}

Bethe-Salpeter-type equations for $I_{n|\bullet}$ can be derived by applying the rule \eqref{eq:treeExpansion_I} to $\frac{\delta^n \psi^\bullet_{\bar\varphi} }{\delta\bar\varphi^n}=\frac{\delta^n}{\delta\bar\varphi^{n-2}}(I_{2|\bullet}\chi^{\bullet|\bullet})$ (see Eq.~\eqref{eq:BS_n=2}). For $n=3$, this gives
\begin{align}\label{eq:BS_n=3}
\frac{\delta^3 \psi^\bullet_{\bar\varphi} }{\delta\bar\varphi^3}=(I_{3|\bullet }+P_3I_{2|\bullet}\chi^{\bullet|\bullet}I_{1|\bullet|\bullet})\chi^{\bullet|\bullet}.
\end{align}

Overall, this formalism makes the analysis of (ir)reducible parts of general 1PI vertices straightforward and systematic: decompositions and BS-type equations for $I_{n|\bullet...}$ follow from the same rule \eqref{eq:treeExpansion_I} and are represented by sums of tree diagrams of a common form. 

The missing piece for getting a closed description is the fully irreducible vertex $I_n$, for which we need to know the structure of the composite effective action. In the next subsection, we show how the Schwinger–Dyson equations for derivatives of $\Gamma$ can be derived to study its structure.

\subsection{Structure of the effective action}\label{subsec:calculationEffectiveAction}
While rules for the diagrammatic content of the 2PI effective action via the Luttinger-Ward functional in terms of the full propagator $G$ are well established \cite{LuttWard1960Ground, Cornwall1974Action2PI}, the corresponding rules for $\Gamma[\bar{\varphi},\psi^{\bullet}]$ for a general $\psi^{\bullet}$ are less straightforward. For the choice $\psi^{\bullet} = G^{\bullet}$, the $n$PI community developed loop expansions of the $n$PI effective action using successive Legendre transformations \cite{Dominicis1964EntropyI, Dominicis1964EntropyII, Berges2004nPI}, while for a local composite field a perturbative expansion via the inversion method has been proposed \cite{Fukuda1995Inversion}. Here, we present a simple recipe for deriving Schwinger-Dyson equations for the vertices built out of $\Gamma[\bar{\varphi},\psi^{\bullet}]$. From these equations, the type of irreducibility and the diagrammatic content of the composite effective action can be deduced.

In direct analogy with the Schwinger–Dyson (SD) equation for a 1PI vertex, we can derive an SD-type equation for the $\psi$-derivatives of the composite field effective action. The starting point is the identity $0=e^{W}\int \mathrm{D}\varphi\,\frac{\delta}{\delta\varphi^a}(e^{-S[\varphi,J_{(\bullet)}]-J_\mathbf{a}\phi^\mathbf{a}[\varphi]})$, from which we obtain, using $\gamma^\mathbf{b}_\mathbf{a}J_\mathbf{b}=-\tfrac{\delta\psi^\mathbf{b}}{\delta\bar\phi^\mathbf{a}}\Gamma_\mathbf{b}$,
\begin{align}\label{eq:SD_1p_Gamma}
    \Gamma_a=\langle\frac{\delta S[\varphi,J_{(\bullet)}]}{\delta \varphi^a}\rangle_{\boldsymbol{\psi}} -\langle\frac{\delta\phi^\mathbf{a}[\varphi]}{\delta\varphi^a}\rangle_{\boldsymbol{\psi}}\frac{\delta\psi^\bullet}{\delta\bar\phi^\mathbf{a}}\Gamma_\bullet,
\end{align}
where expectation values on the right side are considered as functionals of $\boldsymbol{\psi}$. The key idea is to first express these expectation values in terms of connected Green's functions and subsequently rewrite them in terms of 1PI vertices via the standard tree expansion. These vertices are then decomposed in terms of derivatives of the composite effective action $\Gamma[\boldsymbol{\psi}]$ (see Section \ref{subsec:1PIVerticesViaGeneralCompositeFields}).

To illustrate this, let us consider the case of the 1PI effective action, for which the last term does not appear. Expressing the right side in terms of connected Green's functions and relating them to 1PI vertices (via $G_{\bar\varphi}=(-\zeta\Gamma^\mathrm{1PI}_2)^{-1}$, $G^3_{\bar\varphi}=(G_{\bar\varphi})^3\Gamma^\mathrm{1PI}_3$, etc.) leads to the Schwinger-Dyson equation \eqref{eq:1pVertex} written entirely in terms of derivatives of the 1PI effective action. By iterating this equation, starting from the approximation $\Gamma^\mathrm{1PI}\approx S$, one obtains $\Gamma^\mathrm{1PI}$ as a sum of only 1PI diagrams \cite{DeWitt1966Fields}.

For the $n$PI effective action with $\psi=G^\bullet$, the last term also vanishes (see Eq.~\eqref{eq:varphiDerivativeOfGamma}). It then follows that, for interactions of order up to $n+1$, the right side can be expressed in terms of Green's functions up to order $n$, which are precisely the variables of the effective action. This leads to the structure \eqref{effActionStructure}.

In what follows, we use a similar procedure to investigate the structure of the local composite effective action employed in Section~\ref{subsec:ExchangeOfCompositeParticles}.

\subsubsection*{Local composite fields}
Let us return to the choice of a local composite field $\psi^{\bullet}=\bar\phi^{(\bullet)}$, where $\phi^{(\bullet)}[\varphi]$ is defined in Eq.~\eqref{eq:definition_local_phi}. Recovering the composite source in \eqref{eq:definition_local_psi}  gives 
\begin{align}\label{eq:composite_source_via_phi}
    \frac{\delta\Gamma}{\delta\psi^\alpha(\tau)}=-J_{(\alpha)}(\tau)=\Big (\langle\phi^{(\beta)}(\tau)\rangle-\bar\phi^{(\beta)}(\tau)\Big )\chi^{-1}_{0,\beta|\alpha}.
\end{align}

Using \eqref{eq:composite_source_via_phi} and $\frac{\delta\psi^\bullet}{\delta\bar\phi^{(\mathbf{a})}}=\delta^\bullet_{\mathbf{a}}$ in the last term of Eq.~\eqref{eq:SD_1p_Gamma}, we obtain
\begin{align}\label{eq:SDE_1p1PI}
    \frac{\delta\Gamma}{\delta\bar f_l(\tau)}=\langle \frac{\delta S[\varphi]}{\delta f_l(\tau)}\rangle_{\boldsymbol{\psi}}+I^0_{il|\beta}\bar f^\dagger_i(\tau)\Big (\psi^\beta(\tau)-\langle \phi^{(\beta)}(\tau)\rangle \Big )_{\boldsymbol{\psi}},
\end{align}
which leads to Eq.~\eqref{eq:SDE_npGamma}.

For a purely fermionic system with an interaction given by Eq.~\eqref{eq:quartic_fermion_interaction}, we find that $\langle\frac{\delta S_\mathrm{int}}{\delta f_l(\tau)}\rangle$ equals $u_{ijkl}$ contracted with $-\tfrac{1}{2}\langle f^\dagger_i(\tau)f^\dagger_j(\tau)f_k(\tau)\rangle$. This correlator contains the fully disconnected part, $\bar f_i^\dagger(\tau)\bar f_j^\dagger(\tau)\bar f_k(\tau)$, as well as two partially connected ones
\begin{align}
 -\bar f^\dagger_i(\tau)\langle f^\dagger_j(\tau) f_k(\tau)\rangle,\quad -\tfrac{1}{2}\bar f_k(\tau)\langle f^\dagger_i(\tau)f^\dagger_j(\tau)\rangle,
\end{align}
 where we used the anti-symmetry $u_{ijkl}=-u_{jikl}$.
 
 If the coefficients satisfy $U^\alpha_{il} \chi_{0,\alpha|\beta}^{-1}U^\beta_{jk} =-U^\alpha_{jl} \chi_{0,\alpha|\beta}^{-1}U^\beta_{ik}= u_{ijkl}$, the last term in Eq.~\eqref{eq:SDE_1p1PI} cancels the first partially connected part, while the remaining one is eliminated by introducing pairing fields. Taking three additional derivatives of Eq.~\eqref{eq:SDE_1p1PI} w.r.t. fermionic fields yields the following expression for $\Gamma_{ijkl}$,
\begin{multline}\label{eq:SD_I_4}
    \Gamma_{ijkl}(\tau_i,\tau_j,\tau_k,\tau_l)=-2u_{ijkl}\delta^{\tau_l}_{\tau_i}\delta^{\tau_l}_{\tau_j}\delta^{\tau_l}_{\tau_k} \\-\tfrac{1}{2}u_{i'j'k'l}\tfrac{\delta}{\delta \bar f^\dagger_i(\tau_i)}\tfrac{\delta}{\delta \bar f^\dagger_j(\tau_j)}\tfrac{\delta}{\delta \bar f_k(\tau_k)}G_{\boldsymbol{\psi}}^{i'j'k'}(\tau_l,\tau_l,\tau_l),
\end{multline}
where $G^{ijk}(\tau_i,\tau_j,\tau_k)\equiv \langle f^\dagger_i(\tau_i )f^\dagger_j(\tau_j) f_k(\tau_k)\rangle_c$ and the subscript indicates that it is a functional of $\boldsymbol{\psi}$ via the relations $G^{abc}=G^{aa'}G^{bb'}G^{cc'}I_{a'b'c'}$ and $G=(-\zeta I_2)^{-1}$. 

Evaluating the last term in Eq.~\eqref{eq:SD_I_4} naturally generates composite field Green's and vertex functions \footnote{Note that $\frac{\delta}{\delta\psi^a}I_{bc}=\frac{\delta\psi_{\bar\varphi}^{\mathbf{b}}}{\delta\bar\varphi^b}\frac{\delta\psi_{\bar\varphi}^{\mathbf{c}}}{\delta\bar\varphi^c}\Gamma_{a\mathbf{b}\mathbf{c}}$ because $\frac{\delta\psi_{\bar\varphi}^{\mathbf{d}}}{\delta\bar\varphi^d}\Gamma_{\mathbf{d}\bullet}=0$ (see the discussion above Eq.~\eqref{eq:firstDerivatives1PIfromGammaPsi}).}. The first few contributions are shown on the right of Fig.~\ref{fig:SBEformalism}, and they are clearly interaction-irreducible. By further inspection of higher-order terms and the structure of the composite field irreducible vertices, one finds that $\Gamma_{ijkl}$ consists exclusively of interaction-irreducible diagrams.

\section{1PI vertices via Green's functions of composite fields}\label{sec:SymmetricEstimators}

If a non-perturbative method to calculate $n$-point expectation values is available, the standard way to obtain the 1PI vertices is to compute them in terms of the amputated correlation functions using the tree expansion (see Eqs.~\eqref{eq:treeExpansion_G^3}--\eqref{eq:treeExpansion_rule}). When computing a 1PI vertex numerically in this manner, the uncertainty in the inversion of $G$ arising from inexact numerical data contributes for each external index. This can become especially severe in real-frequency calculations \cite{lihm2024symmetric}. To reduce the effect of the amputations, one may instead use an alternative formula (called an estimator) for the 1PI vertex expressed solely in terms of the correlation functions and the self-energy \cite{hafermann2012improved, kaufmann2019symmetric, lihm2024symmetric}. At the end of this section, we present a simple rule (Fig.~\ref{fig:estimators}) to obtain such an estimator from the expression for the 1PI vertex in terms of the amputated correlation functions. As mentioned in the introduction, for theories with only cubic interactions, such as QED, this rule is well-known~\cite{Schwartz2014Quantum}. 

Let us start by recalling Eq.~\eqref{eq:npVertex_fromGamma} for the $n$-point 1PI vertex and writing it in condensed form, extended to fermions:
\begin{align}\label{eq:npVertex_fromGamma'}
    \Gamma^\mathrm{1PI}_n=S_n+\Big (\frac{\delta}{\delta \psi} +{\frac{\delta \mathcal{S}_{\bullet}}{\delta \bar\varphi}} \chi^{\bullet|\bullet}\frac{\delta}{\delta G^{\bullet}} \Big )^{n-1} \frac{\delta \mathcal{S}_{\bullet}}{\delta \bar\varphi} G^\bullet,
\end{align} 
where $\chi^{\bullet|\bullet}\equiv(-\gamma^\bullet_\bullet\Gamma_{\bullet\bullet})^{-1}$. This holds for theories with both cubic and quartic interactions. To simplify the derivations, we assume that for fermionic systems ($\zeta=-1$) only even correlation and vertex functions are nonzero.

The right side of Eq.~\eqref{eq:npVertex_fromGamma'} can be evaluated by first computing all derivatives $\frac{\delta}{\delta\psi}$, which act only on ${\frac{\delta \mathcal{S}_{\bullet}}{\delta \bar\varphi}}$ containing the external index (since $\frac{\delta}{\delta \psi}{\Gamma_{\bullet\bullet\dots}}=\frac{\delta}{\delta \psi}{\Lambda_{\bullet\bullet\dots}}=0$). The remaining derivatives are of the form $\frac{\delta^m \mathcal{S}_{\bullet}}{\delta \bar\varphi^m}\chi^{\bullet|\bullet}\frac{\delta}{\delta G^\bullet}$ and can be treated as follows.

Analogously to $G\frac{\delta}{\delta\bar\varphi}=\frac{\delta}{\delta J_1}$, Eq.~\eqref{eq:K2_general} yields
\begin{align}\label{eq:G^bullet_derivative_via_J_bullet}
\chi^{\bullet|\bullet}\frac{\delta}{\delta G^{\bullet}} &={\frac{\delta G^\bullet}{\delta \bar\phi^{(\bullet)}}}\Big (\frac{\delta \bar{\phi}^{(\bullet)}}{\delta J_{(\bullet)}}-G^{(\bullet)1}G^{-1}\frac{\delta \bar{\phi}^{(\bullet)}}{\delta J_1} \Big )\frac{\delta}{\delta \bar{\phi}^{(\bullet)}}\nonumber\\ &={\frac{\delta G^\bullet}{\delta \bar\phi^{(\bullet)}}} \Big (\frac{\delta}{\delta J_{(\bullet)}}-G^{(\bullet)1}G^{-1}\frac{\delta}{\delta J_1} \Big ),
\end{align}
where we added a zero term ${\frac{\delta G^\bullet}{\delta \bar\phi^{(\bullet)}}}(\frac{\delta \bar{\phi}}{\delta J_{(\bullet)}}-G^{(\bullet)1}G^{-1}\frac{\delta \bar{\phi}}{\delta J_1})\frac{\delta}{\delta \bar{\phi}}$ to the expression and applied the identity $\frac{\delta \bar{\phi}^{\mathbf{a}}}{\delta J_{\mathbf{b}}}\frac{\delta}{\delta \bar{\phi}^{\mathbf{a}}}=\frac{\delta}{\delta J_{\mathbf{b}}}$. Contracting \eqref{eq:G^bullet_derivative_via_J_bullet} with $\frac{\delta^m \mathcal{S}_{\bullet}}{\delta \bar\varphi^m}$ naturally introduces the following composite fields and their corresponding Green's functions.

\begin{figure}
    \centering
    \includegraphics[width=1\linewidth]{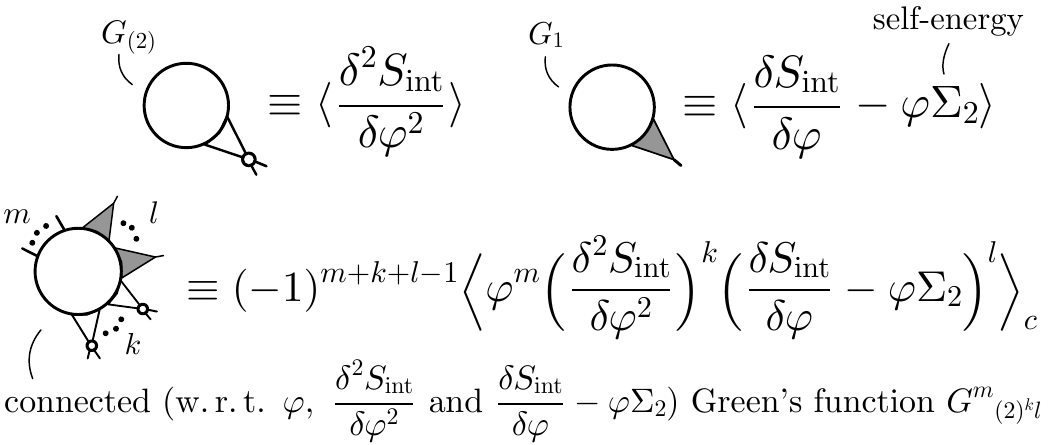}
    
    \caption{Definitions of the composite field Green's functions as given by Eqs.~\eqref{general_G_(n)}, \eqref{general_G_1} and \eqref{eq:definition_G^m_(2)^kn-2k}.}
    \label{fig:compositeFieldGreensFunctions}
\end{figure}

The first composite field is $\frac{\delta}{\delta\varphi^{a_1}}\dots \frac{\delta}{\delta \varphi^{a_m}}S_{\mathrm{int}}$, where $S_\mathrm{int}[\varphi]$ is the interacting part of the action $S[\varphi]$. It naturally appears from the term containing the $J_{(\bullet)}$-derivative:
 \begin{align}\label{general_G_(n)}
   G_{(a_1\dots a_m)}\equiv \frac{\delta^m \mathcal{S}_{\bullet}}{\delta\bar\varphi^{a_1}\dots \delta\bar\varphi^{a_m}}{\frac{\delta G^\bullet}{\delta \bar\phi^{(\bullet)}}}\frac{\delta W}{\delta J_{(\bullet)}}=\langle \frac{\delta^m S_\mathrm{int}[\varphi]}{\delta\varphi^{a_1} \dots\delta\varphi^{a_m}}\rangle,
  \end{align}
where we evaluated $\frac{\delta^m \mathcal{S}_{\bullet}}{\delta\bar\varphi^{a_1}\dots \delta\bar\varphi^{a_m}}\frac{\delta G^\bullet}{\delta \bar\phi^{(b_1\dots b_k)}}=\frac{1}{k!}S_{a_1\dots a_mb_k\dots b_1}[0]\zeta^k$, using Eqs.~\eqref{eq:definition_F}, \eqref{2pCorrelationFunction} and \eqref{eq:3pCorrelationFunction}. In a theory with only cubic and quartic interactions, $G_{(m)}$ vanishes for $m>2$.

The second composite field arises when considering the whole expression~\eqref{eq:G^bullet_derivative_via_J_bullet} contracted with  $\frac{\delta \mathcal{S}_{\bullet}}{\delta \bar\varphi}$: 
\begin{align}
\frac{\delta}{\delta J^1} &\equiv \frac{\delta \mathcal{S}_{\bullet}}{\delta \bar\varphi}{\frac{\delta G^\bullet}{\delta \bar\phi^{(\bullet)}}} \Big (\frac{\delta}{\delta J_{(\bullet)}}-G^{(\bullet)1}G^{-1}\frac{\delta}{\delta J_1} \Big ) \nonumber\\ &=\frac{\delta \mathcal{S}_{\bullet}}{\delta \bar\varphi}{\frac{\delta G^\bullet}{\delta \bar\phi^{(\bullet)}}}\frac{\delta}{\delta J_{(\bullet)}}-\Sigma_2\frac{\delta}{\delta J_1}, \label{eq:definitionJ^1}
\end{align}
where we used the Schwinger-Dyson equation for the self-energy ${\Sigma_2}\equiv-G^{-1}-\zeta S_2[0]$ written in the following form (recalling Eq.~\eqref{general_G_(n)}):
\begin{align}\label{eq:asymetric_self-energy}
   {\Sigma_2}=-\langle\frac{\delta S_\mathrm{int}}{\delta\varphi}\varphi\rangle_c G^{-1}= \frac{\delta \mathcal{S}_{\bullet}}{\delta \bar\varphi}{\frac{\delta G^\bullet}{\delta \bar\phi^{(\bullet)}}}G^{(\bullet)1}G^{-1}.
\end{align}

We use the notation $\frac{\delta}{\delta J^a}$ with a single upper index in $J^a$ because, when applied to the generating functional $W$, the operator in \Eq{eq:definitionJ^1} acts like the derivative with respect to a source term that corresponds to the composite field $(\frac{\delta S_{\mathrm{int}}}{\delta \varphi^a}-\Sigma_{ab}\varphi^b)$. Explicitly,
\begin{align}\label{general_G_1}
 G_a\equiv \frac{\delta W}{\delta J^a}=\langle\frac{\delta S_{\mathrm{int}}}{\delta \varphi^a}-\Sigma_{ab}\varphi^b\rangle.
\end{align}
This also enables us to write \Eq{J_1Derivative} as
\begin{align}\label{eq:dphi_ito_J1}
    \frac{\delta}{\delta \bar\varphi} =  \frac{\delta}{\delta \psi} + \frac{\delta}{\delta J^1}.
\end{align} 

We finally extend the Green's function definition \eqref{eq:correlationFunction} to include the most important composite fields. The general connected (w.r.t. $\varphi, \frac{\delta^2 S_\mathrm{int}}{\delta\varphi^2}$ and $\frac{\delta S_\mathrm{int}}{\delta\varphi}-\varphi\Sigma_2$) Green's function ${G^m}_{(2)^kl}$ is defined as
\begin{align}\label{eq:definition_G^m_(2)^kn-2k}
    {G^m}_{(2)^kl}\equiv(-1)^{m+k+l-1}\Big\langle\varphi^m\Big(\frac{\delta^2 S_{\mathrm{int}}}{\delta\varphi^2}\Big)^k\Big(\frac{\delta S_{\mathrm{int}}}{\delta\varphi}-\varphi\Sigma_2\Big)^{l}\Big\rangle_c,
\end{align}
where the order of the indices matches the order of the fields in the expectation value. For example, ${G^a}_{(bc)}=-\langle\varphi^a\frac{\delta^2 S_\mathrm{int}}{\delta\varphi^b\delta\varphi^c}\rangle_c$.
The diagrammatic definitions of the Green's functions in \Eq{eq:definition_G^m_(2)^kn-2k} are summarized in Fig.~\ref{fig:compositeFieldGreensFunctions}. This general definition is expressive enough to describe both symmetric estimators and asymptotic classes.

For example, using Eq.~\eqref{eq:asymetric_self-energy} we obtain
\begin{align}\label{eq:G_2_via_S_int}
G_{2}=-\langle\frac{\delta S_\mathrm{int}}{\delta\varphi}\frac{\delta S_\mathrm{int}}{\delta\varphi}\rangle_c-\langle\frac{\delta S_\mathrm{int}}{\delta\varphi}\varphi\rangle_c G^{-1}\langle\varphi\frac{\delta S_\mathrm{int}}{\delta\varphi}\rangle_c,
\end{align}
implying that $G_{(2)}+G_2$ coincides with the right side of Eq.~\eqref{eq:selfEnergyEstimatorBosonic} for an even theory. The symmetric self-energy estimator for an even theory can then be written as
\begin{align}\label{eq:symmetric_self-energy}
    \zeta{\Sigma_{ab}}=G_{(ab)}+G_{ab}.
\end{align}

In what follows, we present two approaches for obtaining estimators for 1PI vertices. In Section \ref{subsec:asymptoticClasses}, following the procedure outlined above, we first evaluate the $\psi$-derivatives in Eq.~\eqref{eq:npVertex_fromGamma'} to yield the asymptotic class decomposition. We then derive simple rules to express each class in terms of Green's functions of composite fields. In Section \ref{subsec:estimators}, we show how to get estimators for 1PI vertices directly from their representation in terms of amputated Green's functions, as illustrated in Fig.~\ref{fig:estimators}.  For the remaining sections, we consider a purely bosonic system, so fermionic signs are omitted, and the index order in the Green's functions is irrelevant. It is straightforward to obtain corresponding results for fermions.

\subsection{Estimators for asymptotic classes}\label{subsec:asymptoticClasses}

To generalize the asymptotic class decomposition (see Eqs.~\eqref{eq:3pvertex} and \eqref{eq:Decomposition4p1PIviaClasses}) to 1PI vertices of arbitrary order, we start with \Eq{eq:npVertex_fromGamma'} and exploit the fact that the $\psi$ derivative acts only on $\frac{\delta \mathcal{S}_{\bullet}}{\delta \bar\varphi}$, thereby yielding $\frac{\delta^2 \mathcal{S}_{\bullet}}{\delta \bar\varphi^2}$. Higher derivatives of $\mathcal{S}_{\bullet}$ vanish for theories with at most quartic interactions. As a result, we can group all contributions to the 1PI vertices according to the number of the remaining $G^\bullet$-derivatives as
\begin{align}\label{eq:asymptoticClassDecomposition}
    \Gamma^{\mathrm{1PI}}_{n>2}= S_n + \sum^{n}_{m=\lceil n/2 \rceil} P_{c^n_{n-m}}\mathcal{K}_n^{m-1}, 
\end{align}
where
\begin{align}\label{eq:definition_asymptoticClass}
\mathcal{K}_n^{m-1} &\equiv \Big (\frac{\delta^2 \mathcal{S}_{\bullet}}{\delta \bar\varphi^2} \Big )^{n-m}\Big (\frac{\delta \mathcal{S}_{\bullet}}{\delta \bar\varphi} \Big )^{2m-n}\chi^{\bullet^{m}},\\
\chi^{\bullet^m}&\equiv \Big (\chi^{\bullet|\bullet}\frac{\delta}{\delta G^{\bullet}} \Big )^m \Omega[G^\bullet] \nonumber \\
&= \Big (\chi^{\bullet|\bullet}\frac{\delta}{\delta G^{\bullet}} \Big )^{m-2} \chi^{\bullet|\bullet}.\label{eq:Kn_definition}
\end{align}
The term $\mathcal{K}^{m-1}_n$ represents an asymptotic class of the $n$-point 1PI vertex, depending on $m-1$ external time (or frequency) arguments, assuming time-translational invariance and time-local bare interactions (the same reasoning applies to momentum dependence under the assumption of spatial locality). The counter $c^n_k=
\frac{n!}{2^k k! (n - 2k)!}
$ tracks the number of ways to choose $k$ unordered pairs from $n$ indices, and $\lceil x\rceil$ denotes the ceiling function of $x$.

Equation~\eqref{eq:Kn_definition} has a structure analogous to \Eq{eq:correlationFunction_condensed}. Thus, $\chi^{\bullet^m}$ admits its own tree expansion. For example,
\begin{align}\label{eq:treeExpansion_Kn}
     \chi^{\bullet^3} &=(\chi^{\bullet|\bullet})^3\Gamma_{\bullet^3},\\
     \chi^{\bullet^4} &=(\chi^{\bullet|\bullet})^4\Gamma_{\bullet^4} + P_3(\chi^{\bullet|\bullet})^2 \Gamma_{\bullet^3}\chi^{\bullet|\bullet}\Gamma_{\bullet^3}(\chi^{\bullet|\bullet})^2.
\end{align}

To express $\mathcal{K}^{m-1}_{n}$ in terms of Green's functions of composite fields, we begin from its definition \eqref{eq:definition_asymptoticClass} and use Eq.~\eqref{eq:G^bullet_derivative_via_J_bullet} to calculate $\chi^{\bullet^m}$ as
\begin{align}\label{eq:GreensFunctionReprOf_Kn}
    \chi^{\bullet^m}=\Big (\frac{\delta G^\bullet}{\delta \bar\phi^{(\bullet)}} \Big )^{m-2} \Big(\frac{\delta}{\delta J_{(\bullet)}}-G^{(\bullet)1}G^{-1}\frac{\delta}{\delta J_1}\Big )^{m-2}\chi^{\bullet|\bullet}.
\end{align}

To evaluate derivatives in Eq.~\eqref{eq:GreensFunctionReprOf_Kn}, we derive a simple rule below, Eq.~\eqref{J^1DerivativeOfG_n}, that generates the tree expansion in analogy with Eq.~\eqref{eq:treeExpansion_rule}. 
To this end, let us focus on
\begin{align}
\mathcal{K}_m^{m-1} = \Big (\frac{\delta \mathcal{S}_{\bullet}}{\delta \bar\varphi} \Big)^m\chi^{\bullet^m},    
\end{align}
from which $\chi^{\bullet^m}$ can be obtained by removing $\frac{\delta \mathcal{S}_{\bullet}}{\delta \bar\varphi}$ for each external index (this step is explained below). Using Eqs.~\eqref{eq:GreensFunctionReprOf_Kn} and \eqref{eq:definitionJ^1}, we find
\begin{align}\label{eq:GreensFunctionReprOf_Class_n}
    \mathcal{K}_m^{m-1}= \Big (\frac{\delta}{\delta J^1} \Big)^{m-2}\chi^{\bullet|\bullet} \Big(\frac{\delta \mathcal{S}_{\bullet}}{\delta \bar\varphi} \Big)^2.
\end{align}

For $m=2$, applying Eqs.~\eqref{eq:K2_general}, \eqref{general_G_(n)} and \eqref{eq:G_2_via_S_int} yields
\begin{align}\label{eq:Class1}
    \mathcal{K}_2^1=\chi^{\bullet|\bullet} \Big(\frac{\delta \mathcal{S}_{\bullet}}{\delta \bar\varphi} \Big )^2=G_{2}.
\end{align}

Equation \eqref{eq:GreensFunctionReprOf_Class_n} then reduces to
\begin{align}
    \mathcal{K}_m^{m-1}=\Big (\frac{\delta}{\delta J^1} \Big)^{m-2}G_2.
\end{align}

Since $G^1_1=0$, we have $\frac{\delta}{\delta J^1}G_2=G_3$, and therefore
\begin{align}\label{eq:Class2}
    \mathcal{K}_3^2=G_3.
\end{align}

We obtain the following rule for $\frac{\delta}{\delta J^1}{G^m_n}$:
\begin{multline}\label{J^1DerivativeOfG_n}
   \frac{\delta}{\delta J^1}{G^m_n} ={G^m_{n+1}} -P_n\frac{\delta}{\delta J^1}({\tfrac{\delta \mathcal{S}_{\bullet}}{\delta \bar\varphi}}{\tfrac{\delta G^\bullet}{\delta \bar\phi^{(\bullet)}}}G^{(\bullet)1}G^{-1}){G^{1+m}_{n-1}}\\={G^m_{n+1}}-P_n{G^1_2}G^{-1}{G^{1+m}_{n-1}},
\end{multline}
where only the lower external indices are permuted by $P_n$, and we have used the identity
\begin{align}
\frac{\delta}{\delta J^1}\Big ({\frac{\delta \mathcal{S}_{\bullet}}{\delta \bar\varphi}}{\frac{\delta G^\bullet}{\delta \bar\phi^{(\bullet)}}}\Big )=0.    
\end{align}

Noticing that the recursive relation \eqref{J^1DerivativeOfG_n} effectively takes the same form as Eq.~\eqref{eq:treeExpansion_rule}, we conclude that $\mathcal{K}_m^{m-1}$ is given by the sum of all tree diagrams constructed from $G^k_l$, where the internal upper indices are contracted via $(-G^{-1})$. 
For example, for $\mathcal{K}_4^3$ one finds
\begin{align}\label{eq:Class3}
    \mathcal{K}^3_4 &= G_4- P_3G_2^1 G^{-1} G^1_2.
\end{align}
From the rule \eqref{J^1DerivativeOfG_n}, it follows that each external index in $\mathcal{K}_m^{m-1}$ will appear as a lower index in a Green's function of the form $G^{\dots}_{\dots 1}$, referring to a field $\frac{\delta \mathcal{S}_{\bullet}}{\delta \bar\varphi}$ inside an expectation value (see Eqs.~\eqref{eq:definition_G^m_(2)^kn-2k}, \eqref{general_G_(n)} and \eqref{eq:asymetric_self-energy}). To obtain $\chi^{\bullet^n}$ from $\mathcal{K}_m^{m-1}$, we remove one contraction $\frac{\delta \mathcal{S}_{\bullet}}{\delta \bar\varphi}$ for each external index.
For instance, from $\mathcal{K}_2^{1}=G_2=\chi^{\bullet|\bullet}(\frac{\delta \mathcal{S}_{\bullet}}{\delta \bar\varphi})^2$ we remove two contractions $\frac{\delta \mathcal{S}_{\bullet}}{\delta \bar\varphi}$, recovering $\chi^{\bullet|\bullet}$.

The general asymptotic class $\mathcal{K}^{m-1}_n$ can be obtained from $\chi^{\bullet^m}$ (see definition in Eq.~\eqref{eq:definition_asymptoticClass}), or more simply from $\mathcal{K}^{m-1}_m$ by making the necessary replacements $\frac{\delta \mathcal{S}_{\bullet}}{\delta \bar\varphi}\rightarrow \frac{\delta^2 \mathcal{S}_{\bullet}}{\delta \bar\varphi^2}$. For each such replacement in $G_{\dots 1}^{\dots}$, Eqs.~\eqref{general_G_(n)}--\eqref{general_G_1} imply
\begin{align}\label{eq:ruleGeneralAsymptotic}
    G_{\dots 1}^{\dots }\rightarrow  G_{\dots(2)}^{\dots }-G_{(2)}^1G^{-1}G_{\dots }^{1\dots }.
\end{align}
For example, starting from $\mathcal{K}^1_2=G_{2}=\chi^{\bullet|\bullet}(\frac{\delta \mathcal{S}_{\bullet}}{\delta \bar\varphi})^2$, replacing one $\frac{\delta \mathcal{S}_{\bullet}}{\delta \bar\varphi}\rightarrow \frac{\delta^2 \mathcal{S}_{\bullet}}{\delta \bar\varphi^2}$ (equivalently, applying \eqref{eq:ruleGeneralAsymptotic} to one lower index and using $G^1_1=0$) yields
\begin{align}
G_{2}\rightarrow G_{(2)1}.    
\end{align}
Thus, 
\begin{align}\label{eq:Class13}
    \mathcal{K}^1_3=G_{(2)1}.
\end{align} 
Similarly, from the classes \eqref{eq:Class13} and \eqref{eq:Class2}, we find
\begin{align}\label{eq:Class14_and_Class24}
\mathcal{K}_4^1 &= G_{(2)(2)}-G_{(2)}^1G^{-1}G^1_{(2)},\\
\mathcal{K}_4^2 &= G_{(2)2}-G_{(2)}^1G^{-1}G^1_{2}.\nonumber
\end{align}

\begin{figure*}
    \centering
    
    \includegraphics[width=\linewidth]{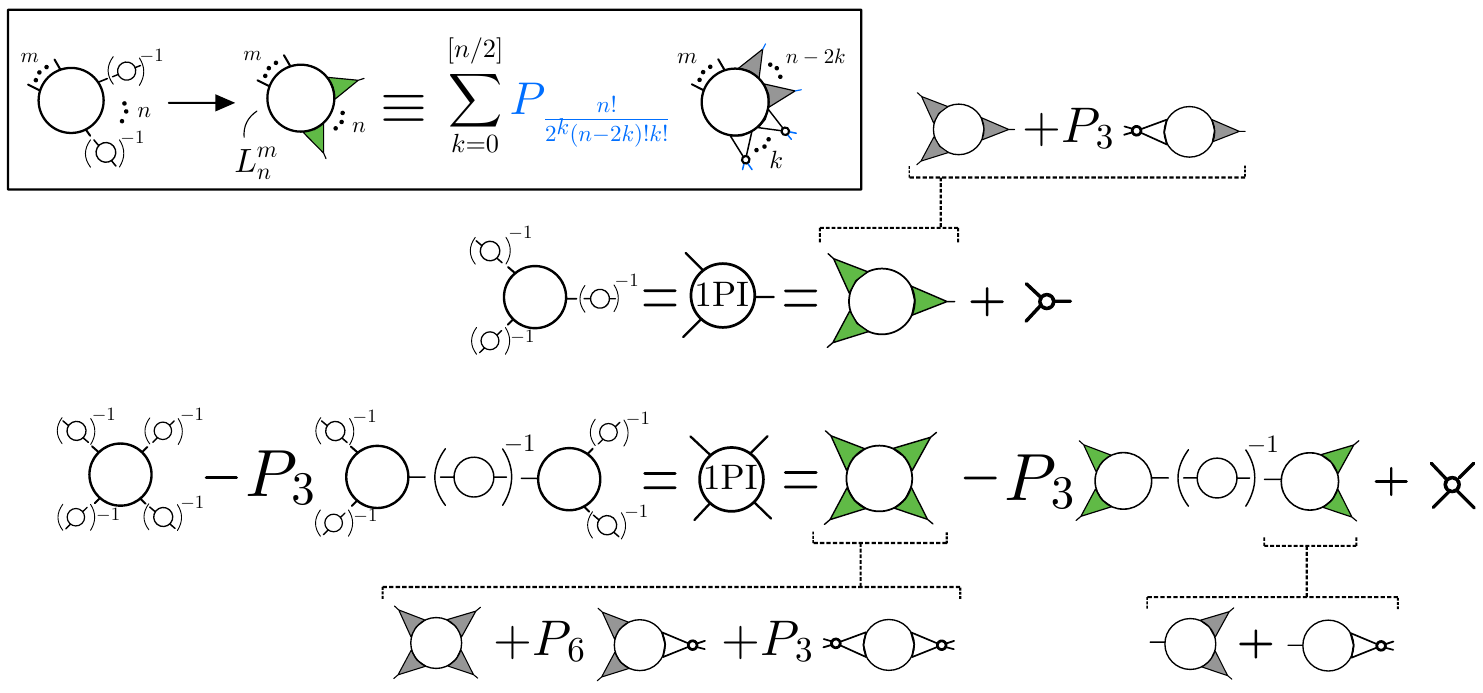}
    
    \caption{Three- and four-point 1PI vertices expressed in terms of the amputated correlation functions (left side) and their corresponding symmetric estimators (right side) (see Eqs.~\eqref{eq:3pVertex_Estimator}, \eqref{eq:estimator_4pVertex}) for a theory with cubic and quartic interactions. To obtain the estimator for an $n$-point vertex from the left-side expression, we add a bare vertex $S_n\equiv(\frac{\delta}{\delta\bar{\varphi}})^nS[\bar{\varphi}]$ and apply the rule in the upper-left corner, which also shows the diagrammatic definition of $L^m_n$ (see \Eq{eq:L^m_n_Definition}).  Further diagrammatic definitions are given in Fig.~\ref{fig:compositeFieldGreensFunctions}.}
    \label{fig:estimators}
\end{figure*}

To obtain the Green's function representation of the 1PI vertices in terms of composite fields, we substitute these results for the asymptotic classes into the decomposition \eqref{eq:asymptoticClassDecomposition}.
From $\Gamma^\mathrm{1PI}_3=S_3+P_3\mathcal{K}_3^1+\mathcal{K}^2_3$ and Eqs.~\eqref{eq:Class1}--\eqref{eq:Class2}, we find
\begin{align}\label{eq:3pVertex_Estimator}
    \Gamma^\mathrm{1PI}_3=S_3+L_3,
\end{align}
where $L_3\equiv P_3G_{(2)1}+G_3$. 
Substituting Eqs.~\eqref{eq:Class3} and \eqref{eq:Class14_and_Class24} into the asymptotic class decomposition of the four-point 1PI vertex~\eqref{eq:Decomposition4p1PIviaClasses} yields
\begin{align}\label{eq:estimator_4pVertex}
    \Gamma^\mathrm{1PI}_4=S_4+L_4-P_3L_2^1G^{-1}L^1_2,
\end{align}
where $L_4=G_4+P_6G_{(2)2}+P_3G_{(2)(2)}$ and $L^1_2=G^1_2+G^1_{(2)}$. For an even theory ($L^1_2=0$) with only quartic interactions, this result reduces to Eq.~(131) of Lihm et al.~\cite{lihm2024symmetric}\footnote{Specifically, we consider the fermionic system from Section~\ref{subsec:ExchangeOfCompositeParticles}. The operator versions of $\frac{\delta S_{\mathrm{int}}}{\delta f_i}$ and $\frac{\delta S_{\mathrm{int}}}{\delta f^\dagger_i}$ then correspond to $\hat q_\sigma=[\hat d_\sigma,\hat H_\mathrm{int}]$ and $(-\hat q^\dagger_\sigma)$ of \cite{lihm2024symmetric}, respectively.}. The estimator in Eq.~\eqref{eq:estimator_4pVertex} also takes a tree expansion form analogous to Eq.~\eqref{eq:Class3} or, equivalently, to the form for $\Gamma^\mathrm{1PI}_n$, when expressed in terms of amputated correlation functions via Eqs.~\eqref{eq:treeExpansion_G^3}--\eqref{eq:treeExpansion_G^4} (see Fig.~\ref{fig:estimators}). We will prove this result for vertices of arbitrary order in the next section.

\subsection{Estimators for 1PI vertices from their expression in terms of amputated Green's functions}\label{subsec:estimators}
Here, we show that the symmetric estimator for a general 1PI vertex can be obtained directly from its representation in terms of amputated correlation functions by applying the following simple rules (see Fig.~\ref{fig:estimators}): 
\begin{enumerate}
    \item Add the bare vertex.
    \item Replace each amputated correlation function containing $n$ external and $m$ internal indices, $G^{m+n}(G)^{-n}$, with $L^m_n$ defined as
\end{enumerate}
\begin{align}\label{eq:L^m_n_Definition}
    L^m_n\equiv G^m_n+{\sum^{\lfloor n/2\rfloor}_{k=1}}P_{c^n_k} G^m_{(2)^kn-2k},
\end{align}
where $c^n_k=\frac{n!}{2^k(n-2k)!k!}$ is the same counter as in Eq.~\eqref{eq:asymptoticClassDecomposition}, and only lower indices are permuted by $P_{c^n_k}$. The sum in Eq.~\eqref{eq:L^m_n_Definition} runs over all possible unique pairings of $n$ indices in $G^m_n$ into composite ones.

For an even theory, from the standard tree expansion in terms of amputated Green's functions,
\begin{align}
    \Gamma^{\text{1PI}}_6 = (G)^{-6}{G^6}-P_{10}(G)^{-3}({G^{4}}G^{-1}{G^4})(G)^{-3},
\end{align} we immediately obtain the estimator for the six-point 1PI vertex
\begin{align}\label{eq:estimator_6pVertex}
    {\Gamma^{\text{1PI}}_6}= {L_6}-P_{10}L_{3}^1G^{-1}L^1_{3},
\end{align}
where ${L_6}={G_6}+P_{15}{G_{(2)4}}+P_{45}{G_{(2)(2)2}}+P_{15}{G_{(2)(2)(2)}}$ and $L^1_3=G^1_3+P_3G^1_{(2)1}$. In the term $P_3 G^1_{(2)1}$, only the lower indices are permuted by $P_{3}$.

We begin the proof with the symmetric estimator for the self-energy, Eq.~\eqref{eq:symmetric_self-energy}, generalized to non-even theories in Appendix~\ref{app:estimator_general_vertex}:
\begin{align}\label{2pVertex}
    {\Gamma_2^{\text{1PI}}}={S_2}+L_2-\frac{1}{2}{S_4}\bar{\varphi}^2,
\end{align}
where $L_2={G_{2}}+{G_{(2)}}$.

We now seek a general formula for the functional derivative $\frac{\delta}{\delta\bar\varphi}L^m_n$. Since we are only interested in an expression for  $\frac{\delta}{\delta\bar\varphi}L^m_n$ in the context of calculating the full 1PI vertex, where $\frac{\delta}{\delta\psi}$ in $\frac{\delta}{\delta \bar{\varphi}}=\frac{\delta}{\delta\psi}+\frac{\delta}{\delta J^1}$ (see \Eq{eq:dphi_ito_J1}) acts only on ${\frac{\delta \mathcal{S}_{\bullet}}{\delta \bar\varphi}}$ containing the external index (see Eq.~\eqref{eq:npVertex_fromGamma'}), we ignore the terms where $\tfrac{\delta}{\delta\psi}$ acts on other parts of the expression when calculating $\frac{\delta}{\delta\bar\varphi}L^m_n$.
We denote equality under this assumption by the symbol $\implies$.

Following a derivation similar to that of Eq.~\eqref{J^1DerivativeOfG_n} (see App.~\ref{app:derivative_of_L}), we obtain
\begin{align}\label{derivativeOfL_n}
    \frac{\delta}{\delta \bar\varphi} L^m_n\implies L^m_{n+1}-P_n L_2^1G^{-1}L^{m+1}_{n-1}.
\end{align}
One can then check that the estimator for the three-point vertex, Eq.~\eqref{eq:3pVertex_Estimator}, follows from Eq.~\eqref{2pVertex} by applying  \eqref{derivativeOfL_n}\footnote{For the last term in Eq.~\eqref{2pVertex}, note that the external indices reside in $S_4$ and $\frac{\delta}{\delta J^1}\bar{\varphi}=0$, so it vanishes upon differentiation.} and using the identity $L^1_1=G^1_1=0$. Applying this rule a second time yields the estimator for $\Gamma^\mathrm{1PI}_4$ given in Eq.~\eqref{eq:estimator_4pVertex}.

For higher-order vertices, we need the derivative of the internal $G^{-1}$,
\begin{align}\label{derivativeOfg-1}
    \frac{\delta}{\delta \bar\varphi}G_{\bar\varphi}^{-1}=-G^{-1} \Big (\frac{\delta}{\delta J^1}G \Big )G^{-1}\implies-G^{-1}L_1^2G^{-1}.
\end{align}
Using this relation and the rule in Eq.~\eqref{derivativeOfL_n}, we can evaluate
\begin{align}
\Gamma_n^\mathrm{1PI}=S_n+\Big (\frac{\delta}{\delta\bar\varphi} \Big )^{n-3}L_3    .
\end{align}

We find that the symmetric estimator for a general 1PI vertex is obtained by adding the bare vertex to the sum of all tree diagrams constructed from $L^m_n$, where the internal indices $m$ are contracted via $(-G^{-1})$.
Thus, a 1PI vertex can be expressed in terms of correlation functions by summing all tree diagrams constructed either from:
\begin{enumerate}
    \item $G^{m+n}(G)^{-n}$ only, using $(-G^{-1})$ as the internal lines, which reproduces the standard representation in terms of amputated Green's functions; or
    \item $L^m_n$ only, using $(-G^{-1})$ as the internal lines, which yields the symmetric estimator.
\end{enumerate}
 This duality implies a simple transformation rule to obtain the estimator from the amputated representation: add the bare vertex and replace each amputated correlation function $G^{m+n}(G)^{-n}$ (having $n$ external and $m$ internal indices) with $L^m_n$ as defined in Eq.~\eqref{eq:L^m_n_Definition}. For the vertices $\Gamma^\mathrm{1PI}_3$ and $\Gamma^\mathrm{1PI}_4$, this correspondence is illustrated in Fig.~\ref{fig:estimators}. This concludes the proof.

\section{Summary and Outlook}
\label{sec:SummaryOutlook}

In this work, we have presented a flexible functional framework based on Legendre transforms and the effective action formalism. This approach naturally unifies and extends several established representations of the one-particle irreducible four-point vertex, including the single-boson exchange, parquet, asymptotic-class decompositions, and symmetric improved estimators. This extension proceeds along two main directions.

First, because the decompositions derived here are all based on the tree expansion (see Fig.~\ref{fig:results}), the formalism allows for the systematic construction of vertices of arbitrary order. This places the various four-point representations within a broader hierarchy and makes higher-order generalizations readily accessible.

Second, the framework naturally incorporates emergent composite degrees of freedom, which are generally non-trivial functions of fundamental field expectation values. Because the vertex building blocks are defined via functional derivatives of the composite effective action $\Gamma[\bar\varphi,\psi^\bullet]$, this flexibility suggests a systematic way to derive self-consistent approximation schemes based on the truncation of $\Gamma[\bar\varphi,\psi^\bullet]$ associated with an emergent effective field theory.

Beyond providing functional foundations for identities previously derived by diagrammatic means, the present approach offers a useful tool for future applications in regimes where diagrammatic counting becomes prohibitive. In particular, our six-point vertex decompositions may prove valuable for studying nonlinear responses and three-particle bound states, where the six-point vertex plays a central role \cite{Kappl2023NonlinearResponses, Rostami2021GaugeNonlinearResponse, Kappl2024ThreeParticleVertex, Eichmann2016Baryons}. Moreover, as illustrated on the right side of Fig.~\ref{fig:SBEformalism}, such nonlinear response functions also appear as building blocks in resummation schemes that go beyond the single-boson exchange approximation while retaining favorable scaling properties. The functional formulation developed here avoids the need for involved combinatorial arguments \cite{RibicPathIntegral}.

Further vertex decompositions may be derived by incorporating higher-order local
composite fields. As a proof of principle, we demonstrate in Appendix~\ref{app:beyond_SBE} that local bilinears and trilinears can serve as composite fields to represent the self-energies and one-particle irreducible vertices in terms of fermion-boson exchange processes. This generalization is especially relevant in quantum chromodynamics, where (i) mesons and baryons emerge as collective excitations associated with local quark bilinear and trilinear fields \cite{GellMann1964Hadrons}, and (ii) parameterizations of quark self-energies and 1PI vertex functions in terms of quark-meson exchanges have already proven effective \cite{Fischer2007Unquenching,Fischer2009GluonSelf, Eichmann2016Baryons, Miramontes2022Meson}.

Ultimately, because the derivations are formulated independently of a specific physical realization, the framework applies equally to quantum field theories across particle physics, condensed matter physics, and beyond.

\begin{acknowledgments}
We thank Jan von Delft, Marcel Gievers, Nepomuk Ritz, Anna Kauch, Tin Ribic, Fabian Kugler, Kilian Fraboulet and Markus Huber for helpful discussions and feedback on the manuscript. OS would further like to thank Valery Gusynin for referring to the literature on functional Legendre transforms. This work is part of the Munich Quantum Valley, supported by the Bavarian state government with funds from the Hightech Agenda Bayern Plus. OS acknowledges funding from
the Ministerium für Kultur und Wissenschaft des Landes
Nordrhein-Westfalen (NRW-Rückkehrprogramm).
\end{acknowledgments}

\appendix
\addcontentsline{toc}{section}{Appendix}

\section{Structure of the 3PI effective action and cluster decomposition formulas}
\subsection{Proof of Eq.~\eqref{effActionStructure}}\label{app:effActionStructure}

To derive Eq.~\eqref{effActionStructure}, we define $\phi^{(a_1\dots a_n)}\equiv\varphi^{a_1}\dots \varphi^{a_n}$ for $n=2,3$ and set $S[\varphi,J_{(\bullet)}]=S[\varphi]$ in the generating functional~\eqref{eq:W_definition}. The two-point correlation function (Eq.~\eqref{eq:2pCorrelationFunction}) can then be written as
\begin{align}\label{2pCorrelationFunction}
G^{ab}=\bar{\varphi}^a\bar{\varphi}^b  -\bar{\phi}^{(ab)}.  
\end{align}
Similarly, for $G^{abc}$ one finds
\begin{align}\label{eq:3pCorrelationFunction}
G^{abc}=2\bar{\varphi}^a\bar{\varphi}^b\bar{\varphi}^c-P_3\bar{\varphi}^{ a} \bar{\phi}^{(bc)}+\bar{\phi}^{(abc)},
\end{align}
where the permutation operator $P_n$ is defined in Sec.~\ref{subsec:1PImethods} (see Eq.~\eqref{eq:treeExpansion_G^4}). From relations \eqref{2pCorrelationFunction}--\eqref{eq:3pCorrelationFunction}, one can express $\bar{\phi}^{\mathbf{a}}$ as a functional of $\bar{\varphi}$ and $G^n$. This allows one to verify Eq.~\eqref{eq:linearRelation_phi_psi} and  $\tfrac{\delta\bar{\phi}^{\mathbf{a}}}{\delta \bar\varphi^d}=\langle \tfrac{\delta{\phi}^{\mathbf{a}}[\varphi]}{\delta \varphi^d}\rangle$.
We also need the relation $\langle\tfrac{\delta{\phi}^{\mathbf{a}}[\varphi]}{\delta \varphi^d}\rangle\gamma^{\mathbf{b}}_{\mathbf{a}} J_{\mathbf{b}}=-\langle\tfrac{\delta S[\varphi]}{\delta \varphi^d}\rangle$ [where the matrix $\gamma^{\mathbf{b}}_{\mathbf{a}}$ is defined in Eq.~\eqref{eq:definition_metric}], which follows from the standard identity $\int \mathrm{D}\varphi\, {\frac{{\delta}}{\delta\varphi^d}(e^{-S[\varphi]-J_{\mathbf{a}}\phi^{\mathbf{a}}[\varphi]})}=0$. We then obtain the Schwinger-Dyson equation (SDE)
\begin{multline}\label{eq:varphiDerivativeOfGamma}
 \tfrac{\delta\Gamma[\bar\varphi,G, G^3]}{\delta\bar\varphi^d}=\tfrac{\delta{\bar\phi}^{\mathbf{a}}}{\delta \bar\varphi^d}\tfrac{\delta\Gamma[\boldsymbol{\bar\phi}]}{\delta \bar\phi^{\mathbf{a}}}=-\tfrac{\delta{\bar\phi}^{\mathbf{a}}}{\delta \bar\varphi^d}\gamma^{\mathbf{b}}_{\mathbf{a}} J_{\mathbf{b}}=\langle \tfrac{\delta S[\varphi]}{\delta \varphi^d}\rangle\\=\tfrac{\delta S[\bar \varphi]}{\delta \bar\varphi^d}-\frac{1}{2}G^{cb}\tfrac{\delta S[\bar \varphi]}{\delta \bar\varphi^b\delta \bar\varphi^c\delta \bar\varphi^d}+\frac{1}{3!}G^{cba}\tfrac{\delta S[\bar \varphi]}{\delta \bar\varphi^a\delta \bar\varphi^b\delta \bar\varphi^c\delta \bar\varphi^d}.
\end{multline}
Formally integrating this equation w.r.t. $\bar\varphi$, with integration constant $\Lambda[G,G^3]$, yields Eq.~\eqref{effActionStructure}.

\subsection{Cluster decomposition formulas}\label{app:cluster_formula}

In this appendix, we decompose multi-particle Green's functions into connected parts and use Bethe-Salpeter-type equations to derive similar decompositions for the vertices $I_{n|...}$ defined in Sec.~\ref{sec:1PIEffectiveActionViaCompositeFields} from the 3PI effective action.

\subsubsection{Decomposition of multi-particle Green's functions}
To decompose $\chi^{n|m}$, defined by Eq.~\eqref{eq:K2_general}, into connected parts, we use $G^{\mathbf{a}\dots(\bullet)}=\tfrac{\delta}{\delta J_\mathbf{a}}\dots \bar\phi^{(\bullet)}$, in which $\bar\phi^{(\bullet)}$ is expressed via connected Green's functions (see Eqs.~\eqref{eq:2pCorrelationFunction} and \eqref{eq:3pCorrelationFunction}). The resulting decompositions can be written more compactly with the help of derivatives $\frac{\delta^m G^n_{\bar\varphi}}{\delta \bar\varphi^m}=(G^{-1}\tfrac{\delta}{\delta J_1})^mG^n$, which are easily evaluated.
For $n=m=2$, we have (see the second row of Fig.~\ref{fig:susceptibilities})
\begin{align}\label{eq:K^2|2_into_connected}
\chi^{ab|cd}=\tfrac{\delta^2 G^{ab}_{\bar\varphi}}{\delta\bar\varphi^{c'}\delta\bar\varphi^{d'}}G^{c'c}G^{d'd}-G^{ad}G^{bc}-\zeta G^{ac}G^{bd},
\end{align}
where
\begin{align}
\tfrac{\delta^2 G^{cd}_{\bar\varphi}}{\delta\bar\varphi^a\delta\bar\varphi^b}=(\Gamma^{\mathrm{1PI}}_{abhk}+\Gamma^{\mathrm{1PI}}_{ahf}G^{fe}\Gamma^{\mathrm{1PI}}_{ebk}+\zeta\Gamma^{\mathrm{1PI}}_{akf}G^{fe}\Gamma^{\mathrm{1PI}}_{ebh})G^{hc}G^{kd}.
\end{align}

For $n=2$ and $m=3$, the decomposition reads
\begin{align}\label{eq:K^3|2_into_connected}
    \chi^{abc|de}=\tfrac{\delta^2 G^{abc}_{\bar\varphi}}{\delta\bar\varphi^{d'}\delta\bar\varphi^{e'}}G^{d'd}G^{e'e}-P_2P_3G^{ad}\tfrac{\delta G^{bc}_{\bar\varphi}}{\delta\bar\varphi^{e'}}G^{e'e},
\end{align}
where the $P_2$ operator permutes the indices $d$ and $e$, and $P_3$ acts on $a,b,c$. 
Finally, for $n=m=3$, one obtains
\begin{align}\label{eq:K^3|3_into_connected}
    \chi^{3|3}(G)^{-3}=\tfrac{\delta^3 G^3_{\bar\varphi}}{\delta\bar\varphi^3}-P_3P_3\delta^1_1\tfrac{\delta^2 G_{\bar\varphi}}{\delta\bar\varphi^2}+6\delta^3_3,
\end{align}
where one $P_3$ operator permutes only the lower indices and the other $P_3$ acts on the upper indices only. For an even theory, this equation (multiplied by $(G)^3$ from the right) is depicted in the upper right of Fig.~\ref{fig:BSEs_nPI} and we obtain for $\tfrac{\delta^3 G^3_{\bar\varphi}}{\delta\bar\varphi^3}$:
\begin{multline}\label{eq:G^3_3}
\tfrac{\delta^3 G^{abc}_{\bar\varphi}}{\delta\bar\varphi^{d'}\delta\bar\varphi^{e'}\delta\bar\varphi^{f'}}G^{d'd}G^{e'e}G^{f'f}=G^{abcdef}\\-G^{ag}\tfrac{\delta^2 G^{bc}_{\bar\varphi}}{\delta\bar\varphi^g\delta\bar\varphi^h}G^{hh'}\tfrac{\delta^2 G^{de}_{\bar\varphi}}{\delta\bar\varphi^{h'}\delta\bar\varphi^k}G^{kf}.
\end{multline}

\subsubsection{Decomposition of irreducible kernels}
Here, we show how to decompose the kernels $I_{1|2|3}$ and $I_{|n|n}\equiv \tfrac{\delta^2\Gamma}{\delta G^n\delta G^n}$ (with $n=2,3$) into connected parts using Bethe-Salpeter-type equations for an even theory. In particular, the decomposition~\eqref{eq:I_1|2|3_decomposition} for $I_{1|2|3}$ is illustrated in the upper right of Fig.~\ref{fig:BSEs_nPI}. These ideas can be applied to other composite field vertices $I_{n|...}$.

Let us start with $I_{|2|2}$. Multiplying both sides of Eq.~\eqref{eq:K^2|2_into_connected} by $\tfrac{\delta^2\Gamma}{\delta G\delta G}$ and substituting Eq.~\eqref{eq:BS_0} for $\tfrac{\delta^2 G_{\bar\varphi}}{\delta\bar\varphi^2}$ yields
\begin{align}\label{eq:I_|2|2_into_connected}
    \frac{\delta^2\Gamma}{\delta G^{aa'}\delta G^{bb'}}G^{a'c}G^{b'd}=-\tfrac{1}{2}I_{aa'|bb'}G^{a'c}G^{b'd}+\delta^{cd}_{ab},
\end{align}
which shows that $I_{2|2}$ is the connected part of $\tfrac{\delta^2\Gamma}{\delta G\delta G}$.
  
The decomposition for $I_{|3|3}$ is similarly derived from \eqref{eq:K^3|3_into_connected}. First, from $\frac{\delta\chi^{ab|cde}_{\bar\varphi}}{\delta\bar\varphi^f}=\chi^{ab|a'b'}I_{f|a'b'|c'd'e'}\chi^{c'd'e'|cde}$ and \eqref{eq:K2_general} we obtain
\begin{align}\label{eq:I_1|2|3_via_amputations}
I_{1|2|3}=I_{|2|2}\frac{\delta \chi^{2|3}_{\bar\varphi}}{\delta\bar\varphi}I_{|3|3}, 
\end{align}
where we employed $\Gamma_{\bullet\bullet}=I_{|\bullet|\bullet}$. Using Eqs.~\eqref{eq:K^3|2_into_connected} and \eqref{eq:K^3|3_into_connected} we get $G\frac{\delta \chi^{3|2}_{\bar\varphi}}{\delta\bar\varphi}=\chi^{3|3}+P_3G\chi^{2|2}$, which can be substituted into Eq.~\eqref{eq:I_1|2|3_via_amputations} to yield 
\begin{align}\label{eq:I_|3|3_decomposition}
    3G^{aa'}\tfrac{\delta^2\Gamma}{\delta G^{a'bc}\delta G^{def}}=-G^{aa'}I_{a'|bc|def}-\tfrac{1}{3}P_3\delta^a_d\tfrac{\delta^2\Gamma}{\delta G^{bc}\delta G^{ef}}.
\end{align}
We see that $I_{1|2|3}$ is the connected part of $\tfrac{\delta^2\Gamma}{\delta G^3\delta G^3}$. 

Let us further decompose the kernel  $I_{1|2|3}$ so the Bethe-Salpeter equations \eqref{eq:BS_0} and \eqref{eq:BS_n=3} (also shown in Fig.~\ref{fig:BSEs_nPI}) involve $I_{2|2}$ and $I_{3|3}$ only. First, from $I_{|3|3}\frac{\delta^3 G^3_{\bar\varphi}}{\delta\bar\varphi^3}=-I_{3|3}-P_3I_{1|\bullet|3}\frac{\delta^2 G^\bullet_{\bar\varphi}}{\delta\bar\varphi^2}$ (see Eq.~\eqref{eq:BS_n=3}) and $I_{|3|3}\frac{\delta^3 G^3_{\bar\varphi}}{\delta\bar\varphi^3}=-\frac{1}{6}P_6(G)^{-3}+3P_3I_{|3|1\bullet} \frac{\delta^2 G^\bullet_{\bar\varphi}}{\delta\bar\varphi^2}-6I_{|3|3}$ (see Eq.~\eqref{eq:K^3|3_into_connected}) we obtain
\begin{align}
I_{3|3}+P_3(I_{1|\bullet|3}+3I_{|3|1\bullet})\tfrac{\delta^2 G^\bullet_{\bar\varphi}}{\delta\bar\varphi^2}=6I_{|3|3}+\frac{1}{6}P_6(G)^{-3},
\end{align}
where $P_3$ permutes indices $1$ and $2$.
Next, we substitute \eqref{eq:I_|3|3_decomposition} and use Eqs.~\eqref{eq:BS_0} and \eqref{eq:K^2|2_into_connected} to find $I_{1\bullet |3}-\frac{1}{3}(P_3-1)P_3G^{-1}I_{\bullet|2}=-2I_{1|\bullet|3}$, where the leftmost $P_3$ operator permutes $\bullet=2$ with one index in $G^{-1}$ and the second $P_3$ permutes $2$ with another index in $G^{-1}$. This yields the decomposition for $I_{1|2|3}$, which, after contraction with $\tfrac{\delta^2 G^\bullet_{\bar\varphi}}{\delta\bar\varphi^b\delta\bar\varphi^c}$, becomes (see the upper right of Fig.~\ref{fig:BSEs_nPI})
\begin{align}\label{eq:I_1|2|3_decomposition}
\tfrac{\delta^2 G^{b'c'}_{\bar\varphi}}{\delta\bar\varphi^b\delta\bar\varphi^c}I_{a|b'c'|def}=\tfrac{1}{3}P_3\Gamma^\mathrm{1PI}_{fbch} G^{hh'}I_{ah'|de}-\tfrac{1}{2}\tfrac{\delta^2 G^{b'c'}_{\bar\varphi}}{\delta\bar\varphi^b\delta\bar\varphi^c}I_{ab'c'|def},
\end{align}
where $P_3$ permutes the indices $d,e,f$ only.

\section{Fermion-boson exchanges}\label{app:beyond_SBE}
In this appendix, we consider the general system with fermions $f_i(\tau), f_j^\dagger(\tau)$, introduced at the beginning of Sec.~\ref{subsec:ExchangeOfCompositeParticles}. In addition to the composite field given by Eq.~ \eqref{eq:definition_local_phi}, we include $\phi^{(\alpha i)}(\tau,\tau')\equiv \phi^{\alpha}(\tau) f_i(\tau')$ and its complex conjugate. Clearly, its propagator corresponds to the simultaneous exchange of a boson and a fermion, which appears, for example, in the tree expansion of the six-point 1PI vertex. For a mixed theory (like QCD), four-point 1PI vertices also acquire an analogous exchange term if two of the external indices are bosonic (e.g., gluonic).

To check whether we can simplify the structure of the composite effective action and 1PI vertices with this new field, we introduce the corresponding sources $J_{(\alpha i)}(\tau,\tau')$ and $ J^{\dagger}_{(\alpha i)}(\tau, \tau')$ into $W[J_\mathbf{a}]$ using the following transformation:
\begin{align}\nonumber
    J_{(\alpha)}(\tau)\rightarrow  J_{(\alpha)}(\tau) +\int \mathrm{d}\tau '[&J_{(\alpha i)}(\tau,\tau') f_i(\tau')\\&+J^{\dagger}_{(\alpha i)}(\tau,\tau') f^{\dagger}_i(\tau')].
\end{align}
In addition to the composite variable $\psi^\alpha(\tau)$ given by Eq.~\eqref{eq:definition_local_psi}, we define
\begin{align}\label{eq:definition_bilocal_psi}
    \psi^{\alpha i}(\tau,\tau')\equiv \frac{\delta}{\delta J_{(\alpha)}(\tau)}\bar f_i(\tau')=\bar\phi^{(\alpha i)}(\tau,\tau')- \psi^\alpha(\tau)\bar f_i(\tau'),
\end{align}
where $\bar\phi^{(\alpha i)}(\tau,\tau')=\frac{\delta W}{\delta J_{(\alpha i)}(\tau,\tau')}$. In condensed notation, $\psi^{\bullet}$ denotes a vector of the bilinear field $\psi^\alpha(\tau)$, the trilinear field $\psi^{\alpha i}(\tau, \tau')$ and its complex conjugate.

Before discussing 1PI vertices, let us show that such a choice naturally expresses the self-energy via fermion-boson exchange. To this end, we note that in the self-energy decomposition \eqref{eq:secondDerivative1PIfromGammaPsi}, $\Gamma^\mathrm{1PI}_{ab}=\Gamma_{ab}+\tfrac{\delta\psi^\bullet_{\bar\varphi}}{\delta\bar\varphi^a}\Gamma_{\bullet b}$, only the trilinear component $\psi^\bullet_{\bar\varphi}=\psi^{\alpha j}_{\bar\varphi}(\tau_1,\tau_2)$ in the last term contributes for $\bar\varphi^a=\bar f_i(\tau)$. Fermion-boson exchange appears in its derivative, 
\begin{align}
    \tfrac{\delta \psi^{\alpha j}_{\bar\varphi}(\tau_1,\tau_2)}{\delta\bar f_i(\tau)}=\int \mathrm{d}\tau_3 \mathrm{d}\tau_4 I_{ki|\beta}(\tau_3,\tau|\tau_4)\chi^{\beta|\alpha}(\tau_4,\tau_1)G^{jk}(\tau_2,\tau_3),
\end{align}
which is derived by differentiating $\psi_{\bar\varphi}^{\alpha j}(\tau_1,\tau_2)=\tfrac{\delta \bar\phi^{(\alpha)}(\tau_1)}{\delta J_j(\tau_2)}$ w.r.t. $\bar f_i(\tau)$ at $\psi^\bullet=\psi^\bullet_{\bar\varphi}$, using $\tfrac{\delta}{\delta J_1}=G\tfrac{\delta}{\delta\bar\varphi}$ and $\tfrac{\delta^2 \psi^\bullet_{\bar\varphi}}{\delta\bar\varphi^a\delta\bar\varphi^b}=\chi^{\bullet|\bullet}I_{ab|\bullet}$.

However, it is known that for a purely fermionic system with only quartic interactions, there is an analogous representation for the self-energy in which $\Gamma_{\bullet b}$ is a bare vertex. In this case, $\Gamma[\bar\varphi,\psi^\bullet]$ must have a simple $\bar\varphi$-dependence, which implies a simple explicit $\bar\varphi$-dependence of $\Gamma^\mathrm{1PI}[\bar\varphi]$ when expressed as the inverse Legendre transform of $\Gamma[\bar\varphi,\psi^\bullet]$.

\subsection*{Structure of the effective action and 1PI vertices}
To explore the structure of $\Gamma[\bar\varphi,\psi^\bullet]$, we evaluate its derivative as in Eq.~\eqref{eq:SDE_1p1PI}:
\begin{multline}\label{eq:f_derivative_of_Gamma_local_trilinear}
    \frac{\delta\Gamma}{\delta \bar f^\dagger_i(\tau)}=\langle \frac{\delta S[\varphi]}{\delta f^\dagger_i(\tau)}\rangle+I^0_{il|\beta}\langle \phi^{(\beta)}(\tau) f_l(\tau)\rangle\\+I_{il|\beta}^0\Big (\psi^{l\beta}(\tau, \tau)-\psi^\beta(\tau)\bar f_l(\tau)\Big ).
\end{multline}
For a purely fermionic system with quartic interactions, the choice $U^\alpha_{il} \chi_{0,\alpha|\beta}^{-1}U^{\beta}_{jk} =-U^\alpha_{jl} \chi_{0,\alpha|\beta}^{-1}U^{\beta}_{ik} =\tfrac{1}{2}u_{ijkl}$ allows one to cancel the first and second terms completely. Similarly to the 3PI action, we then integrate Eq.~\eqref{eq:f_derivative_of_Gamma_local_trilinear} and apply an inverse Legendre transform to obtain
\begin{align}\label{eq:1PI_via_local_Gamma}
    \Gamma^\mathrm{1PI}[\bar\varphi]=S_0[\bar\varphi]+\Omega[ \psi^\bullet_{\bar\varphi}],
\end{align}
where $S_0$ is the non-interacting part of the classical action and $\Omega\equiv \Lambda-\psi^\bullet \tfrac{\delta\Lambda}{\delta \psi^\bullet}$ with $\Lambda[\psi^\bullet]$ being the $\bar\varphi$-independent part of $\Gamma[\bar\varphi,\psi^\bullet]$.

This result is directly analogous to Eq.~\eqref{eq:1PIfromGamma}. As in the case of the 3PI action, it implies that 1PI vertices can be expressed via the derivatives of $\Gamma$ w.r.t. the composite fields as in Eqs.~\eqref{eq:3pvertex} and \eqref{eq:parquet}. Moreover, it allows us to relate different building blocks appearing in the decompositions of 1PI vertices. In particular, similarly to the relation \eqref{eq:I_4_via_I_3|3}, we have $I_{abcd}=I_{abc|\bullet}\tfrac{\delta\psi^{\bullet}_{\bar\varphi}}{\delta\bar\varphi^d}$, where $I_{abc|\bullet }$ satisfies Eq.~\eqref{eq:BS_n=3}.

In QCD, the structure~\eqref{eq:1PI_via_local_Gamma} does not apply: gluons cannot be integrated out to yield an effective four-point quark interaction. Nevertheless, Eq.~\eqref{eq:f_derivative_of_Gamma_local_trilinear} can be used to simplify the quark-field dependence partially if, for example, $I^0_{ij|\alpha}$ is a bare quark-gluon vertex and $\chi_0^{\alpha|\beta}$ is generalized to a bare gluon propagator \footnote{In this case, one can simplify Eq.~\eqref{eq:f_derivative_of_Gamma_local_trilinear} as follows. Using the Schwinger-Dyson equation $-J_\alpha(\tau')=\langle\tfrac{\delta S_0[\varphi]}{\delta A^\alpha(\tau')}\rangle+\langle \tfrac{\delta S_\mathrm{int}[\varphi]}{\delta A^\alpha(\tau')}\rangle$ for the gluon field $A^\alpha(\tau)$ in $0=\tfrac{\delta J_{\alpha}(\tau')}{\delta J_l(\tau)}$, we get an exact identity, $\langle A^\alpha (\tau'')f_l(\tau)\rangle_c=\int \mathrm{d}\tau'\chi_0^{\alpha|\beta} (\tau'',\tau')\langle \tfrac{\delta S_\mathrm{int}[\varphi]}{\delta A^\beta(\tau')}f_l(\tau)\rangle_c$, which implies that
\begin{align}
 \langle\tfrac{\delta S_\mathrm{int}}{\delta f^\dagger_i(\tau)}\rangle=-I^0_{il|\beta} \langle \phi^{(\beta)} (\tau)f_l(\tau)\rangle_c +\dots
\end{align}
where the ellipsis denotes terms with bare gluon vertices. Substituting this into the right side of Eq.~\eqref{eq:f_derivative_of_Gamma_local_trilinear}, we find that the contribution involving the expectation value of the quark trilinear, $\langle \phi^{(\beta)} (\tau)f^\dagger_j(\tau)\rangle$, is canceled.
}. It is therefore natural to explore whether a self-consistent scheme can be derived in the spirit of the 3PI calculations in Ref.~\cite{Williams2016QCD}. The advantage of our framework over standard $n$PI approaches is the capacity to account for higher-order exchange terms via lower-order building blocks, which are associated with renormalized propagators and interactions of emergent degrees of freedom (i.e., mesons and baryons).

\section{Estimators for the self-energy and general vertices}\label{app:estimator_general_vertex}
In this appendix, we derive the symmetric estimator for the self-energy~\eqref{2pVertex} and then prove Eq.~\eqref{derivativeOfL_n}. Throughout this section, we set $J_{(\bullet)}=0$, where $\bullet=(2,3)$.

\subsection{Symmetric self-energy estimator}
Using the relation in Eq.~\eqref{eq:linearRelation_phi_psi}, one can rewrite Eq.~\eqref{eq:1pVertex} as
\begin{align}\label{eq:1pVertex'}
    {\Gamma_1^{\text{1PI}}}={S_1}+(\bar\phi^{(\bullet)}-{\phi^{(\bullet)}}[\bar\varphi])\frac{\delta G^\bullet}{\delta \bar\phi^{(\bullet)}}\frac{\delta\mathcal{S}_\bullet}{\delta \bar\varphi}.
\end{align}

Applying $\frac{\delta}{\delta\bar\varphi}$ to both sides of Eq.~\eqref{eq:1pVertex'} yields
\begin{align}\label{selfEnergyEstimators}
    {\Gamma_2^{\text{1PI}}}={S_2}+\frac{\delta}{\delta\bar\varphi}\Big ((\bar\phi^{(\bullet)}-{\phi^{(\bullet)}})\frac{\delta G^\bullet}{\delta \bar\phi^{(\bullet)}}\frac{\delta\mathcal{S}_\bullet}{\delta \bar\varphi}\Big ) .
\end{align}

To obtain the symmetric estimator, we write $\frac{\delta}{\delta\bar\varphi}=\frac{\delta}{\delta {\psi}}+\frac{\delta}{\delta J^1}$ for the last two terms in Eq.~\eqref{selfEnergyEstimators}, which leads to
\begin{align}\label{almost2pVertex}
    {\Gamma_2^{\text{1PI}}}={S_2}+{G_{(2)}}+{G_{2}}-{\phi^{(\bullet)}}\frac{\delta G^\bullet}{\delta \bar\phi^{(\bullet)}}\frac{\delta^2\mathcal{S}_\bullet}{\delta \bar\varphi^2}.
\end{align}
Here, we used that $\frac{\delta}{\delta\psi}$ acts only on $\frac{\delta\mathcal{S}_\bullet}{\delta \bar\varphi}$ with the external index (see Eq.~\eqref{eq:npVertex_fromGamma'}), and that $\frac{\delta}{\delta J^1}(\frac{\delta\mathcal{S}_\bullet}{\delta \bar\varphi}\frac{\delta G^\bullet}{\delta \bar\phi^{(\bullet)}})=0$, so $\frac{\delta}{\delta J^1}(\frac{\delta\mathcal{S}_\bullet}{\delta \bar\varphi}\frac{\delta G^\bullet}{\delta \bar\phi^{(\bullet)}}{\phi^{(\bullet)}})=0$ and $\frac{\delta}{\delta J^1}(\frac{\delta\mathcal{S}_\bullet}{\delta \bar\varphi}\frac{\delta G^\bullet}{\delta \bar\phi^{(\bullet)}}{\bar\phi^{(\bullet)}})=G_2$. 

Using the identity ${S_2}-\frac{\delta^2\mathcal{S}_\bullet}{\delta \bar\varphi^2}\frac{\delta G^\bullet}{\delta \bar\phi^{(\bullet)}}{\phi^{(\bullet)}}={S_2[0]}+{\bar{\varphi}}{S_3[0]}$, one then obtains the symmetric self-energy estimator given in Eq.~\eqref{2pVertex}.

\subsection{Proof of Eq.~\eqref{derivativeOfL_n}}\label{app:derivative_of_L}
For the derivative $\frac{\delta}{\delta \psi}G_n$, we have
\begin{multline}\label{phiDerivativeOfG_n}
   \frac{\delta}{\delta \psi}G_n\implies P_n\frac{\delta^2\mathcal{S}_\bullet}{\delta \bar\varphi^2}\frac{\delta G^\bullet}{\delta \bar\phi^{(\bullet)}}(G^{(\bullet)}_{n-1}-G^{(\bullet)1}G^{-1} G^1_{n-1})\\=P_n{G_{(2)n-1}}-P_n G_{(2)}^1G^{-1}G^1_{n-1}.
\end{multline}
In total, using also Eq.~\eqref{J^1DerivativeOfG_n},
\begin{align}\label{J_1DerivativeOfG_n}
    \frac{\delta}{\delta\bar\varphi}{G_{n}}\implies G_{n+1}+P_n{G_{(2)n-1}}-P_n L_2^1G^{-1}G^1_{n-1},
\end{align}
where $L_2^1=G_{(2)}^1+G_2^1$. Note that this formula can easily be generalized to cases where additional upper indices or lower composite ones $(2)$ are added to $G_n$, since the derivative $\frac{\delta}{\delta\psi}$ will not act on $\frac{\delta^2\mathcal{S}_\bullet}{\delta \bar\varphi^2}$.

From Eq.~\eqref{J_1DerivativeOfG_n}, it is straightforward to prove Eq.~\eqref{derivativeOfL_n} by applying definition \eqref{eq:L^m_n_Definition}.

\section{Table of Symbols}
\label{App:TableOfSymbols}
\begin{widetext}
\begin{longtable}{@{} l p{0.55\linewidth} l @{}}
\toprule
Symbol & Meaning & First use / Definition \\
\midrule
\endfirsthead
\toprule
Symbol & Meaning &  First use / Definition \\
\midrule
\endhead
\midrule
\multicolumn{3}{r}{\emph{continued on next page}}\\
\endfoot
\bottomrule
\endlastfoot

\multicolumn{3}{@{}l}{\textbf{Index conventions}}\\
\midrule
\sym{a,b,c,\dots}{DeWitt multi-indices (including discrete and/or continuous quantum numbers) of the fundamental field $\varphi^a $.}{Above Eq.~\eqref{eq:correlationFunction}}

\sym{1,2,3,\dots}{Condensed notation indicating the number of indices, e.g., $G^2 \equiv G^{ab}$, $G^4 \equiv G^{abcd}$.}{Above Eq.~\eqref{eq:correlationFunction_condensed}}

\sym{\mathbf{a},\mathbf{b},\mathbf{c},\dots}{Bold indices collecting fundamental and composite components: $\mathbf{a}=(a,(\bullet))$.}{Above Eq.~\eqref{eq:Ja_barvarphia}}

\sym{(\bullet)}{Multi-index for the composite field $\phi^{(\bullet)}[\varphi]$ and its corresponding source $J_{(\bullet)}$ in the generating functional.}{Eq.~\eqref{eq:Ja_barvarphia}}

\sym{\bullet}{Multi-index for the composite field $\psi^\bullet$, defined as a function of $\bar\phi^{(\bullet)}$ and $\bar\varphi$.}{Below Eq.~\eqref{eq:K2}}

\sym{\bullet|\bullet|\ldots}{Multiple composite indices indicating derivatives w.r.t.\ several composite fields.}{Eq.~\eqref{eq:definition_I_bullet_n}}

\sym{\bullet^{n}}{Notation indicating that an object possesses $n$ generic composite indices.}{\Eq{eq:definition_asymptoticClass}}

\sym{\bar{\varphi}}{Expectation value of the fundamental field $\varphi$; as a subscript, it indicates an implicit functional dependence on $\bar{\varphi}$, e.g., $\psi_{\bar{\varphi}}^{\bullet}$.}{Eq.~\eqref{eq:1PI_via_Gamma}}

\multicolumn{3}{@{}l}{\textbf{Fields, expectation values and sources}}\\
\midrule
\sym{\varphi, \varphi^a}{Fundamental field variable in the path integral, where $\varphi \equiv \varphi^a$.}{\Eq{eq:Def_Expval}}

\sym{\phi^{\mathbf{a}}}{Generalized field including fundamental and composite components.}{\Eq{eq:W_definition}}

\sym{\bar{\phi}^\mathbf{a}}{Expectation value of $\phi^\mathbf{a}$ at $J_{(\bullet)}=0$.}{Below Eq.~\eqref{eq:1PcorrelationFunctions}}

\sym{\psi^\bullet}{Function of $\bar\phi^{(\bullet)}$ and $\bar\varphi$ chosen as the independent variable of the effective action $\Gamma[\bar\varphi,\psi^\bullet]$.}{Below \Eq{eq:K2}}

\sym{\boldsymbol{\bar{\phi}}}{Vector $\bar{\phi}^{\mathbf{a}}$ with a suppressed index: $\boldsymbol{\bar{\phi}}\equiv\begin{pmatrix}
    \bar\varphi,\bar{\phi}^{(\bullet)}
\end{pmatrix}$.}{Above \Eq{eq:1PcorrelationFunctions}}

\sym{\boldsymbol{\psi}}{Vector $\psi^{\mathbf{a}}$ with a suppressed index: $\boldsymbol{\psi}\equiv\begin{pmatrix}
    \psi,\psi^\bullet
\end{pmatrix}$.}{Below \Eq{eq:K2}}

\sym{J_a}{Source coupled linearly to the fundamental field as $J_a\varphi^a$.}{Eq.~\eqref{eq:GeneratingFunctionalW}}

\sym{J_{(\bullet)}}{Source coupled linearly to the composite field as $J_{(\bullet)}\phi^{(\bullet)}[\varphi]$.}{\Eq{eq:Ja_barvarphia}}

\sym{\frac{\delta}{\delta J^1}}{Derivative operator defined as $\frac{\delta}{\delta J^1}\equiv \frac{\delta \mathcal{S}_{\bullet}}{\delta \bar\varphi}{\frac{\delta G^\bullet}{\delta \bar\phi^{(\bullet)}}}(\frac{\delta}{\delta J_{(\bullet)}}-G^{(\bullet)1}G^{-1}\frac{\delta}{\delta J_1})$.}{Eqs.~\eqref{eq:definitionJ^1},\eqref{general_G_1}}

\sym{\psi}{Notation for the expectation value $\bar\varphi$ used to distinguish the variable of $\Gamma[\psi=\bar\varphi,\psi^\bullet]$ from the variable of $\Gamma^{\mathrm{1PI}}[\bar\varphi]$.}{Above Eq.~\eqref{eq:derivativeOfG^bullet'}}
\multicolumn{3}{@{}l}{\textbf{Permutation and statistics}}\\
\midrule

\sym{\zeta}{Statistical factor: $\zeta=+1$ for bosons and $\zeta=-1$ for fermions.}{Above Eq.~\eqref{eq:GeneratingFunctionalW}}

\sym{\gamma^{\mathbf{b}}_\mathbf{a}}{Matrix-valued sign factor $
\gamma^{\mathbf{b}}_\mathbf{a} =\zeta^{N_\mathbf{a}}\delta^{\mathbf{b}}_\mathbf{a}$,
where $N_\mathbf{a}$ is the number of field components denoted by $\mathbf{a}$.}{\Eq{eq:definition_metric}}

\sym{P_{n}}{Sum over all $n$ distinct permutations of the external indices $a,b,\dots$ in the associated term (with a factor $\zeta$ inserted if the permutation involves an interchange of an odd number of indices).}{Below \Eq{eq:treeExpansion_G^4}}

\multicolumn{3}{@{}l}{\textbf{Connected Green's functions and propagators}}\\
\midrule

\sym{G^{a_{1}\dots a_{n}}}{Connected $n$-point Green’s function: $G^{a_{1}\dots a_{n}} \equiv \frac{\delta^n W}{\delta J_{a_1}\cdots\delta J_{a_n}}$.}{\Eq{eq:correlationFunction}}

\sym{G^{n}}{Condensed notation for the connected $n$-point function: $G^n \equiv G^{a_{1}\dots a_{n}}$.}{\Eq{eq:correlationFunction_condensed}}

\sym{G^{ab},G^2,G}{Two-point connected Green’s function (full propagator): $G^{ab}=G^{2}=G$.}{\Eq{eq:2pCorrelationFunction}}

\sym{G^{-1}}{Inverse full propagator (matrix inverse of $G^{ab}$).}{\Eq{eq:1PI_Invertibility}}

\sym{G^\bullet}{Vector of connected two- and three-point Green's functions, where $\bullet=(2,3)$.}{Below \Eq{eq:K2}}

\sym{\chi^{\bullet|\bullet}}{Composite field propagator, defined as the inverse of $-\gamma^\bullet_\bullet\Gamma_{\bullet\bullet}$.}{\Eq{eq:K2_general}}

\sym{G_1}{Condensed notation for $G_1\equiv {G_a}\equiv \frac{\delta W}{\delta J^a}=\langle\frac{\delta S_{\mathrm{int}}}{\delta \varphi^a}-\Sigma_{ab}\varphi^b\rangle$.}{\Eq{general_G_1}}

\sym{G_{(n)}}{Condensed notation for $ {G_{(a_1...a_n)}} =\langle \frac{\delta^n S_{\mathrm{int}}[\varphi]}{\delta \varphi^{a_1}\dots \delta\varphi^{a_n}}\rangle$.}{\Eq{general_G_(n)}}

\sym{{G^m}_{(2)^kl}}{Connected Green's function w.r.t. $m$ fundamental fields $\varphi$, $k$ composite fields $\frac{\delta^2 S_{\mathrm{int}}}{\delta \varphi^2}$ and $l$ composite fields $\frac{\delta S_{\mathrm{int}}}{\delta \varphi}-\varphi\Sigma_2$.}{\Eq{eq:definition_G^m_(2)^kn-2k}}

\multicolumn{3}{@{}l}{\textbf{Effective actions and generating functionals}}\\
\midrule
\sym{W[J]}{Generating functional of connected Green's functions.}{Eq.~\eqref{eq:GeneratingFunctionalW}}

\sym{S[\bar{\varphi}]}{Classical action.}{\Eq{eq:classicalAction}}

\sym{\Gamma^{\mathrm{1PI}}[\bar\varphi]}{One-particle irreducible effective action (Legendre transform of $W[J]$).}{\Eq{eq:1PI_Definition}}

\sym{{\Gamma}[\bar\varphi,G,G^3]}{Three-particle irreducible (3PI) effective action.}{\Eq{effActionStructure}}

\sym{\Lambda[G,G^{3}]}{Generalization of the Luttinger-Ward functional for the 3PI effective action.}{\Eq{effActionStructure}}

\sym{\Omega[G,G^{3}]}{Legendre transform of $\Lambda$ with respect to $G,G^3$.}{\Eq{eq:OmegafromLambda}}

\sym{\Gamma[\boldsymbol{\bar{\phi}}]}{Composite effective action: $\Gamma[\boldsymbol{\bar{\phi}}] \equiv \Gamma\big[\bar\varphi,{\bar{\phi}^{(\bullet)}}\big]$.}{\Eq{eq:Gamma_Definition}}

\sym{\Gamma[\bar\varphi, \psi^\bullet]}{Composite effective action after a change of basis from $\bar\phi^{(\bullet)}$ to $\psi^\bullet$: $\Gamma[\boldsymbol{\psi}] \equiv \Gamma[\bar\varphi, \psi^\bullet] \equiv \Gamma\big[\bar\varphi,{\bar{\phi}^{(\bullet)}}[\boldsymbol{\psi}]\big]$.}{Eq.~\eqref{eq:Gamma_change_of_basis}}

\multicolumn{3}{@{}l}{\textbf{Vertices, asymptotic classes and estimators}}\\
\midrule

\sym{S_{a_{1}\dots a_{n}}[0]}{Bare vertex functions.}{\Eq{eq:classicalAction}}

\sym{S_{\mathrm{int}}}{Interacting part of the classical action.}{Below Eq.~\eqref{eq:K^3|3}}

\sym{\mathcal{S}_{a_{1}\dots a_{n}}[\bar\varphi]}{Derivatives of the classical action: $\mathcal{S}_{a_{1}\dots a_{n}}=\zeta^n S_{a_n\dots a_1}(-1)^{n-1}/n!$.}{\Eq{eq:definition_F}}

\sym{\Gamma^{\mathrm{1PI}}_{a_{1}\dots a_{n}}}{$n$-point vertex defined by $n$ functional derivatives of $\Gamma^{\mathrm{1PI}}$.}{\Eq{eq:1PIvertexDefinition}}

\sym{\Gamma_{\mathbf{a}_{1}\dots \mathbf{a}_{n}}}{$n$-point vertex defined by $n$ functional derivatives of either $\Gamma[\bar{\boldsymbol{\phi}}]$ or $\Gamma[\boldsymbol{\psi}]$, depending on the context.}{Eq.~\eqref{eq:CompositeVertexDefinition}}

\sym{\Sigma_{ab}}{Self-energy: $\Sigma_2=-G^{-1}-\zeta S_2[0]$.}{Above \Eq{eq:asymetric_self-energy}}

\sym{I_{a_{1}\dots a_{n}}}{Fully irreducible vertex w.r.t. the composite fields.}{\Eq{eq:definition_In}}

\sym{I_{a_{1}\dots a_{n}|\bullet}}{Irreducible vertex w.r.t. the composite fields. }{\Eq{eq:definition_I_bullet_n}}

\sym{I_{k|m}}{Condensed notation for $I_{a_{1}\dots a_{k}|\bullet}$ with $\bullet=b_1\dots b_m$, defined from the 3PI effective action. }{\Eq{eq:definitionI_22}}

\sym{\mathcal{K}^{m}_{n}}{Asymptotic class of an $n$-point 1PI vertex, depending on $m$ independent frequencies.}{\Eq{eq:definition_asymptoticClass}}

\sym{\chi^{\bullet^m}}{Composite field kernel with $m$ composite indices, admitting a tree expansion analogous to connected Green’s functions.}{\Eq{eq:Kn_definition}}

\sym{L^m_n}{Building blocks of symmetric estimators that replace amputated Green’s functions.}{\Eq{eq:L^m_n_Definition}}

\end{longtable}
\end{widetext}

\FloatBarrier

\bibliography{bibliography}

\end{document}